\documentclass[twocolumn,prd,preprintnumbers,superscriptaddress,altaffilletter,nofootinbib]{revtex4-1}

\usepackage{amsmath}
\usepackage{amssymb}
\usepackage{amsfonts}
\usepackage{graphicx}
\usepackage[colorlinks,citecolor=blue,urlcolor=red,bookmarks=false,hypertexnames=true]{hyperref}
\usepackage{lineno}
\usepackage[usenames,dvipsnames]{xcolor}
\usepackage{xspace}
\usepackage{caption}
\usepackage{subcaption}
\usepackage{acronym}
\usepackage{tikz}
\usetikzlibrary{arrows,shapes,trees,decorations.pathreplacing}
\usepackage{orcidlink}
\usepackage{enumitem}

\allowdisplaybreaks

\allowdisplaybreaks

\begin{document}

\title{Template bank for sub solar mass compact binary mergers in the fourth observing run of Advanced LIGO, Advanced Virgo, and KAGRA}

\author{Chad Hanna$^*$}
\affiliation{Department of Physics, The Pennsylvania State University, University Park, PA 16802, USA}
\affiliation{Institute for Gravitation and the Cosmos, The Pennsylvania State University, University Park, PA 16802, USA}
\affiliation{Department of Astronomy and Astrophysics, The Pennsylvania State University, University Park, PA 16802, USA}
\affiliation{Institute for Computational and Data Sciences, The Pennsylvania State University, University Park, PA 16802, USA}

\author{James Kennington$^*$ \orcidlink{0000-0002-6899-3833}}
\affiliation{Department of Physics, The Pennsylvania State University, University Park, PA 16802, USA}
\affiliation{Institute for Gravitation and the Cosmos, The Pennsylvania State University, University Park, PA 16802, USA}

\author{Wanting Niu$^*$ \orcidlink{0000-0003-1470-532X}}
\affiliation{Department of Physics, The Pennsylvania State University, University Park, PA 16802, USA}
\affiliation{Institute for Gravitation and the Cosmos, The Pennsylvania State University, University Park, PA 16802, USA}

\author{Shio Sakon$^*$ \orcidlink{0000-0002-5861-3024}}
\affiliation{Department of Physics, The Pennsylvania State University, University Park, PA 16802, USA}
\affiliation{Institute for Gravitation and the Cosmos, The Pennsylvania State University, University Park, PA 16802, USA}

\author{Divya Singh$^*$ \orcidlink{0000-0001-9675-4584}}
\affiliation{Department of Physics, The Pennsylvania State University, University Park, PA 16802, USA}
\affiliation{Institute for Gravitation and the Cosmos, The Pennsylvania State University, University Park, PA 16802, USA}
\affiliation{Department of Physics, University of California, Berkeley, CA 94720, USA}

\author{Shomik Adhicary \orcidlink{0009-0004-2101-5428}}
\affiliation{Department of Physics, The Pennsylvania State University, University Park, PA 16802, USA}
\affiliation{Institute for Gravitation and the Cosmos, The Pennsylvania State University, University Park, PA 16802, USA}

\author{Pratyusava Baral \orcidlink{0000-0001-6308-211X}}
\affiliation{Leonard E.\ Parker Center for Gravitation, Cosmology, and Astrophysics, University of Wisconsin-Milwaukee, Milwaukee, WI 53201, USA}

\author{Amanda Baylor \orcidlink{0000-0003-0918-0864}}
\affiliation{Leonard E.\ Parker Center for Gravitation, Cosmology, and Astrophysics, University of Wisconsin-Milwaukee, Milwaukee, WI 53201, USA}

\author{Kipp Cannon \orcidlink{0000-0003-4068-6572}}
\affiliation{RESCEU, The University of Tokyo, Tokyo, 113-0033, Japan}

\author{Sarah Caudill}
\affiliation{Department of Physics, University of Massachusetts, Dartmouth, MA 02747, USA}
\affiliation{Center for Scientific Computing and Data Science Research, University of Massachusetts, Dartmouth, MA 02747, USA}

\author{Bryce Cousins \orcidlink{0000-0002-7026-1340}}
\affiliation{Department of Physics, University of Illinois, Urbana, IL 61801 USA}
\affiliation{Department of Physics, The Pennsylvania State University, University Park, PA 16802, USA}
\affiliation{Institute for Gravitation and the Cosmos, The Pennsylvania State University, University Park, PA 16802, USA}

\author{Jolien D. E. Creighton \orcidlink{0000-0003-3600-2406}}
\affiliation{Leonard E.\ Parker Center for Gravitation, Cosmology, and Astrophysics, University of Wisconsin-Milwaukee, Milwaukee, WI 53201, USA}

\author{Becca Ewing}
\affiliation{Department of Physics, The Pennsylvania State University, University Park, PA 16802, USA}
\affiliation{Institute for Gravitation and the Cosmos, The Pennsylvania State University, University Park, PA 16802, USA}

\author{Heather Fong}
\affiliation{Department of Physics and Astronomy, University of British Columbia, Vancouver, BC, V6T 1Z4, Canada}
\affiliation{RESCEU, The University of Tokyo, Tokyo, 113-0033, Japan}
\affiliation{Graduate School of Science, The University of Tokyo, Tokyo 113-0033, Japan}

\author{Richard N. George \orcidlink{0000-0002-7797-7683}}
\affiliation{Center for Gravitational Physics, University of Texas at Austin, Austin, TX 78712, USA}

\author{Patrick Godwin \orcidlink{0000-0002-7489-4751}}
\affiliation{LIGO Laboratory, California Institute of Technology, MS 100-36, Pasadena, California 91125, USA}
\affiliation{Department of Physics, The Pennsylvania State University, University Park, PA 16802, USA}
\affiliation{Institute for Gravitation and the Cosmos, The Pennsylvania State University, University Park, PA 16802, USA}

\author{Reiko Harada}
\affiliation{RESCEU, The University of Tokyo, Tokyo, 113-0033, Japan}
\affiliation{Graduate School of Science, The University of Tokyo, Tokyo 113-0033, Japan}

\author{Yun-Jing Huang \orcidlink{0000-0002-2952-8429}}
\affiliation{Department of Physics, The Pennsylvania State University, University Park, PA 16802, USA}
\affiliation{Institute for Gravitation and the Cosmos, The Pennsylvania State University, University Park, PA 16802, USA}

\author{Rachael Huxford}
\affiliation{Department of Physics, The Pennsylvania State University, University Park, PA 16802, USA}
\affiliation{Institute for Gravitation and the Cosmos, The Pennsylvania State University, University Park, PA 16802, USA}

\author{Prathamesh Joshi \orcidlink{0000-0002-4148-4932}}
\affiliation{Department of Physics, The Pennsylvania State University, University Park, PA 16802, USA}
\affiliation{Institute for Gravitation and the Cosmos, The Pennsylvania State University, University Park, PA 16802, USA}

\author{Soichiro Kuwahara}
\affiliation{RESCEU, The University of Tokyo, Tokyo, 113-0033, Japan}
\affiliation{Graduate School of Science, The University of Tokyo, Tokyo 113-0033, Japan}

\author{Alvin K. Y. Li \orcidlink{0000-0001-6728-6523}}
\affiliation{RESCEU, The University of Tokyo, Tokyo, 113-0033, Japan}
\affiliation{Graduate School of Science, The University of Tokyo, Tokyo 113-0033, Japan}

\author{Ryan Magee \orcidlink{0000-0001-9769-531X}}
\affiliation{LIGO Laboratory, California Institute of Technology, Pasadena, CA 91125, USA}

\author{Duncan Meacher \orcidlink{0000-0001-5882-0368}}
\affiliation{Leonard E.\ Parker Center for Gravitation, Cosmology, and Astrophysics, University of Wisconsin-Milwaukee, Milwaukee, WI 53201, USA}

\author{Cody Messick \orcidlink{0000-0002-8230-3309}}
\affiliation{Leonard E.\ Parker Center for Gravitation, Cosmology, and Astrophysics, University of Wisconsin-Milwaukee, Milwaukee, WI 53201, USA}

\author{Soichiro Morisaki \orcidlink{0000-0002-8445-6747}}
\affiliation{Institute for Cosmic Ray Research, The University of Tokyo, 5-1-5 Kashiwanoha, Kashiwa, Chiba 277-8582, Japan}

\author{Debnandini Mukherjee  \orcidlink{0000-0001-7335-9418}}
\affiliation{NASA Marshall Space Flight Center, Huntsville, AL 35811, USA}
\affiliation{Center for Space Plasma and Aeronomic Research, University of Alabama in Huntsville, Huntsville, AL 35899, USA}

\author{Alexander Pace \orcidlink{0009-0003-4044-0334}}
\affiliation{Department of Physics, The Pennsylvania State University, University Park, PA 16802, USA}
\affiliation{Institute for Gravitation and the Cosmos, The Pennsylvania State University, University Park, PA 16802, USA}

\author{Cort Posnansky \orcidlink{0009-0009-7137-9795}}
\affiliation{Department of Physics, The Pennsylvania State University, University Park, PA 16802, USA}
\affiliation{Institute for Gravitation and the Cosmos, The Pennsylvania State University, University Park, PA 16802, USA}

\author{Anarya Ray \orcidlink{0000-0002-7322-4748}}
\affiliation{Leonard E.\ Parker Center for Gravitation, Cosmology, and Astrophysics, University of Wisconsin-Milwaukee, Milwaukee, WI 53201, USA}
\affiliation{Center of Interdisciplinary Education and Research in Astrophysics, Northwestern University, IL 60201, USA}

\author{Surabhi Sachdev \orcidlink{0000-0002-0525-2317}}
\affiliation{School of Physics, Georgia Institute of Technology, Atlanta, GW 30332, USA}
\affiliation{Leonard E.\ Parker Center for Gravitation, Cosmology, and Astrophysics, University of Wisconsin-Milwaukee, Milwaukee, WI 53201, USA}

\author{Stefano Schmidt \orcidlink{0000-0002-8206-8089}}
\affiliation{Nikhef, Science Park 105, 1098 XG, Amsterdam, The Netherlands}
\affiliation{Institute for Gravitational and Subatomic Physics (GRASP), Utrecht University, Princetonplein 1, 3584 CC Utrecht, The Netherlands}

\author{Urja Shah \orcidlink{0000-0001-8249-7425}}
\affiliation{School of Physics, Georgia Institute of Technology, Atlanta, GW 30332, USA}

\author{Ron Tapia}
\affiliation{Department of Physics, The Pennsylvania State University, University Park, PA 16802, USA}
\affiliation{Institute for Computational and Data Sciences, The Pennsylvania State University, University Park, PA 16802, USA}

\author{Leo Tsukada  \orcidlink{0000-0003-0596-5648}}
\affiliation{Department of Physics, The Pennsylvania State University, University Park, PA 16802, USA}
\affiliation{Institute for Gravitation and the Cosmos, The Pennsylvania State University, University Park, PA 16802, USA}
\affiliation{Department of Physics and Astronomy, University of Nevada, Las Vegas, 4505 South Maryland Parkway, Las Vegas, NV 89154, USA}
\affiliation{Nevada Center for Astrophysics, University of Nevada, Las Vegas, NV 89154, USA}

\author{Koh Ueno \orcidlink{0000-0003-3227-6055}}
\affiliation{RESCEU, The University of Tokyo, Tokyo, 113-0033, Japan}

\author{Aaron Viets \orcidlink{0000-0002-4241-1428}}
\affiliation{Concordia University Wisconsin, Mequon, WI 53097, USA}

\author{Leslie Wade}
\affiliation{Department of Physics, Hayes Hall, Kenyon College, Gambier, Ohio 43022, USA}

\author{Madeline Wade \orcidlink{0000-0002-5703-4469}}
\affiliation{Department of Physics, Hayes Hall, Kenyon College, Gambier, Ohio 43022, USA}

\author{Zach Yarbrough \orcidlink{0000-0002-9825-1136}}
\affiliation{Department of Physics and Astronomy, Louisiana State University, Baton Rouge, LA 70803, USA}

\author{Noah Zhang \orcidlink{0009-0003-3361-5538}}
\affiliation{School of Physics, Georgia Institute of Technology, Atlanta, GW 30332, USA}

\def\thefootnote{*}\footnotetext{These authors contributed equally to this work.}
\def\thefootnote{\arabic{footnote}}
\date{\today}
\setcounter{footnote}{0}

\begin{abstract}
Matched-filtering searches for gravitational-wave signals from compact binary mergers employ template banks which are a collection of modeled waveforms described by unique intrinsic parameters.
We present two banks designed for low-latency and archive sub-solar mass (SSM) searches in data from the fourth observing run of LIGO-Virgo-KAGRA, and demonstrate the efficacy of the banks via simulated signals. 
Further, we introduce several new methodological developments to the geometric placement algorithm~\cite{Hanna:2023, manifold}, including a more stable metric-estimation technique and a boundary-padding procedure. Together, these enhancements enable the method to work for exceedingly low component masses necessary for SSM bank production, representing the first successful application of the manifold-method to lower-mass regions of the compact binary parameter space.
The archive search bank contains a total of $3,452,006$ templates, and covers a mass parameter space of $0.2$ to $10\ M_\odot$ in the larger component and $0.2$ to $1.0\ M_\odot$ in the smaller component, the spin parameter space of $-0.9$ to $0.9$ for masses above $0.5$ $M_\odot$ and $-0.1$ to $0.1$ for masses below $0.5$ $M_\odot$, and the mass ratio parameter space of $1$ to $10$. The PSD used was from a week of the first half of the fourth observing run of Advanced LIGO, Advanced Virgo, and KAGRA, and the low frequency cutoff was set to $45$ Hz with a maximum waveform duration of $128$ seconds. The bank simulations performed using {\fontfamily{qcr}\selectfont SBank} have shown that the banks presented in this paper have sufficient efficacy for use in their respective searches.

\end{abstract}
\maketitle

\acrodef{LSC}[LSC]{LIGO Scientific Collaboration}
\acrodef{LVC}[LVC]{LIGO Scientific and Virgo Collaboration}
\acrodef{LVK}[LVK]{LIGO Scientific, Virgo and KAGRA Collaboration}
\acrodef{aLIGO}{Advanced Laser Interferometer Gravitational-Wave Observatory}
\acrodef{aVirgo}{Advanced Virgo}
\acrodef{LIGO}[LIGO]{Laser Interferometer Gravitational-Wave Observatory}
\acrodef{IFO}[IFO]{interferometer}
\acrodef{LHO}[LHO]{LIGO-Hanford}
\acrodef{LLO}[LLO]{LIGO-Livingston}
\acrodef{O2}[O2]{second observing run}
\acrodef{O1}[O1]{first observing run}
\acrodef{O3}[O3]{third observing run}
\acrodef{O3a}[O3a]{first half of the third observing run}
\acrodef{O3b}[O3b]{second half of the third observing run}
\acrodef{O4a}[O4a]{first half of the fourth observing run}
\acrodef{O4}[O4]{fourth observing run}
\acrodef{NSF}[NSF]{National Science Foundation}

\acrodef{SSM}[SSM]{subsolar-mass}
\acrodef{BH}[BH]{black hole}
\acrodef{BBH}[BBH]{binary black hole}
\acrodef{BNS}[BNS]{binary neutron star}
\acrodef{IMBH}[IMBH]{intermediate-mass black hole}
\acrodef{NS}[NS]{neutron star}
\acrodef{BHNS}[BHNS]{black hole--neutron star binaries}
\acrodef{NSBH}[NSBH]{neutron star--black hole binary}
\acrodef{PBH}[PBH]{primordial black hole}
\acrodef{CBC}[CBC]{compact binary coalescence}
\acrodef{GW}[GW]{gravitational wave}
\acrodef{GWH}[GW]{gravitational-wave}
\acrodef{DBH}[DBH]{dispasstive black hole binaries}

\acrodef{SNR}[SNR]{signal-to-noise ratio}
\acrodef{FAR}[FAR]{false alarm rate}
\acrodef{PSD}[PSD]{power spectral density}
\acrodef{GR}[GR]{general relativity}
\acrodef{NR}[NR]{numerical relativity}
\acrodef{PN}[PN]{post-Newtonian}
\acrodef{EOB}[EOB]{effective-one-body}
\acrodef{ROM}[ROM]{reduced-order model}
\acrodef{IMR}[IMR]{inspiral--merger--ringdown}
\acrodef{EOS}[EoS]{equation of state}
\acrodef{FF}[FF]{fitting factor}
\acrodef{FT}[FT]{Fourier Transform}

\acrodef{LAL}[LAL]{LIGO Algorithm Library}
\acrodef{GWTC}[GWTC]{Gravitational Wave Transient Catalog}

\newcommand{\PN}[0]{\ac{PN}\xspace}
\newcommand{\BBH}[0]{\ac{BBH}\xspace}
\newcommand{\BNS}[0]{\ac{BNS}\xspace}
\newcommand{\BH}[0]{\ac{BH}\xspace}
\newcommand{\NR}[0]{\ac{NR}\xspace}
\newcommand{\GW}[0]{\ac{GW}\xspace}
\newcommand{\SNR}[0]{\ac{SNR}\xspace}
\newcommand{\SSM}[0]{\ac{SSM}\xspace}
\newcommand{\aLIGO}[0]{\ac{aLIGO}\xspace}
\newcommand{\PSD}[0]{\ac{PSD}\xspace}
\newcommand{\GR}[0]{\ac{GR}\xspace}
\newcommand{\EOS}[0]{\ac{EOS}\xspace}
\newcommand{\LVC}[0]{\ac{LVC}\xspace}


\newcommand{\GSTLAL}{GstLAL\xspace}
\newcommand{\IMRPHENOMD}{IMRPhenomD\xspace}
\newcommand{\MANIFOLD}{{\fontfamily{qcr}\selectfont manifold}\xspace}
\newcommand{\SBANK}{{\fontfamily{qcr}\selectfont SBank}\xspace}

\section{Introduction}

The first \ac{GW} detection of a \ac{BBH} merger, GW150914~\cite{Abbott:2016, Abbott:2016b, LIGOScientific:2018mvr}, by Advanced \ac{LIGO} ~\cite{Harry:2010, Aasi:2015, Abbott:2019b, Abbott:2020c, Capote:2024rmo} and \ac{aVirgo} ~\cite{TheVirgo:2014hva, TheVirgo:2019}, brought about the era of \ac{GW} astronomy which has been furthered by subsequent detections of more \ac{BBH} mergers, \ac{BNS} mergers~\cite{Abbott:2017} \cite{Abbott:2020} and \ac{NSBH} mergers \cite{Abbott:2021c}. 
These observations have provided insight into the mass spectrum and population properties for merging compact objects along with the equation of state of dense matter \cite{Abbott:2023b} \cite{Abbott:2018b}. 

Since the \ac{O1}, the \ac{LVK} ~\cite{Abbott:2020b, KAGRAscience, KAGRA:2020} has published over $90$ \ac{GW} events from modelled and unmodelled searches for \acp{CBC} ~\cite{theligoscientificcollaboration2021gwtc3}.
The \GSTLAL-based compact binary search pipeline~\cite{GstLAL, Sachdev:2019, Hanna:2020, Cannon:2021, Ewing:2024, Tsukada:2024, Sakon:2024, Ray:2023} is one of the matched-filtering~\cite{Sathyaprakash:1991, Dhurandhar:1992mw, Owen:1995tm, Owen:1998dk} pipelines~\cite{Usman:2016, Nitz:2017, Nitz:2018, DalCanton:2021, Adams:2016, Aubin:2021, Andres:2022, Alléné:2025} which runs low-latency searches for \acp{CBC} by correlating the data with banks of templates that model the \ac{GW} signals expected from these binaries.
Beyond the searches for \acp{CBC} with component masses greater than a solar mass, searches for compact binaries with at least one component below $1$ $M_\odot$ have also been carried out in previous observing runs using data from Initial \ac{LIGO}~\cite{Abbott:2005pf, Abbott:2008ssm}, and Advanced \ac{LIGO} and \ac{aVirgo}~\cite{Abbott:2018, Abbott:2019, Abbott:2022, Abbott:2023, Nitz:2021a, Nitz:2021b, Nitz:2021c, Nitz:2022, Phukon:2021}, henceforth \ac{LIGO} and Virgo respectively, as well as in the ongoing observing run.

Since stellar processes are expected to produce only super-solar mass compact objects, searches for \acp{GW} from mergers of \ac{SSM} components are motivated by formation channels that extend beyond the standard model, including dark matter models that predict the existence of compact objects with masses below $1$ $M_\odot$ and \acp{PBH} that formed from over-densities in the early universe. 
Although the mass spectrum of \acp{PBH} can coincide with the spectrum of stellar \acp{BH}, the existence of \acp{BH} below $1$ $M_\odot$ is not excluded~\cite{Carr:2020, Carr:2021}.
The first \ac{BBH} event detected by \ac{LIGO} was hypothesized to be of primordial origin probing the long-standing question about the nature of dark matter~\cite{Bird:2016, Clesse:2016, Sasaki:2017}.
Other scenarios like the cooling and gravitational collapse of dissipative dark matter halos are predicted to produce \ac{SSM} \acp{BH} with appreciable detectable merger rates~\cite{Shandera:2018, Ryan:2022}.
\ac{GW} observations or lack thereof, especially of \ac{SSM} mergers, can constrain \ac{PBH} and \ac{DBH} populations and their dark matter abundances~\cite{Shandera:2018, Carr:2021, Abbott:2022, Abbott:2023,Singh:2021}. 
Other exotic objects like boson stars~\cite{Mielke:2000} and formation channels like imploding neutron stars~\cite{Kouvaris:2010vv, McDermott:2012, Bramante:2014zca, Singh:2022wvw, Bhattacharya:2023, Dasgupta:2021} from the dark matter sector can populate the \ac{BH} mass spectrum around a solar mass and also fall within the parameter space of \ac{SSM} searches. 
While there have been no detections of \ac{GW}s from \ac{SSM} systems, the most recent \ac{SSM} search by the \ac{LVK} using data from \ac{O3b}~\cite{Abbott:2023} have provided upper limits on the merger rate of compact binaries with at least one \ac{SSM} component.

\Ac{LVK}'s matched-filtering pipelines have operated super-solar and \ac{SSM} searches using template banks produced by stochastic template placing algorithms in previous observing runs ~\cite{Abbott:2021b, Abbott:2021a, theligoscientificcollaboration2021gwtc3, Abbott:2018, Abbott:2019, Abbott:2022, Abbott:2023}.
In \ac{O4}, individual \ac{LVK} matched-filtering pipelines produced independet template banks using \acp{PSD} that better reflect \ac{O4} detector sensitivities.
\GSTLAL used a binary-tree approach template placement algorithm ~\cite{manifold, Hanna:2023}, \MANIFOLD, to produce its template banks for \ac{O4} ~\cite{Sakon:2024}, including the \ac{SSM} template banks, to improve the computational efficiency of producing the template bank and downstream data products derived from the template bank.  
Producing \ac{SSM} template banks with \MANIFOLD required development work described in this paper, as the low-mass region was not investigated with \MANIFOLD previously. 
In this paper, we motivate using \MANIFOLD for the \ac{O4} \ac{SSM} template bank generation in section ~\ref{section:motivation}, present the methods to produce template banks that extend its component masses down to $0.2$ $M_\odot$ in section ~\ref{section:methods} and appendix section ~\ref{supp:ellipse}, describe the parameter space covered by the \ac{O4} \GSTLAL low-latency \ac{SSM} search in section ~\ref{section:design}, illustrate the template bank effectualness test results in section ~\ref{section:results}, and provide the template bank parameter descriptions and bank effectualness test results of the archival \ac{SSM} search in the appendix section ~\ref{supp:ll_bank}.
Through the template bank effectualness tests, we show that the \ac{SSM} template banks are capable of detecting potentially electromagnetically-bright systems with low masses, which provides opportunities for multimessnger detections when \ac{SSM} searches are deployed.  

\section{Motivation}
\label{section:motivation}

In this section, we motivate the generation of new template banks with a computationally efficient method of template bank generation, \MANIFOLD ~\citep{Hanna:2023}, for \GSTLAL's searches in \ac{O4} by addressing the following; the use of noise curves that better reflects \ac{O4} detector sensitivity, the computational efficiency of generating template banks, the computational efficiency of producing data products derived from the template banks, and the need for low-latency \ac{SSM} template banks. 
Here, \ac{SSM} binaries consist of at least one \ac{SSM} component. 

Firstly, \GSTLAL template banks for \ac{O4} were newly generated instead of reusing template banks from previous observing runs, as the sensitivities of the detectors have changed compared to \ac{O3} and because the sensitivities of the detectors dictate the discreteness of the template placement. 
Unlike previous observing runs where a common template bank was used across pipelines for each observing run ~\cite{Abbott:2021b, Abbott:2021a, theligoscientificcollaboration2021gwtc3, Abbott:2018, Abbott:2019, Abbott:2022, Abbott:2023}, pipelines produced independent template banks in \ac{O4}. 
\GSTLAL's super-solar mass template bank ~\cite{Sakon:2024} and the low-latency \ac{SSM} template bank (see appendix section ~\ref{supp:ll_bank}) used the projected \ac{O4} \ac{PSD}, as they were generated prior to the start of \ac{O4}. 
The \ac{SSM} archival search (i.e., ``offline" search), template bank for \GSTLAL was generated using a portion of the \ac{O4a} \ac{PSD} as it was produced after obtaining \ac{O4a} data. 

Secondly, \MANIFOLD was used to generate \GSTLAL's \ac{O4} template banks due to its computational efficiency in producing templates. 
As shown in ~\citet{Sakon:2024}, \MANIFOLD reduces the computational cost of generating template bank significantly compared to traditional stochastic template placement methods, e.g., from $\mathcal{O}\left(\rm{week}\right)$ to $\mathcal{O}\left(10 \rm{minutes}\right)$ for super-solar mass template banks. 
Such abilities to rapidly generate template banks enable developers to test and tune template bank configurations in shorter time scales than using stochastic methods. 

Thirdly, the generation of downstream data products that are derived from the template bank, such as the population model ~\citep{fong2018simulations} and $p_{\rm astro}$ model ~\citep{Ray:2023}, are computationally efficient with \MANIFOLD as the template volume elements needed to compute the population model and $p_{\rm astro}$ model are stored in the process of generating the template banks. 
The binary tree algorithm of \MANIFOLD enables computationally efficient checkerboarding of the template bank, which is a procedure that splits the template bank into two nearly identical halves such that two instances of analyses can be run on separate high-throughput computing clusters to mitigate failures ~\cite{Sakon:2024}. 

Fourthly, a computationally efficient \ac{SSM} template bank for low-latency \ac{SSM} searches is needed in \ac{O4} since for the first time in the history of \ac{GW} detections, the \ac{LVK} is searching for \ac{SSM} systems in low-latency. 
Due to the number of available computational resources that \GSTLAL has for low-latency analyses, \GSTLAL used a limited-parameter space template bank that extends down to $0.5$ $M_\odot$ and spins less than $0.3$ in magnitude for its low-latency analysis whereas its archival search template bank extended down to component masses of $0.2$ $M_\odot$ and spins spanned up to $0.9$ in magnitude, which matches \ac{SSM} template banks of previous observing runs. 
Template banks extending down to low-mass regions have not been generated with \MANIFOLD before and the original version of \MANIFOLD described in ~\citet{Hanna:2023} required new methods to produce a \ac{SSM} template bank with minimum component masses of $0.2$ $M_\odot$ while limiting the number of templates to $\mathcal{O}\left(10^6\right)$ and overcoming numerical and computational issues.
Such new methods, namely the metric estimation stability improvement and the boundary enhancement are illustrated in section ~\ref{section:methods}.  
Due to these reasons, all template banks used by \GSTLAL in \ac{O4} have been generated with \MANIFOLD, in addition to having consistency and compatibility across the \ac{O4} template banks. 

\section{Design}
\label{section:design}

The offline \ac{SSM} template bank covers the same parameter space as the \ac{SSM} template bank used in the \ac{O3} ~\cite{Abbott:2022, Abbott:2023}.
This ensures consistency with previous searches where the target search space was determined with consideration for the computational cost of \ac{SSM} searches.
Refer to App. ~\ref{supp:ll_bank} for the design and bank validity test results of the low-latency \ac{SSM} template bank.

\begin{table}
\begin{center}
	\begin{tabular}{ l | l }
		\hline
		Parameter & Offline SSM \\
		\hline
		\hline
		Primary mass, $m_1$ & $\in [0.2, 10 M_\odot]$  \\
		Secondary Mass, $m_2$ & $\in [0.2, 1 M_\odot]$  \\
		Mass ratio, $q=m_1/m_2$ & $\in [1, 10]$  \\
		Dimensionless spin, $s_{\rm {i,z}}$, for $m_i \in [0.2, 0.5 M_\odot]$ & $\lvert s_{\rm {i,z}} \rvert <  0.1$ \\
		Dimensionless spin, $s_{\rm {i,z}}$, for $m_i > 0.5 M_\odot$ & $\lvert s_{\rm {i,z}} \rvert <  0.9$ \\
		Lower frequency cut-off & 45 Hz \\
		Higher frequency cut-off & 1024 Hz \\
		Waveform approximant & IMRPhenomD \\
		Minimum match & 96.5 $\%$ \\ 
		Maximum waveform duration & 128 seconds \\
		\hline
		Total number of templates & 3452006 \\
		\hline
	\end{tabular}
	\caption{Parameter space of the \ac{O4} offline \ac{SSM} template bank.}
	\label{table:design_offline_bank} 
\end{center}
\end{table}

\begin{figure*}[htp]
    \centering
    \begin{minipage}{\textwidth}
        \begin{subfigure}[b]{8.6cm}
            \centering
            \includegraphics[width=8.6cm]{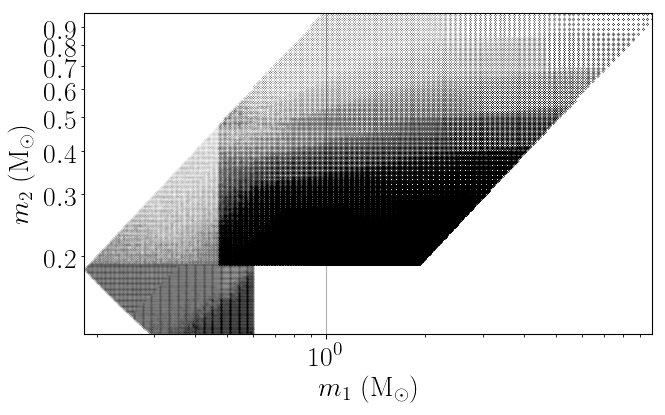}
            \caption{\protect Template placement in the $m_1$-$m_2$ space.} 
            \label{fig:template_m1_m2}
        \end{subfigure}
        \hfill
        \begin{subfigure}[b]{8.6cm}
            \centering
            \includegraphics[width=8.6cm]{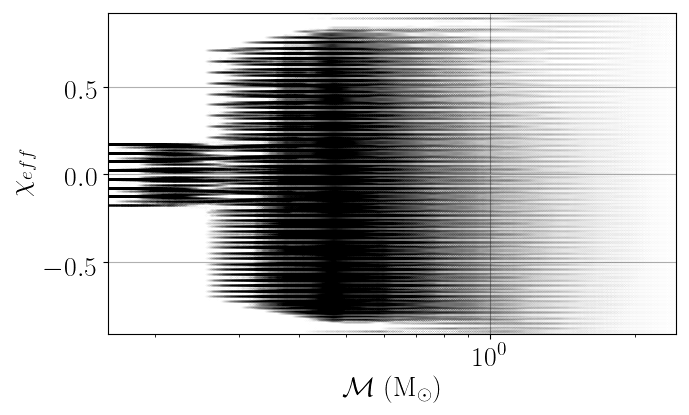}
            \caption{\protect Template placement in the $\mathcal{M}$-$\chi_{\rm {eff}}$ space.}
            \label{fig:template_mchirp_chi}
        \end{subfigure}   
            \caption{\protect Template placement of the \ac{O4} offline \ac{SSM} template bank.}
            \label{fig:template_placement}
    \end{minipage}
\end{figure*}

Table ~\ref{table:design_offline_bank} lists the parameter and configuration 
choices that were used to generate the template bank. 
Following previous \ac{LVK} \ac{SSM} searches, we define \ac{SSM} binaries as those that contain at least one component whose mass is below $1$ $M_\odot$. 
The templates are limited to spin-aligned systems to reduce computational costs and preserve consistency with previous searches 
\footnote{There have been studies on the impact of precession searches on the search sensitivies using the \GSTLAL algorithm and a modified version of \MANIFOLD ~\citep{Schmidt:2024cd, Schmidt:2024ab}. These studies showed an exponential increase in the number of templates. The \ac{SSM} searches presented in this paper exclude precession due to limited computational resources and to maintain consistency with the previous \ac{LVK} \ac{SSM} searches.}.
The $z$-components contribute to the effective spin, $\chi_{\rm {eff}}$: 
\begin{equation}
        \chi_{\rm {eff}} = \frac{m_1 \times s_{1, z} + m_2 \times s_{2, z}}{m_1 + m_2} . 
        \label{eq:chi_eff}
\end{equation}
The \MANIFOLD method ~\cite{Hanna:2022} assigns the same $z$-component spins for both components 
in the binary such $s_{1,z} = s_{2, z}$, and the value of $\chi_{\rm {eff}}$ is 
treated as $s_{\rm {i,z}}$.
The restricted component spins for low mass binaries was chosen to primarily reduce computational costs of filtering data by restricting the size of the bank~\cite{Abbott:2022, Abbott:2023}.
Since low-mass templates have long waveforms that are inspiral-dominated, we 
restrict the lower frequency cutoff to $45$ Hz for computational efficiency and to be consistent with previous \ac{LVK} \ac{SSM} searches ~\citep{Abbott:2018, Abbott:2019, Abbott:2022, Abbott:2023}.  
We further set a maximum duration cutoff ~\cite{Sakon:2024} at $128$ seconds, which is a feature to set a limit on the template duration by truncating the low-frequency end for to reduce memory usage.
Depending on the intrinsic parameters, the lower frequency cutoff on some templates in the bank can be higher than $45$ Hz.
We apply this feature both in the process of computing matches for template placement ~\cite{Hanna:2023} and in the process of performing SVD waveform ~\cite{Sakon:2024}, whose impact on SNR recovery is discussed in section~\ref{section:results}.
$128$ second is chosen as a number of the power of two closest to the duration of BNS waveforms that are detectable in the \ac{LIGO} sensitivity band ~\citep{Abbott:2017}.
We estimated the \ac{SNR} loss due to the maximum duration cutoff by comparing the optimal \acp{SNR} obtained with and without the maximum duration requirement. 
The optimal \ac{SNR} without the maximum duration cutoff was computed using $f_{\rm low} = 45$ Hz.
The \ac{SNR} loss increasingly affects systems with $\mathcal{M} \lesssim 0.4 M_\odot$, reaching a loss of nearly $16 \%$ for systems with $m_{\rm i} = 0.2 M_\odot$ $\left({\rm i} = 1, 2 \right)$. 
Templates with $m_{\rm i} \gtrsim 0.5 M_\odot$ are not impacted by the maximum duration cutoff.
More details on this feature can be found at ~\cite{Sakon:2024}.
The minimum match ~\cite{Owen:1995tm} was set to $96.5 \%$ which ensures that no more than approximately $10$ $\%$ of astrophysical signals can be missed while limiting the total number of templates in the template bank.
The detector's noise \ac{PSD} used to generate the template bank was obtained 
towards the end of \ac{O4a}, specifically, a week starting on December $29$, 
$2023$, at $00:00:00$ UTC, to accurately reflect the sensitivity of the detectors in \ac{O4a}.
The waveform approximant used to generate templates was \IMRPHENOMD 
~\cite{Khan:2016, Husa:2016}, which is a computationally efficient phenomenological 
waveform model that combines analytic \ac{PN} and \ac{EOB} methods for the inspiral 
portion of the waveforms and uses \ac{NR} simulations for the merger and ringdown 
portion of the waveform of spin-aligned \acp{BBH} in the frequency domain.    
As a result, the \ac{O4} offline \ac{SSM} template bank has $3452006$ templates. 
Fig. \ref{fig:template_placement} shows the placement of the templates of the 
\ac{O4} offline \ac{SSM} template bank 
\footnote{The templates extend beyond the designed $m_1$-$m_2$ or $\mathcal{M}$-
$\chi_{\rm {eff}}$ parameter space as a result of \MANIFOLD needing extra templates 
around the edges of the designed region to mitigate edge effects. Refer to 
~\cite{manifold} for details.}.
We plot the bank in the space of component masses, $m_1$-$m_2$ as well as in the 
parameter space of chirp mass, $\mathcal{M}$ (eq.~\ref{eq:mchirp}) and effective spin, $\chi_{\rm {eff}}$ (eq.~\ref{eq:chi_eff}) which are well-measured from \ac{GW} signals.
\begin{equation}
	\mathcal{M} = \frac{\left(m_1 m_2\right)^{3/5}}{\left(m_1 + m_2\right)^{1/5}} . 
	\label{eq:mchirp}
\end{equation}

\section{Methods}
\label{section:methods}

As mentioned above, gravitational wave searches rely on template banks, which are collections of waveforms generated from specific physical parameters. Using geometric techniques, we reframe the bank construction problem to be the efficient covering of a subset of the intrinsic parameter space \cite{Owen:1998dk, Hanna:2023}. Specifically, an ideal bank may be viewed as a network of points in the physical parameter space distributed such that for any signal within the specified constraints, at least one template in the bank is sufficiently ``near'' the signal to produce a \SNR above a target threshold \cite{Hanna:2023}. Following standard geometric approaches \cite{Owen:1998dk}, we define ``nearness'' using the intrinsic geometry of the parameter space, which will be detailed further in \ref{subsection:methods-manifold}. These techniques have been demonstrated to be effective for standard astrophysical parameters for compact binary systems \cite{Sakon:2024}; however, in the lower-mass regions of parameter space necessary for a \ac{SSM} search, problems arise in two areas of the treebank algorithm. First, numerical estimation of the manifold metric on which the geometric methods rely is more prone to instability. Second, there is insufficient template density near the boundary of the constraint sub-manifold leading to lower matches with extremal intrinsic parameters. We present two enhancements to the existing tree-bank algorithm that were necessary to produce a bank for the offline \ac{SSM} search, namely, boundary enhancement and neighborhood metric estimation.

\subsection{Geometric Techniques and \MANIFOLD}
\label{subsection:methods-manifold}

We begin our detailed discussion of methodology with a brief review of the geometric techniques used in \MANIFOLD that are essential in the production of this bank\cite{Hanna:2023}. The eponymous manifold $\mathfrak{M}$ is the set of possible waveforms for a given astrophysical event, such as CBC, which we parameterize using intrinsic-parameter coordinates such that a point $p$ in $\mathfrak{M}$ is equivalent to a choice of waveform parameters $\vec{\lambda}$. 
Following standard waveform analysis, we define a measure of similarity of waveforms, called the \textit{match}, which is essentially a correlation of the waveforms weighted inversely by detector noise and maximized over time-shifts and phases between the waveforms\cite{Hanna:2023}. The match is computed as
\begin{equation}
	\langle h | g \rangle \equiv  \max_{t, \phi} \left| \int_{f_{\mathrm{low}}}^{f_{\mathrm{high}}} \frac{\tilde{h}(f) \tilde(g)^*(f)}{S_n (f)} df \right| ,
	\label{eq:match}
\end{equation}
where $h$ and $g$ are complex-valued time-domain waveforms containing both the sine and cosine phases, $\tilde{h}$ is the Fourier transform of $h$, $\tilde{g}^*$ is the complex-conjugate of $\tilde{g}$, and $S_n$ is the \PSD. We define the \textit{mismatch}, denoted $\delta^2$ to be 
\begin{equation}
	\delta^2(h, g) \equiv 1 - \langle h | g \rangle.
	\label{eq:mismatch}
\end{equation}
Noting that the mismatch functions similarly enough to a metric distance, we then define a Riemannian metric via the standard Fisher method \cite{Owen:1998dk}
\begin{equation}
	g_{ij} = \partial_i \partial_j \delta^2,
	\label{eq:metric}
\end{equation}
which makes clear the notion of one signal being geometrically ``near" to another, with the local identification of the distance $d$ and the mismatch $\delta^2$
\begin{equation}
	d(h, g) \equiv \int_\gamma \sqrt{g_{\gamma(s)}\left[\gamma'(s), \gamma'(s)\right]} ds \approx \sqrt{\delta^2(h, g)},
	\label{eq:distance}
\end{equation}
where $\gamma$ is some smooth path connecting $h$ and $g$ in $\mathfrak{M}$ that minimizes the above integral.
In regions $\mathcal{U}\subset \mathfrak{M}$ where the metric can be assumed to be constant, the geometric approach allows for fast computation of the mismatch as $\delta^2(h, g) = g_{ij} \Delta \vec{\lambda}^i \Delta \vec{\lambda}^j$\cite{Hanna:2023}.

Further, the metric $g$ also permits a local notion of \textit{volume}. We define two types of volumes that are essential to the treebank algorithm: region volume and template volume. 
First, the volume of a for a small region, denoted  $V_\mathcal{U}$, is computed following the standard Riemannian definition. For hyperrectangular regions $\mathcal{U}$ where the metric can be assumed to be constant, the volume may be efficiently computed using \MANIFOLD as $V_\mathcal{U} = \sqrt{\left|g \right|}\prod_k \Delta \lambda^k$.
Second, the \textit{template volume} $V_{\delta^2}$ is defined in terms of a desired maximum mismatch between nearby templates\cite{Hanna:2023}, $\delta^2_{\mathrm{max}}$. Assuming an $n$-dimensional parameter space, the template volume is given by $V_{\delta^2} = \left[2 \sqrt{\delta^2_{\mathrm{max}} / n}\right]^n$.

The treebank algorithm used by \MANIFOLD produces a hyperrectangular partition of the parameter space in which each hyperrectanglular region represents a template in the bank whose parameters are the center of the region\cite{Hanna:2023}. The algorithm begins with one large hyperrectangle, iteratively splitting into smaller hyperrectangles following a binary tree technique, in which a region that is ``too large" is split along its longest side according to the local metric. The notion of ``too large" is defined where the volume of a hyperrectangular region $R$ exceeds the desired template volume\cite{Hanna:2023}, or $V_R > V_{\delta^2}$. 
As a result, for each template $h$ in the bank, there is a region $\mathcal{U}_h \subset \mathfrak{M}$ such that $h$ is the only template contained in $\mathcal{U}_h$ and $V_{\mathcal{U}_h} < V_{\delta^2}$\footnote{Note that the set $\left\{\mathcal{U}_h \right\}$ for all $h$ in the bank constitutes a covering of the subset of $\mathfrak{M}$ specified by the bank constraints, and is the precise sense in which bank construction is a covering problem.}.  

\subsection{Metric Estimation in \SSM Region}
A principal assumption of the \MANIFOLD method is the ability to estimate the local metric anywhere in the physically valid subset of the parameter space. Further, it is assumed that this metric has a Riemannian signature, or equivalently, all positive eigenvalues. These assumptions can break down in a region of the parameter space for several reasons, including decreased differentiability of the waveform function, degeneracy of one or more of the chosen coordinates, and others. 

For the \SSM region, exceedingly low masses caused both degeneracy of the $\chi_{\rm{eff}}$ parameter as well as reduced smoothness of the waveform approximant \IMRPHENOMD, resulting in more frequent failure of the numerical estimation routines. These failures were detected by checking for negative eigenvalues. To remedy these failures, we generalized the notion of the local metric being estimated at a point to being estimated within a neighborhood of a point. We define the neighborhood of a point $p$ in $\mathfrak{M}$ to be a local region $\mathcal{U}_p$ bounded by a maximum mismatch $\delta^2_{\mathcal{U}}$, such that for any other point $q$ in $\mathcal{U}_p$, $\delta^2(p, q) < \delta^2_{\mathcal{U}}$. When it may be assumed the metric is locally constant, the neighborhood $\mathcal{U}_p$ takes the shape of a hyperellipsoid centered at $p$, for more on this construction see App. \ref{supp:ellipse}. 

\begin{figure}[!h]
    \centering
    \begin{tikzpicture}[x=0.75pt,y=0.75pt,yscale=-0.8,xscale=0.8]
    \draw  [color={rgb, 255:red, 0; green, 0; blue, 0 }  ,draw opacity=0 ][fill={rgb, 255:red, 0; green, 0; blue, 0 }  ,fill opacity=1 ] (271,231) .. controls (271,228.24) and (273.24,226) .. (276,226) .. controls (278.76,226) and (281,228.24) .. (281,231) .. controls (281,233.76) and (278.76,236) .. (276,236) .. controls (273.24,236) and (271,233.76) .. (271,231) -- cycle ;
    \draw  [color={rgb, 255:red, 0; green, 0; blue, 0 }  ,draw opacity=0 ][fill={rgb, 255:red, 0; green, 0; blue, 0 }  ,fill opacity=1 ] (221,198) .. controls (221,195.24) and (223.24,193) .. (226,193) .. controls (228.76,193) and (231,195.24) .. (231,198) .. controls (231,200.76) and (228.76,203) .. (226,203) .. controls (223.24,203) and (221,200.76) .. (221,198) -- cycle ;
    \draw  [color={rgb, 255:red, 0; green, 0; blue, 0 }  ,draw opacity=0 ][fill={rgb, 255:red, 255; green, 26; blue, 0 }  ,fill opacity=1 ] (321,162) .. controls (321,159.24) and (323.24,157) .. (326,157) .. controls (328.76,157) and (331,159.24) .. (331,162) .. controls (331,164.76) and (328.76,167) .. (326,167) .. controls (323.24,167) and (321,164.76) .. (321,162) -- cycle ;
    \draw  [color={rgb, 255:red, 0; green, 0; blue, 0 }  ,draw opacity=0 ][fill={rgb, 255:red, 0; green, 0; blue, 0 }  ,fill opacity=1 ] (274,129) .. controls (274,126.24) and (276.24,124) .. (279,124) .. controls (281.76,124) and (284,126.24) .. (284,129) .. controls (284,131.76) and (281.76,134) .. (279,134) .. controls (276.24,134) and (274,131.76) .. (274,129) -- cycle ;
    \draw  [color={rgb, 255:red, 0; green, 0; blue, 0 }  ,draw opacity=0 ][fill={rgb, 255:red, 0; green, 223; blue, 75 }  ,fill opacity=1 ] (441,84) .. controls (441,81.24) and (443.24,79) .. (446,79) .. controls (448.76,79) and (451,81.24) .. (451,84) .. controls (451,86.76) and (448.76,89) .. (446,89) .. controls (443.24,89) and (441,86.76) .. (441,84) -- cycle ;
    \draw    (327.5,149) .. controls (328.47,137.36) and (317.21,128.54) .. (297.37,128.94) ;
    \draw [shift={(295.5,129)}, rotate = 357.27] [fill={rgb, 255:red, 0; green, 0; blue, 0 }  ][line width=0.08]  [draw opacity=0] (12,-3) -- (0,0) -- (12,3) -- cycle    ;
    \draw  [dash pattern={on 0.84pt off 2.51pt}]  (318,154) .. controls (294.11,137.42) and (250.74,168.38) .. (238.39,187.55) ;
    \draw [shift={(237.5,189)}, rotate = 300.07] [fill={rgb, 255:red, 0; green, 0; blue, 0 }  ][line width=0.08]  [draw opacity=0] (12,-3) -- (0,0) -- (12,3) -- cycle    ;
    \draw    (315,164) .. controls (297.54,161.09) and (277.73,194.88) .. (274.73,214.25) ;
    \draw [shift={(274.5,216)}, rotate = 276.01] [fill={rgb, 255:red, 0; green, 0; blue, 0 }  ][line width=0.08]  [draw opacity=0] (12,-3) -- (0,0) -- (12,3) -- cycle    ;
    \draw    (336,169) .. controls (368.51,177.87) and (440.79,143.07) .. (446.77,99.97) ;
    \draw [shift={(447,98)}, rotate = 95.19] [fill={rgb, 255:red, 0; green, 0; blue, 0 }  ][line width=0.08]  [draw opacity=0] (12,-3) -- (0,0) -- (12,3) -- cycle    ;
    \draw   (184.98,235.14) .. controls (162.35,194.19) and (209.31,124.9) .. (289.86,80.38) .. controls (370.42,35.85) and (454.07,32.95) .. (476.71,73.89) .. controls (499.34,114.84) and (452.38,184.13) .. (371.82,228.65) .. controls (291.26,273.18) and (207.61,276.08) .. (184.98,235.14) -- cycle ;
    \draw  [dash pattern={on 4.5pt off 4.5pt}]  (396.5,92) -- (353.85,138.53) ;
    \draw [shift={(352.5,140)}, rotate = 312.51] [color={rgb, 255:red, 0; green, 0; blue, 0 }  ][line width=0.75]    (10.93,-3.29) .. controls (6.95,-1.4) and (3.31,-0.3) .. (0,0) .. controls (3.31,0.3) and (6.95,1.4) .. (10.93,3.29)   ;
    \draw (321,170.4) node [anchor=north west][inner sep=0.75pt]    {$p$};
    \draw (256,236.4) node [anchor=north west][inner sep=0.75pt]    {$q_{m-1}$};
    \draw (205,205.4) node [anchor=north west][inner sep=0.75pt]    {$q_{i}$};
    \draw (257,111.4) node [anchor=north west][inner sep=0.75pt]    {$q_{1}$};
    \draw (454,81.4) node [anchor=north west][inner sep=0.75pt]    {$q_{m}$};
    \draw (382,67.4) node [anchor=north west][inner sep=0.75pt]    {$g_{\mu \nu }( q_{m})$};
    \draw (333,143.4) node [anchor=north west][inner sep=0.75pt]    {$g_{\mu \nu }( p)$};
    \end{tikzpicture}
    \caption{Schematic representation of the bounded random sampling approach to solving metric estimation problems. The outer bounding ellipse represents a level surface of mismatch at p, using a previously evaluated metric from beyond the ellipse. Points $q_k$, $k\in \{1, ..., m\}$ are randomly sampled within the ellipse, the first $m-1$ iterations of which fail to resolve the error. Point $q_m$ represents the first randomly sampled point to solve the error in metric evaluation. Last, the assumption is made $g_{\mu\nu}(p)\approx g_{\mu\nu}(q_m)$, which resolves the estimation error.}
    \label{fig:random-selection}
\end{figure}
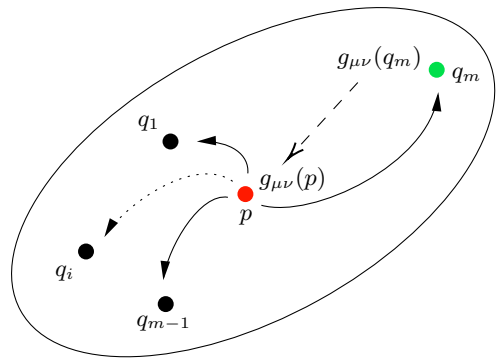

The procedure for mismatch-neighborhood metric estimation at a point $p$, depicted in Fig. \ref{fig:random-selection}, amounts to the attempted metric estimation at each point $q_i$ in a sequence of randomly selected points sampled uniformly over the neighborhood $\mathcal{U}_p$, terminating either on successful estimation of $g$ or after a maximum number of attempts. Specifically, the sampling algorithm can be described by the following steps:

\begin{enumerate}
	\item Metric evaluation fails at a point $p$, detected by the presence of negative eigenvalues in the metric $g_{ij}$. 
	\item A hyperellipsoid of maximum mismatch centered at $p$ is determined using a reference metric. 
	\item A point $q_i$ is randomly sampled from within the hyperellipsoid, so that $q_i$ is ``nearby" to $p$. 
	\item Metric estimation is attempted at $q_i$. 
	\item Repeat Steps 3-4 until the metric is successfully estimated at some point $q_n$, or until a maximum number of attempts is exceeded. 
	\item Finally, we use the metric $g(q_n)$ for the neighborhood by associating $g(p)\approx g(q_n)$.	
\end{enumerate}
Where the reference metric in Step 2 is produced by a general estimate at the beginning of the treebank algorithm and succeeded in each iteration of splitting by the metric of the parent hyperrectangle.

\subsection{Boundary Enhancement}

Another principal assumption of the \MANIFOLD technique is that templates within the original boundaries (the original hyperrectangle) are sufficient to recover signals in the same region. This assumption holds manifestly true in the bulk of the targeted search subset of the parameter space; however, it does not hold as well near the boundary of the given space. This agrees with the notion that a signal is best recovered when it is ``surrounded" by templates \cite{Owen:1998dk}, which emerges naturally from stochastic bank generation methods, as they often place templates outside the bounds of target signals \cite{Sathyaprakash:1991}. The templates that are placed ``outside" the target signal constraint space serve to improve the fitting factor to the most extreme targeted signals, e.g. those whose parameters are very near the maximum targeted values.

\begin{figure}
	\centering
		\includegraphics[width=\linewidth]{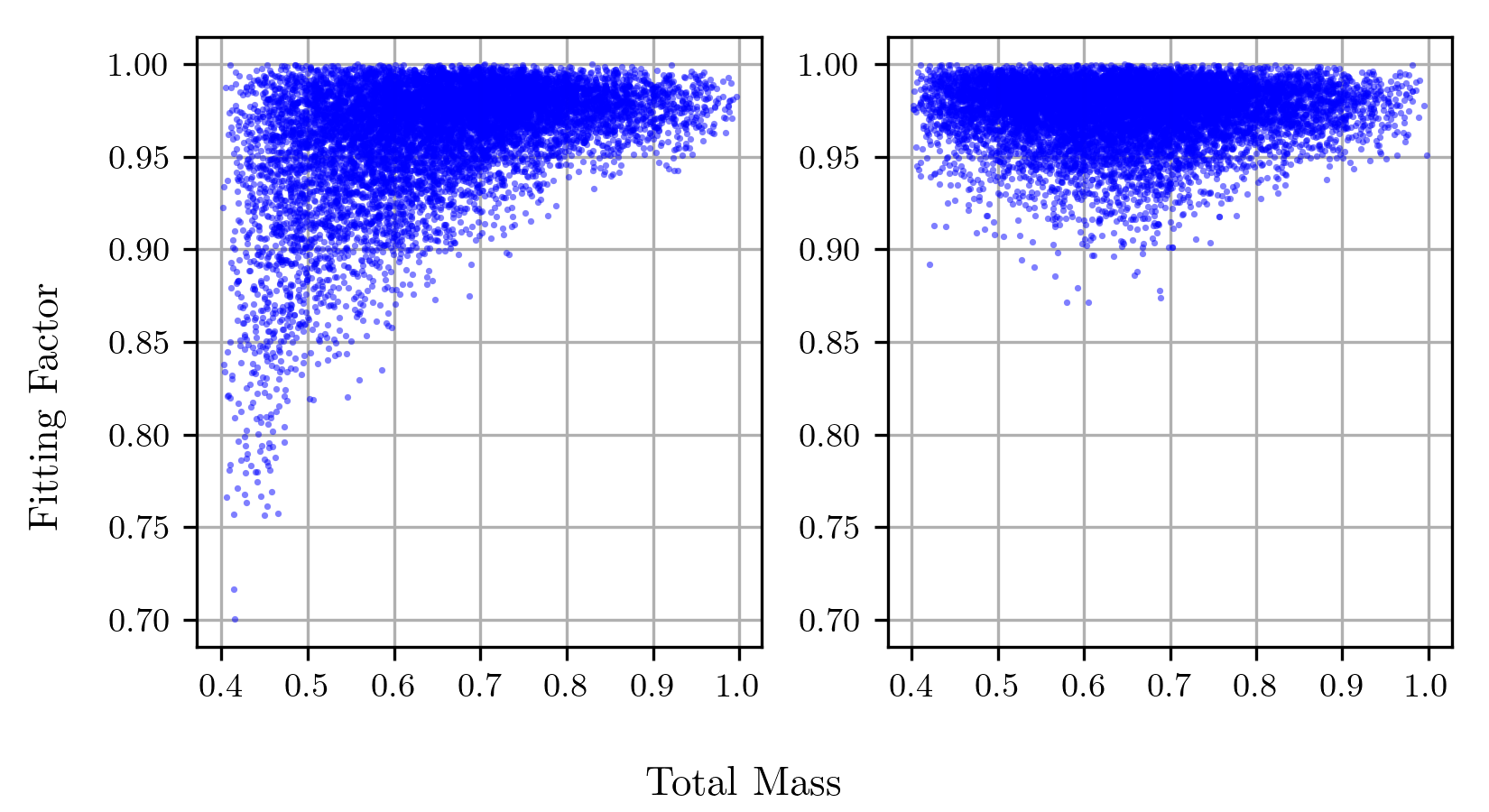}
		\caption{Simulations of signal recovery for low mass events. The original bank (left) clearly has a lower recovery at the low mass boundary. The combined bank (right) has much greater performance at the low mass boundary.}
		\label{fig:methods_boundary}
\end{figure}

The \MANIFOLD method used in~\cite{Sakon:2024} resulted in an insufficiently dense \SSM template bank near the low-mass boundary of the constraint sub-manifold, both internally and externally to the constraint manifold, reducing the efficacy of the generated template bank in this parameter space.
To improve the performance in these critical regions, we implemented a padding technique in which multiple banks with different constraint parameters can be combined along certain boundaries. The final bank was the union of two banks, the original bank as described above covering the targeted \SSM parameter spaces, as well as a smaller ``marginal" bank focused solely on the low-mass boundary. The process for finalizing a choice of padding-constraints, which were used to produce the marginal bank, was iterative, balancing the increase in number of templates against the performance of bank simulations along the low mass boundary. Specifically, the steps were (1) widen the constraints to cover more overlap at the boundary, (2) evaluate the signal recovery near the boundary, and (3) if the signal recovery was insufficient, repeat. In general, the more widely the constraints were padded, the more the improvement in the recovery of signals near the original boundary. We deemed an increase in the number of templates up to roughly 30\% (on the order of 500,000 additional templates) to be an acceptable cost for improved coverage. We stopped widening the padding once the bank’s performance was sufficient – specifically, when a significant majority of simulated signals near the original boundary achieved match values above 95\%. At that point, further increasing the bank yielded diminishing returns. Fig. \ref{fig:methods_boundary} shows the resulting impact of the boundary padding, in that signal recovery near the low mass boundary was dramatically improved.

\section{Results}
\label{section:results}

\label{subsection:bank_simulation}

Bank simulation is a method of computing the \ac{FF} between the template waveforms and simulated \ac{GW} waveforms, to assess the effectualness of template banks against \ac{GW} signals.
The \ac{FF} is given in ~\cite{Privitera:2014, Apostolatos:1995} as:
\begin{equation}
	FF \left(\hat{u}_s , u \right) = \max_{k} \langle \hat{u}_k | \hat{u}_s \rangle 
	\label{eq:fitting_factor}
\end{equation}
where $\langle \hat{u}_k | \hat{u}_s \rangle$ is defined in Eq. \ref{eq:match}, $u_s$ is the waveform of a simulated signal, $\hat{u}_i$ are normalized waveforms such that $\langle \hat{u}_i | \hat{u}_i \rangle = 1$, and $k$ is the bank index such that $u = \{u_k\}_{k=1}^{N}$ and $N$ is the number of templates in a bank.

To perform bank simulations with the presented banks, we use the \SBANK program, which simulates \ac{GW} signals using \ac{LAL} ~\cite{lalsuite} and computes the \ac{FF} to find the template with the highest match for the signal. Previous work showed ~\cite{Sakon:2024} that \SBANK bank simulation tests have been used to assess template banks generated with \MANIFOLD, and that template banks generated with \MANIFOLD are as effective as template banks generated with \SBANK. 

In this paper, we present the bank simulation results of the \ac{O4} offline \ac{SSM} bank using the \SBANK software ~\cite{Ajith:2014, Capano:2016, Harry:2009, Privitera:2014}. We use the waveform approximant, the \ac{PSD}, and the low-frequency cutoff described in Section ~\ref{section:design} for simulation studies presented here.
It must be noted that \MANIFOLD can apply a maximum duration when computing templates in its waveform computation, while \SBANK currently does not support this feature. 
Therefore, the bank simulation test below cannot reflect the SNR loss due to this duration cutoff in our templates.
To estimate the bank performance while accounting for the maximum waveform duration cut applied during the template placement, we set a $f_{low}$ cutoff higher than $45$ $Hz$ for different mass regions. See Table ~\ref{table:banksim_params}.

We ran two bank simulation tests on the template bank as the following: 
\begin{enumerate} 
	\item Assess the effectualness of the template bank by using simulated signals that span the template bank parameter space. (Section ~\ref{subsubsection:results_banksim_tempbankregion}.)
	\item Assess the effectualness of the template bank against electromagnetically-bright events, i.e., \acp{BNS} and \acp{NSBH}. (Section ~\ref{subsubsection:results_banksim_bnsnsbh}.) 
\end{enumerate}

\subsubsection{Simulation studies in the template bank space}
\label{subsubsection:results_banksim_tempbankregion}

\begin{figure*}[htbp]
	\centering
	\begin{subfigure}[b]{0.24\textwidth}
		\includegraphics[height=3cm]{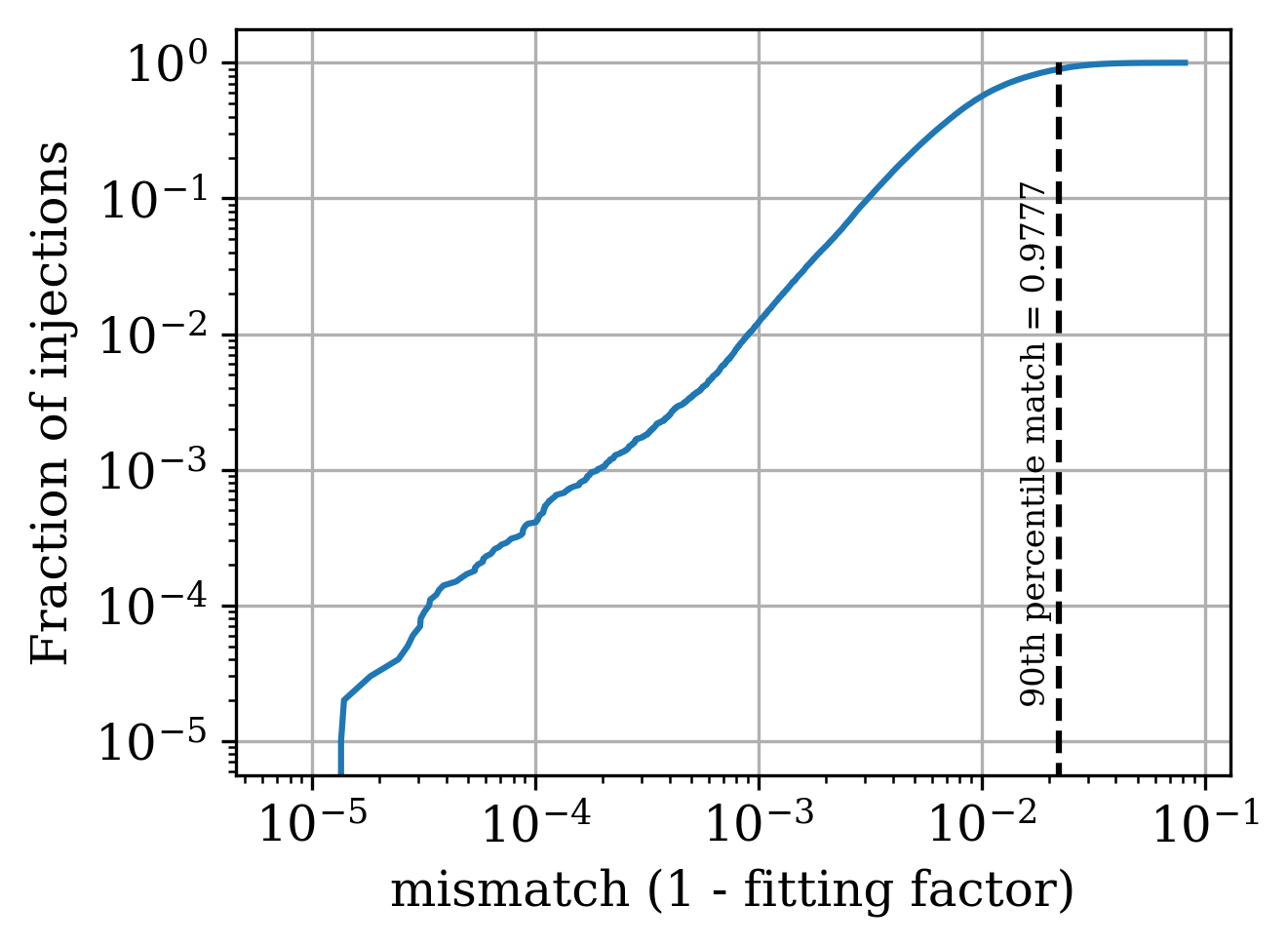}
		\caption{Cumulative histogram of mismatches.}
		\label{fig:results_banksim_lowlow_cumulative}
	\end{subfigure}%
	\begin{subfigure}{0.24\textwidth}
		\includegraphics[height=3cm]{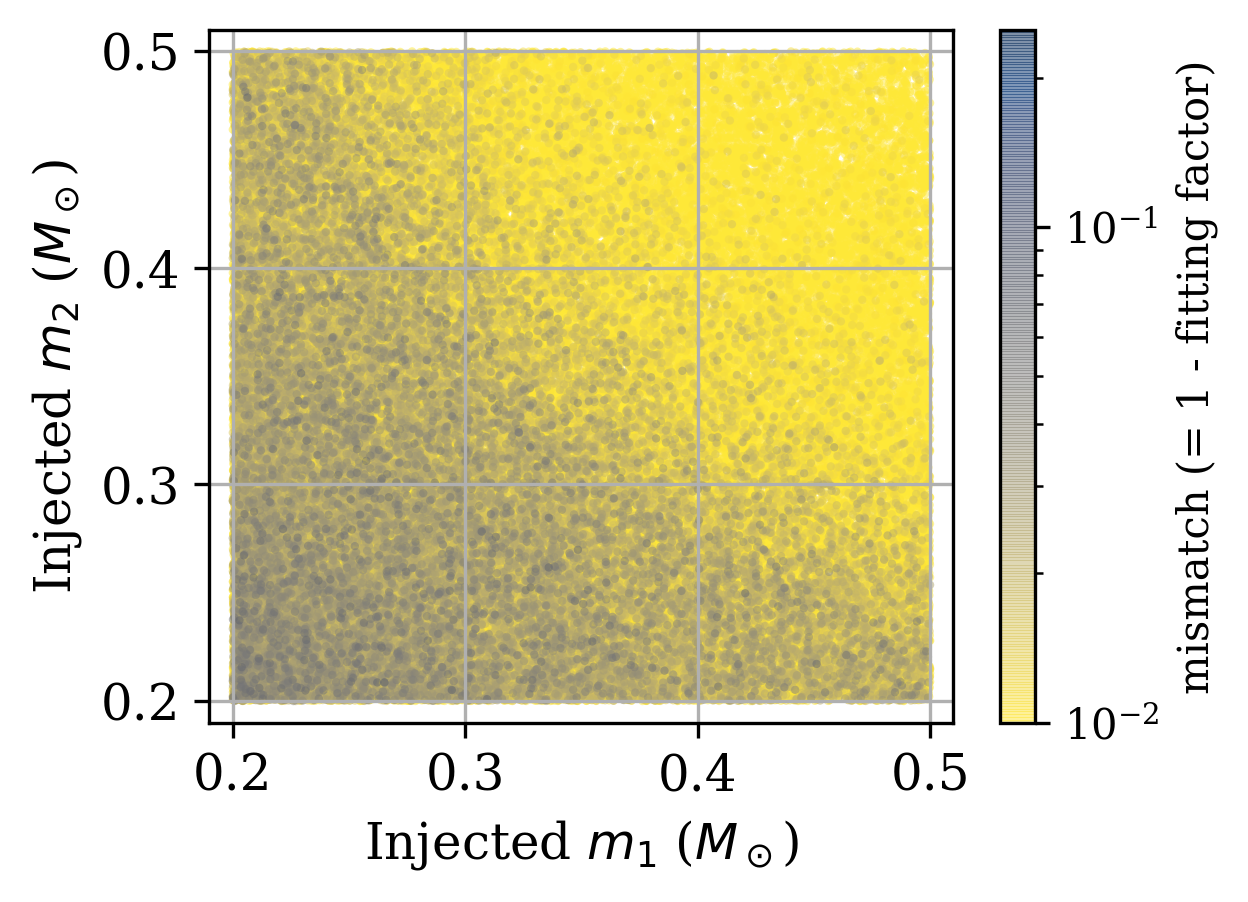}
		\caption{\protect Mismatches in the $m_1$-$m_2$ plane.}
		\label{fig:results_banksim_lowlow_m1m2}
	\end{subfigure}
	\begin{subfigure}{0.24\textwidth}
		\includegraphics[height=3cm]{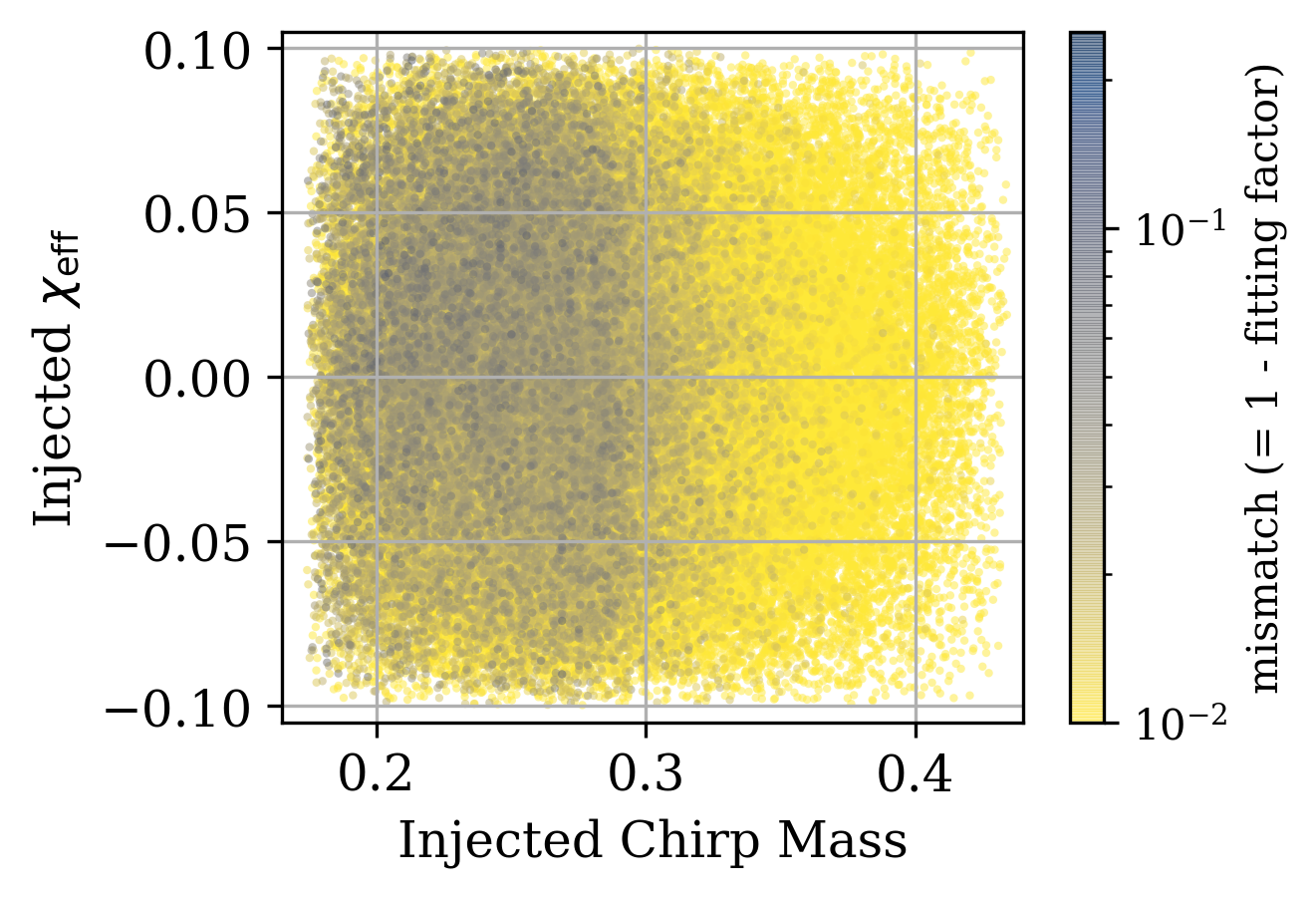}
		\caption{\protect Mismatches in the $\mathcal{M}$-$\chi_{\rm {eff}}$ plane.}
		\label{fig:results_banksim_lowlow_mcchi}
	\end{subfigure}%
	\begin{subfigure}{0.24\textwidth}
		\includegraphics[height=3cm]{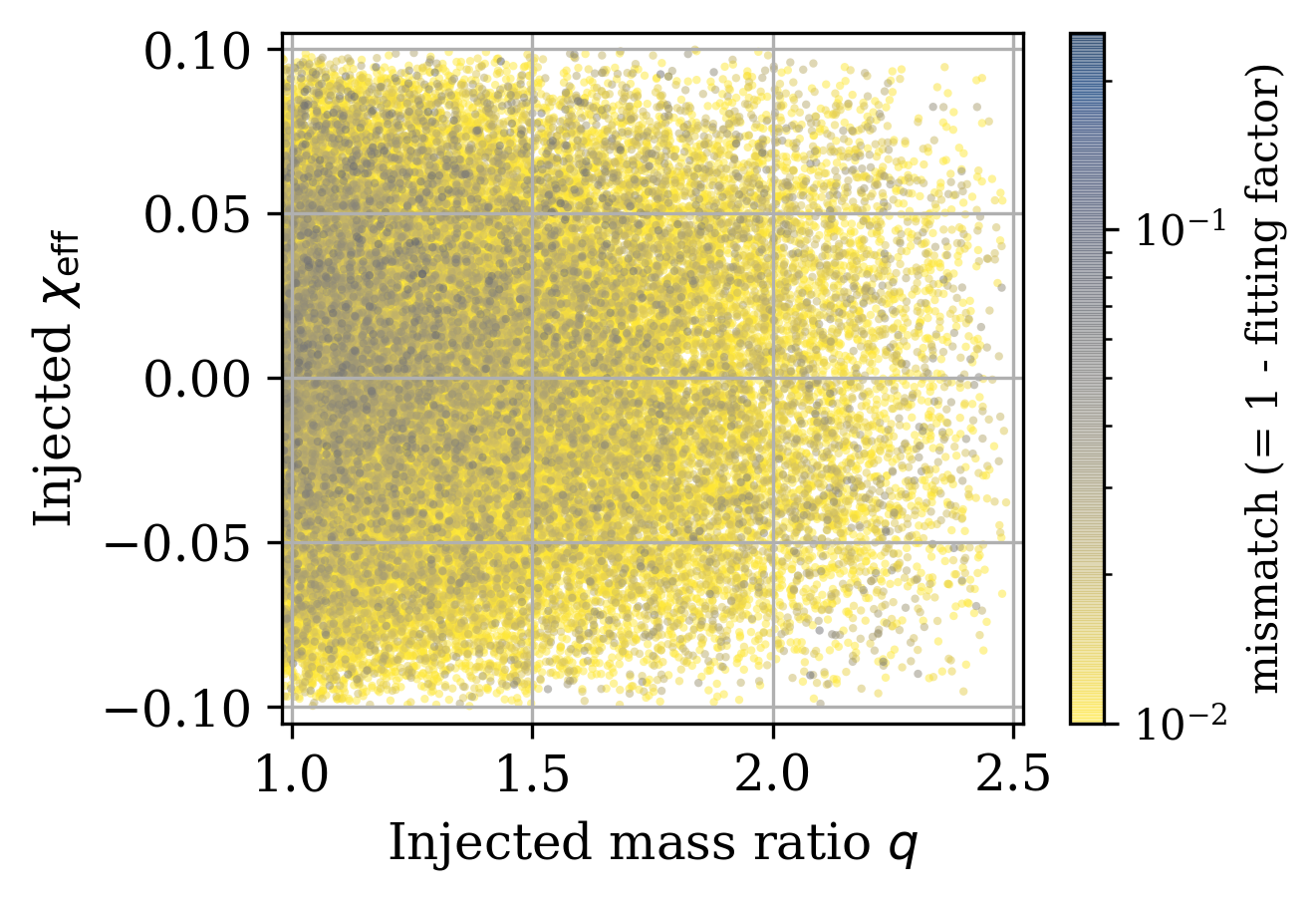}
		\caption{\protect Mismatches in the $q$-$\chi_{\rm {eff}}$ plane.}
		\label{fig:results_banksim_lowlow_qchi}
	\end{subfigure}
	\caption{\protect Plots for simulated signals in the ``low-low" region. For purposes of visually presenting the mismatches, mismatches smaller than $10^{-2}$ have been mapped to $10^{-2}$. (a) shows that $90 \%$ of the simulated signals are recovered with a match $\geqslant$ $97.77 \%$.}
	\label{fig:results_banksim_lowlow}
\end{figure*}

\begin{figure*}[htbp]
	\centering
	\begin{subfigure}{0.24\textwidth}
		\includegraphics[height=3cm]{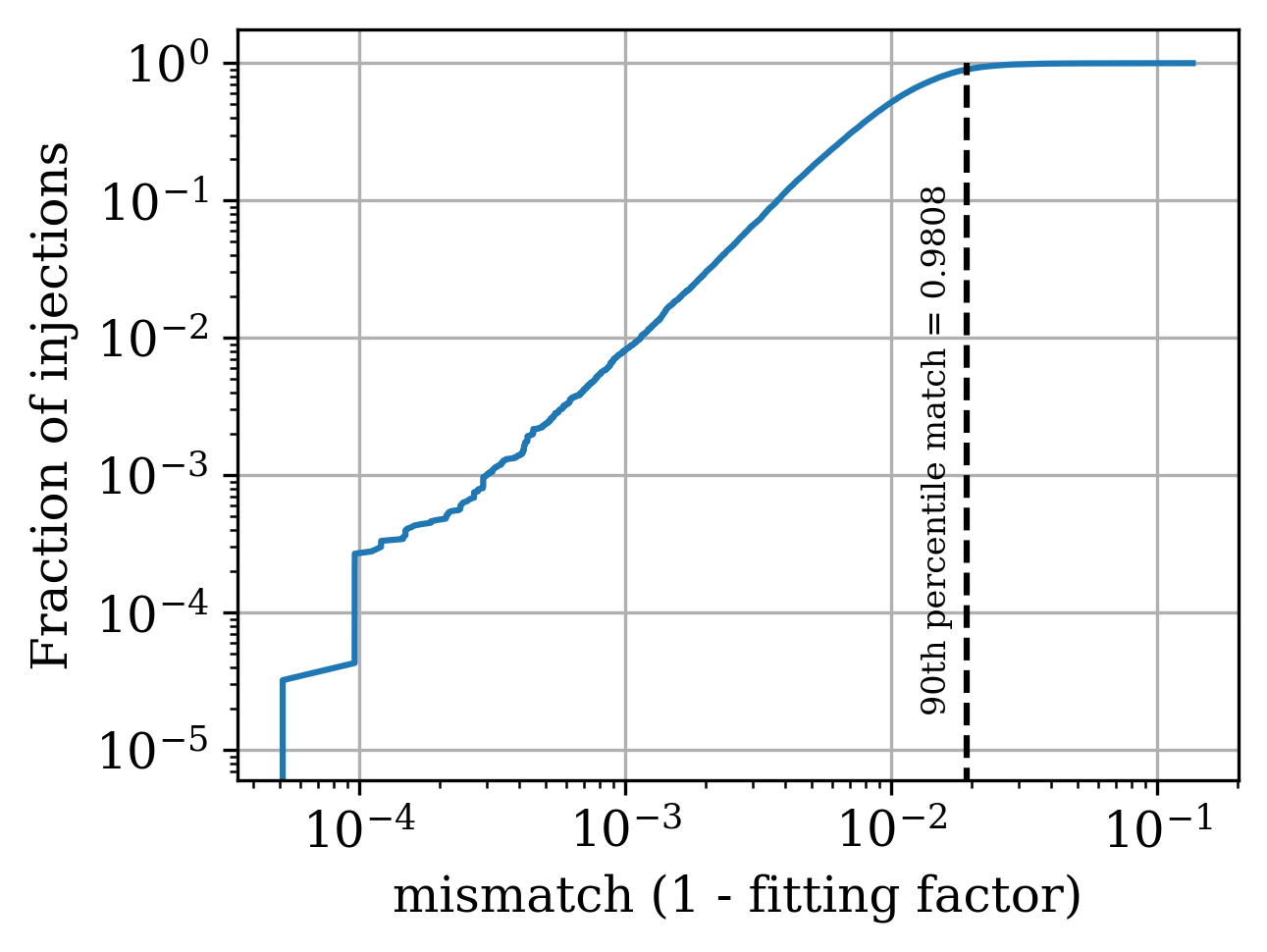}
		\caption{\protect Cumulative histogram of mismatches.}
		\label{fig:results_banksim_highlow_cumulative}
	\end{subfigure}%
	\begin{subfigure}{0.24\textwidth}
		 \includegraphics[height=3cm]{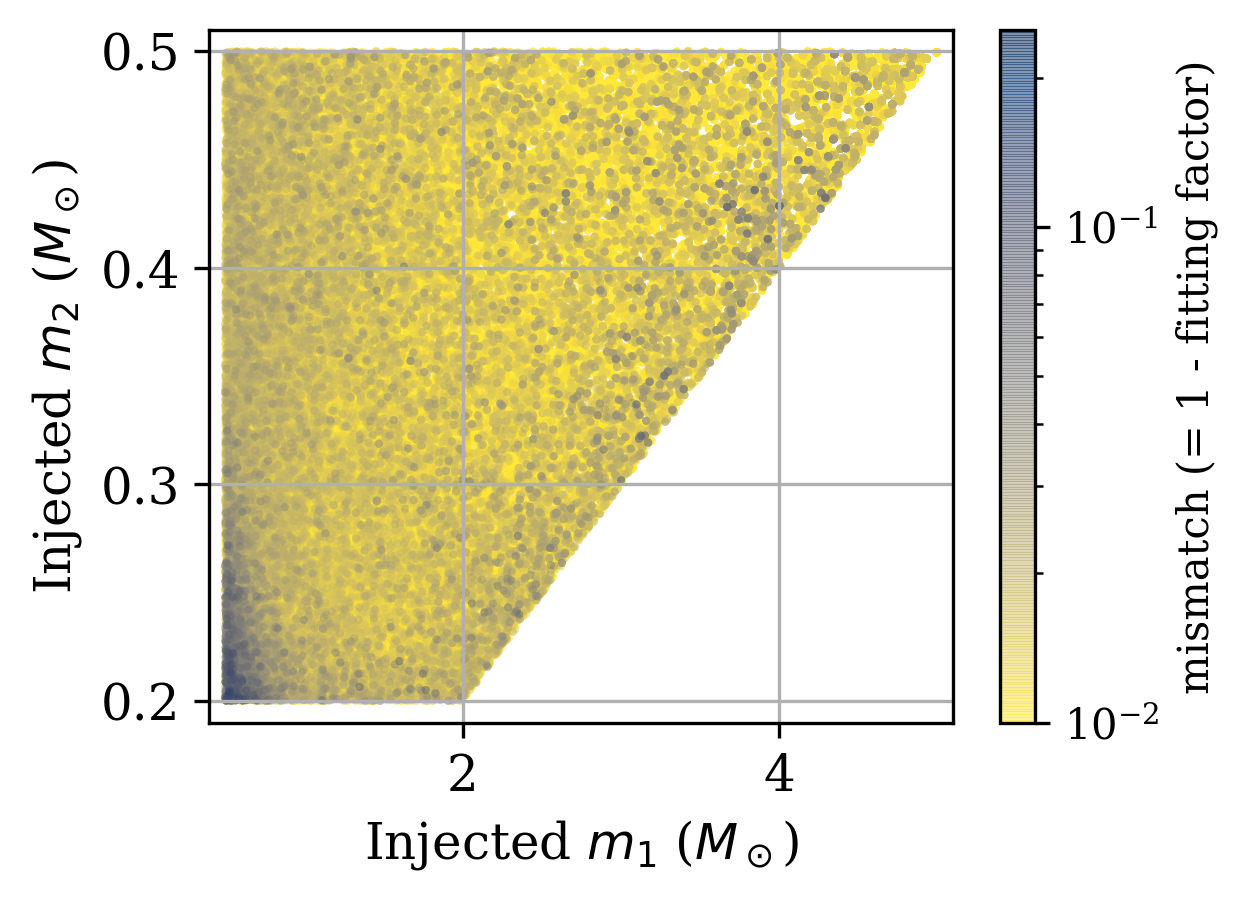}
		\caption{\protect Mismatches in the $m_1$-$m_2$ plane.}
		\label{fig:results_banksim_highlow_m1m2}
	\end{subfigure}
	\begin{subfigure}{0.24\textwidth}
		\includegraphics[height=3cm]{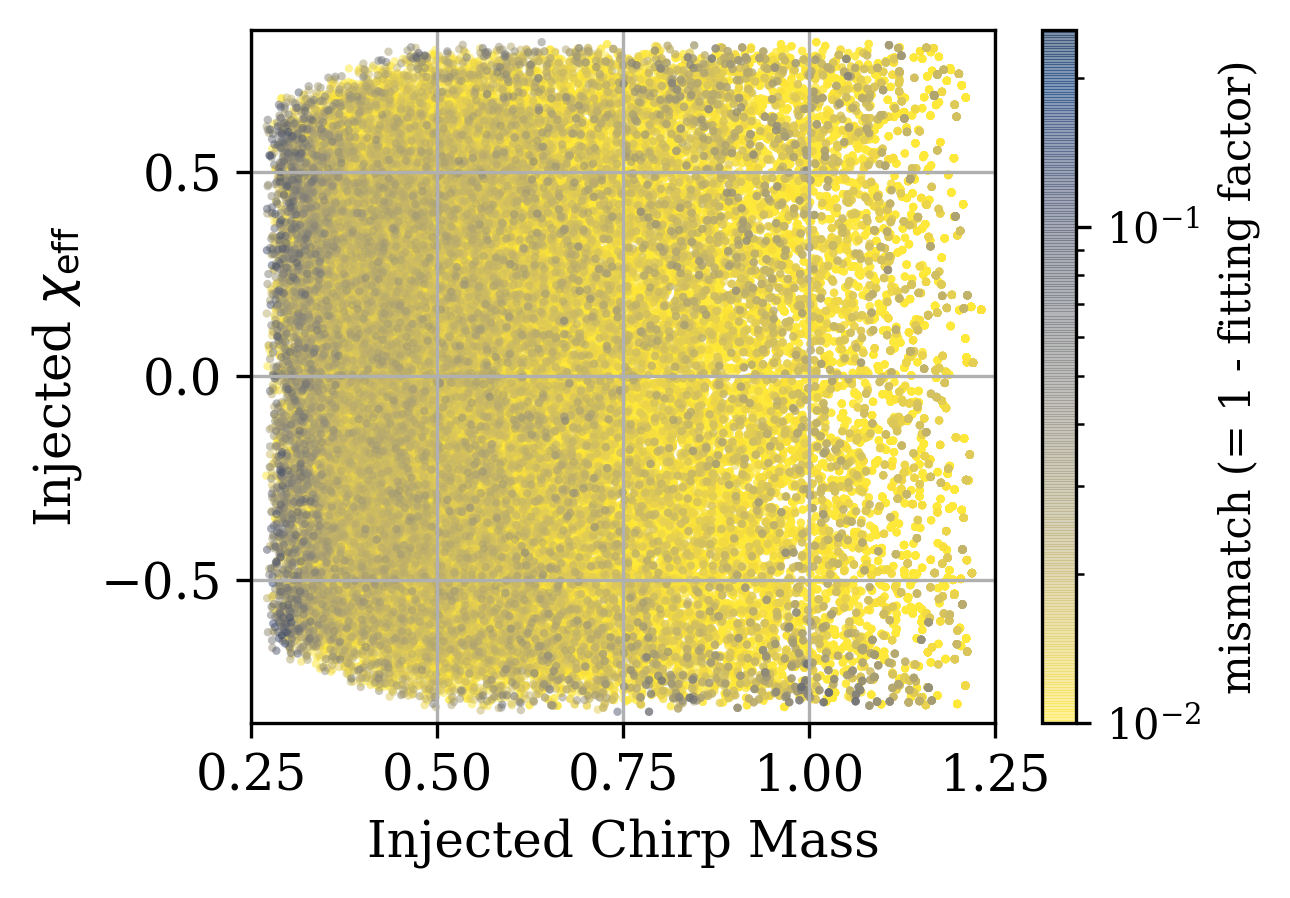}
		\caption{\protect Mismatches in the $\mathcal{M}$-$\chi_{\rm {eff}}$ plane.}
		\label{fig:results_banksim_highlow_mcchi}
	\end{subfigure}%
	\begin{subfigure}{0.24\textwidth}
		\includegraphics[height=3cm]{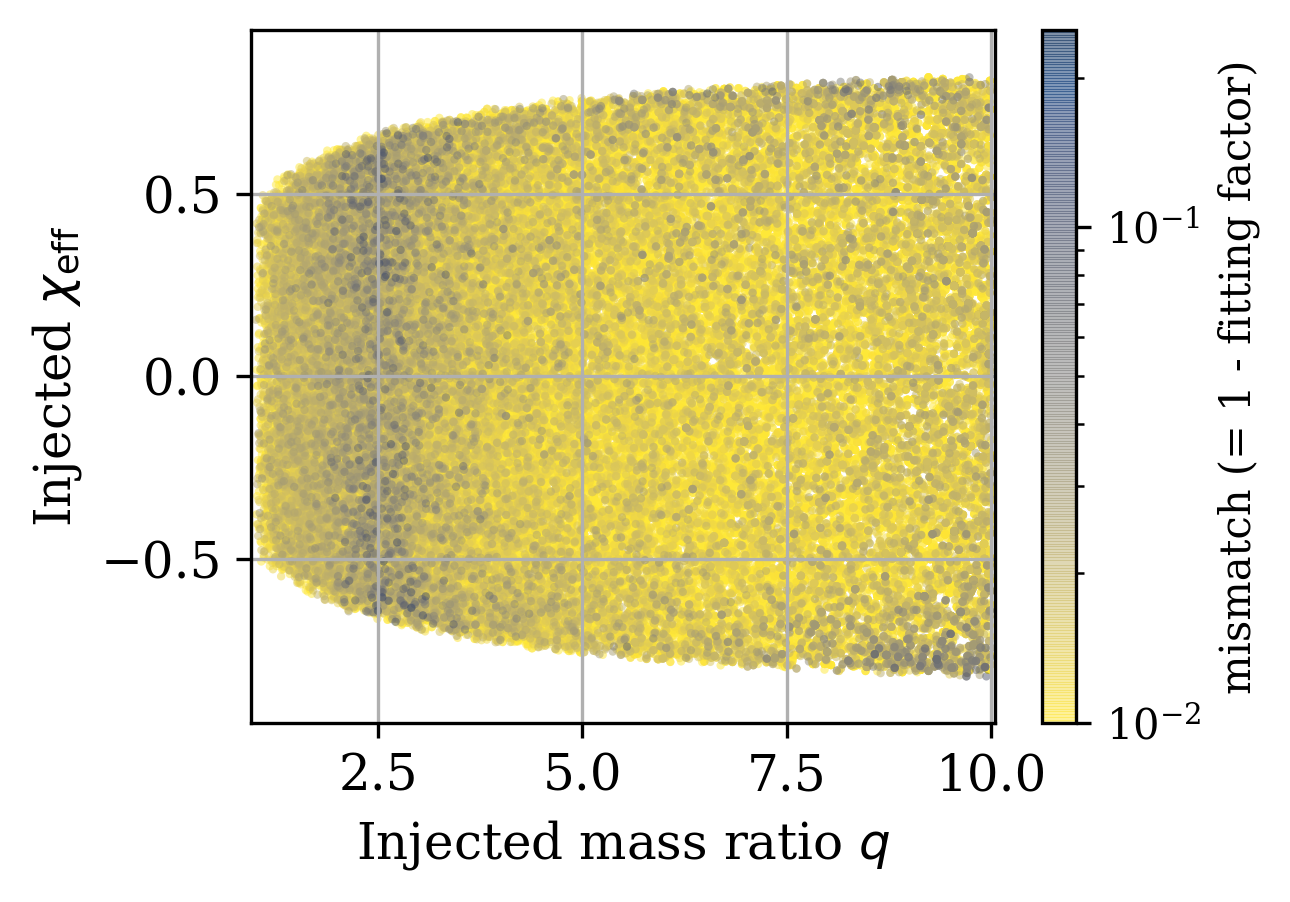}
		\caption{\protect Mismatches in the $q$-$\chi_{\rm {eff}}$ plane.}
		\label{fig:results_banksim_highlow_qchi}
	\end{subfigure}
	\caption{\protect Plots for the simulated signals in the ``high-low" region. For purposes of visually presenting the mismatches, mismatches smaller than $10^{-2}$ have been mapped to $10^{-2}$. (a) shows that $90 \%$ of the simulated signals are recovered with a match $\geqslant$ $98.08 \%$.}
	\label{fig:results_banksim_highlow}
\end{figure*}

\begin{figure*}[!htbp]
	\centering
	\begin{subfigure}{0.24\textwidth}
		\includegraphics[height=3cm]{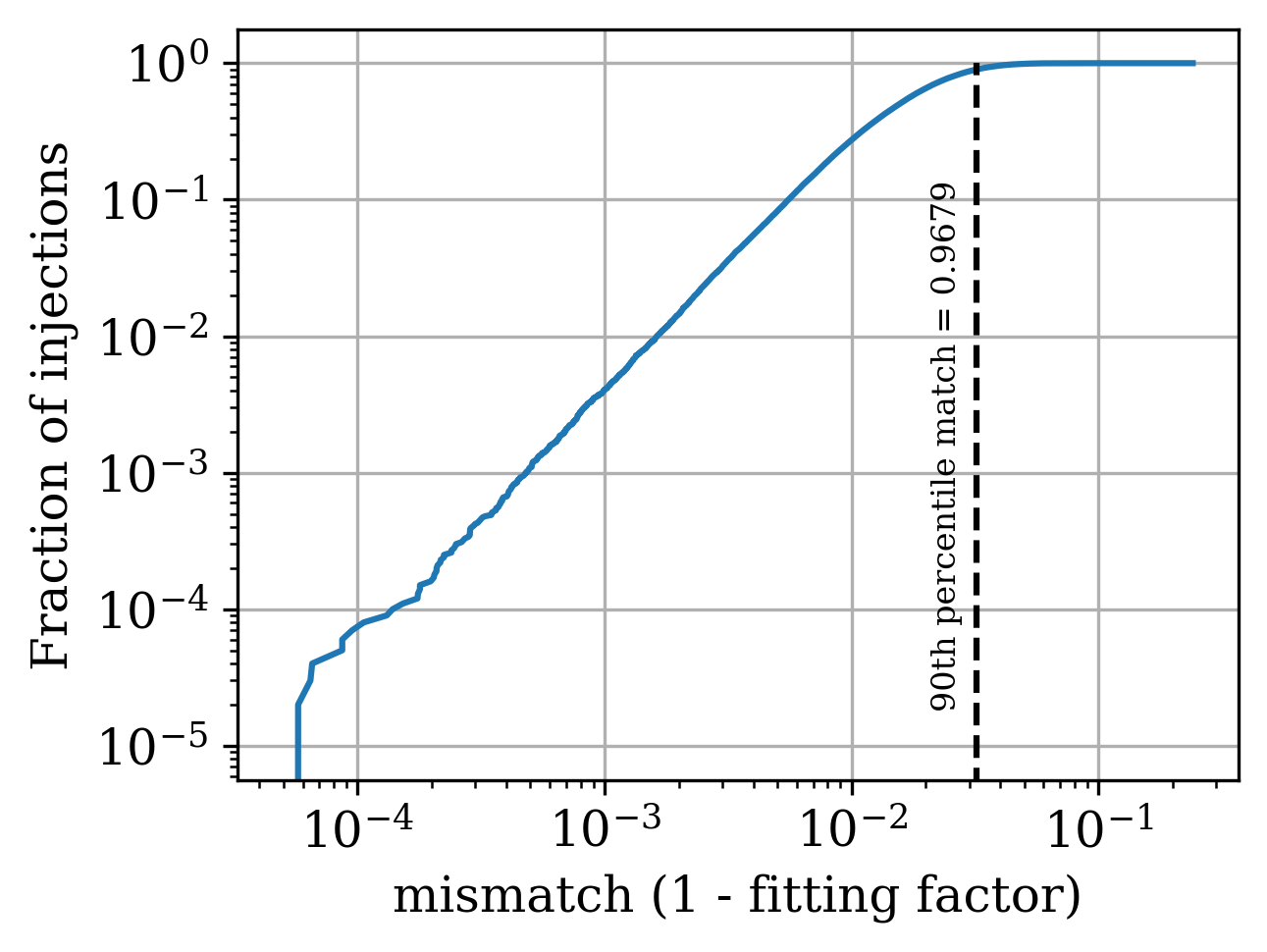}
		\caption{\protect Cumulative histogram mismatches.}
		\label{fig:results_banksim_highhigh_cumulative}
	\end{subfigure}%
	\begin{subfigure}{0.24\textwidth}
		 \includegraphics[height=3cm]{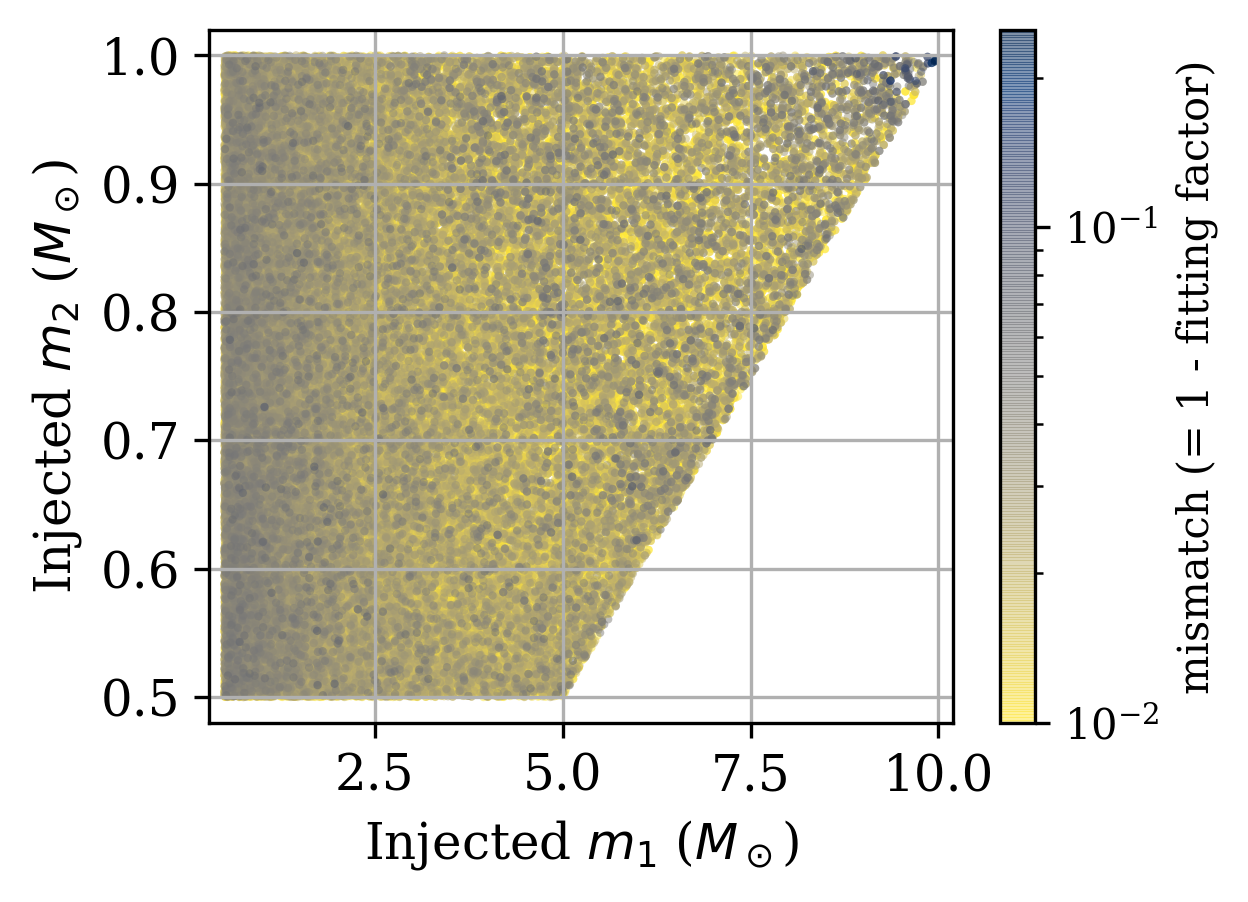}
		\caption{\protect Mismatches in the $m_1$-$m_2$ plane.}
		\label{fig:results_banksim_highhigh_m1m2}
	\end{subfigure}
	\begin{subfigure}{0.24\textwidth}
		\includegraphics[height=3cm]{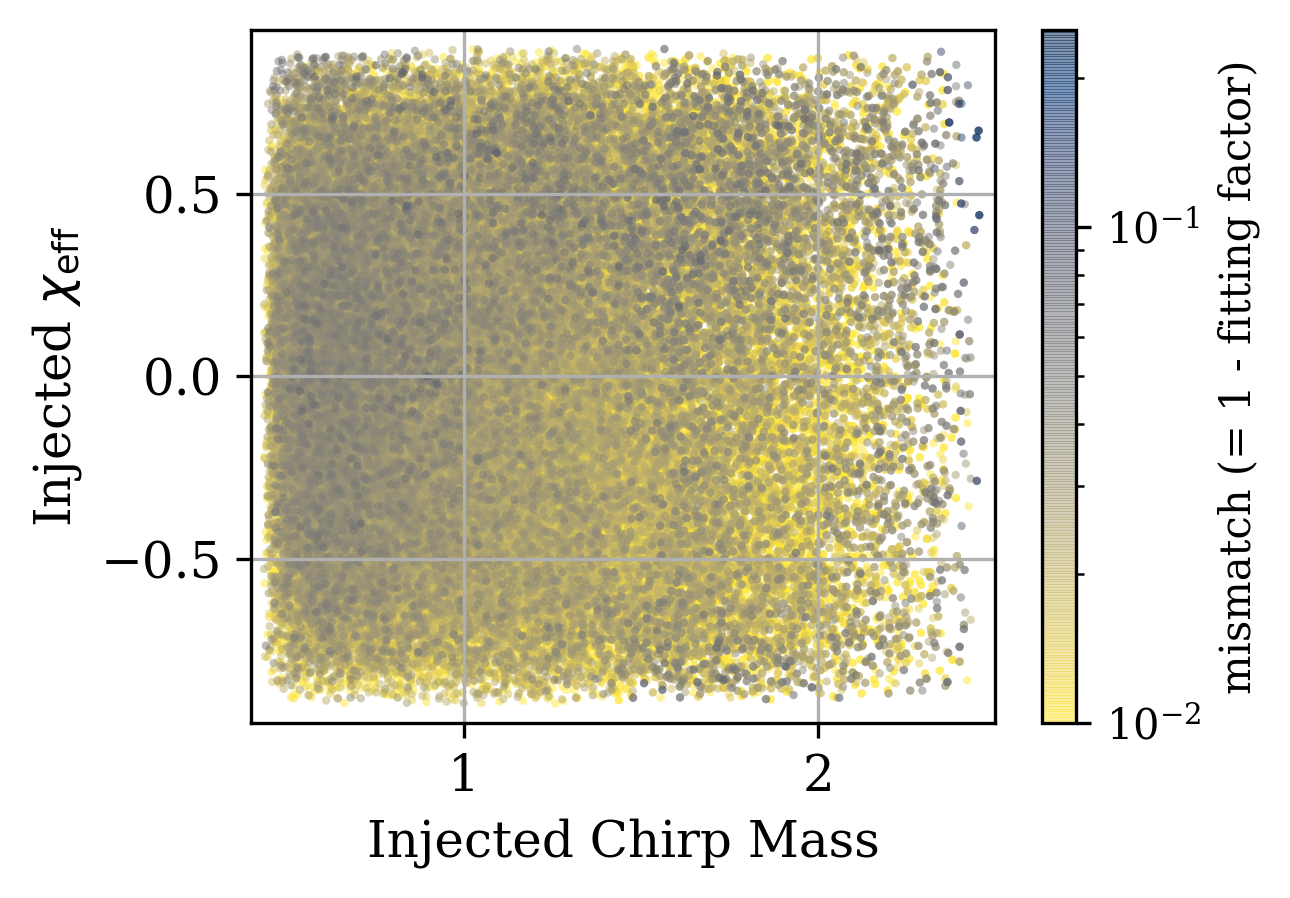}
		\caption{\protect Mismatches in the $\mathcal{M}$-$\chi_{\rm {eff}}$ plane.}
		\label{fig:results_banksim_highhigh_mcchi}
	\end{subfigure}%
	\begin{subfigure}{0.24\textwidth}
		\includegraphics[height=3cm]{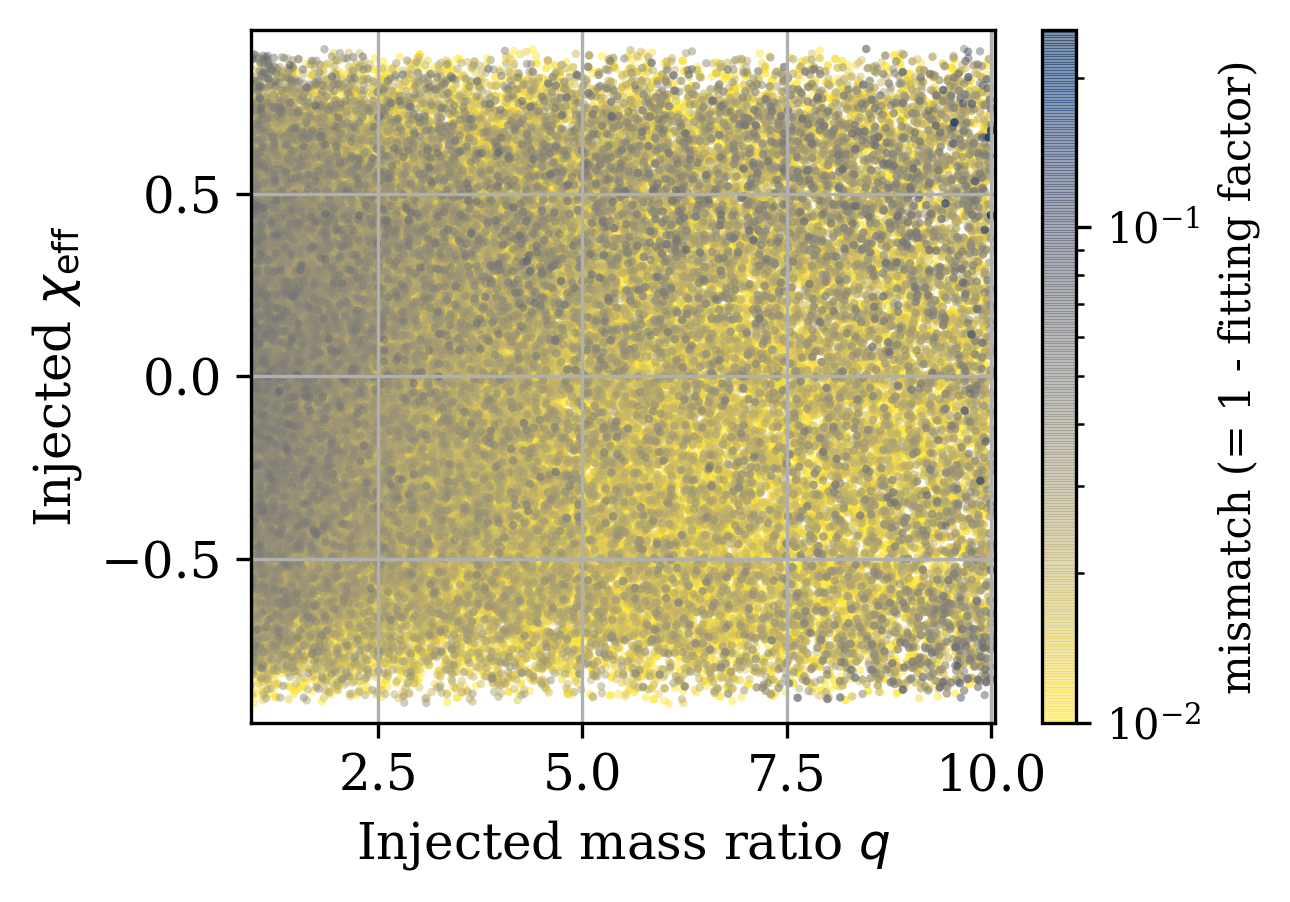}
		\caption{\protect Mismatches in the $q$-$\chi_{\rm {eff}}$ plane.}
		\label{fig:results_banksim_highhigh_qchi}
	\end{subfigure}
	\caption{\protect Plots for the simulated signals in the ``high-high" region. For purposes of visually presenting the mismatches, mismatches smaller than $10^{-2}$ have been mapped to $10^{-2}$. (a) shows that $90 \%$ of the simulated signals are recovered with a match $\geqslant$ $96.79 \%$.}
	\label{fig:results_banksim_highhigh}
\end{figure*}


The simulated signals used for efficacy tests presented in this section cover the same mass and spin parameter space as the template bank that is described in Section ~\ref{section:design}.  
We choose to sample component masses from a log distribution since lower mass regions are more densely populated with templates. The template bank parameter space was divided into three regions for generating simulated signals, the details of which are summarized in Table \ref{table:banksim_params}. 

\begin{table}
\begin{center}
	\begin{tabular}{ l | c | c | c}
		\hline
		Parameter & Low-Low & High-Low & High-High \\
		\hline
		\hline
		$m_1$ ($M_\odot$) & $\left[0.2, 0.5 \right]$ & $\left[0.5, 10.0 \right]$ & $\left[0.5, 10.0 \right]$  \\
		$m_2$ ($M_\odot$) & $\left[0.2, 0.5 \right]$ & $\left[0.2, 0.5 \right]$ & $\left[0.5, 1.0 \right]$  \\
		$s_{1,z}$  & $\left[-0.1, 0.1\right]$ & $\left[-0.9, 0.9\right]$ & $\left[-0.9, 0.9\right]$  \\
		$s_{2,z}$  & $\left[-0.1, 0.1\right]$ & $\left[-0.1, 0.1\right]$ & $\left[-0.9, 0.9\right]$  \\
		$f_{low}$ cutoff & $75$Hz & $55$Hz & $45$Hz \\
		$f_{high}$ cutoff & $1024$Hz & $1024$Hz & $1024$Hz \\
		Count & 100,000 & 93,157 & 99,975  \\
		\hline
	\end{tabular}
	\caption{Parameter space constraints for simulated signals during bank testing for the three regions ``low-low", ``high-low", and ``high-high".}
	\label{table:banksim_params} 
\end{center}
\end{table}

Figs. ~\ref{fig:results_banksim_lowlow}, ~\ref{fig:results_banksim_highlow}, 
and ~\ref{fig:results_banksim_highhigh} show the mismatches between the templates and the simulated signals in the three regions described above. $90$ $\%$ of the ``low-low" simulated signals have a match of $97.77$ $\%$ or higher, $98.08$ $\%$ or higher for the ``high-low" simulated signals, and $96.79$ $\%$ or higher for the ``high-high" simulated signals. 
When the results of these three sets are combined, $90$ $\%$ of the 
simulated signals have a match of $97.42$ $\%$ or higher with templates in the bank. The low mismatches, i.e., higher \acp{FF}, of the ``low-low" simulated signals are due to an over-density of templates in the extreme low-mass regions as a result of the padding procedure described in Section \ref{section:methods}.  
In the ``high-high" region, the relatively higher mismatches are a combined result of no padding and approaching a boundary of the template bank caused by the discontinuous spin constraints between components' low and high masses. This particular boundary is naturally visible in the $\mathcal{M}-\chi_{\rm {eff}}$ plots, as in Fig. \ref{fig:template_mchirp_chi}, and can be visualized as a vertical line separating the high and low spin regions.  
Despite the higher mismatches, the simulation studies show that the bulk of the ``high-high" region is also densely populated with templates. 
The distribution of mismatches on combining the ``low-low", ``high-low" and ``high-high" show that the templates populate the search parameter space as designed and their density is sufficient to detect \ac{GW} signals within the subset of parameter space of the \ac{SSM} search. 

In \ac{O4}, the \GSTLAL pipeline runs the \ac{BNS}/\ac{NSBH}/\ac{BBH} search as well as the \ac{SSM} search, so multiple banks can detect the same \ac{GW} signals that are at the edges of the template banks. We perform a bank simulation test to assess the effectualness of the search against such signals that lie at the edge of the \ac{O4} offline \ac{SSM} template bank using the ``high-high" set of simulated signals.
For this study, we combine the \ac{O4} \ac{BNS}/\ac{NSBH}/\ac{BBH} template bank described in ~\cite{Sakon:2024} and the \ac{O4} offline \ac{SSM}.
 
Compared to the mismatches for the ``high-high'' signals with the \ac{SSM} template bank, the combined bank gives a slightly higher match of $97.14 \%$ or higher for $90 \%$ of the simulated signals. The mismatches only decrease as the total mass increases as a result of having higher mass templates in the combined bank. This further illustrates that the \ac{SSM} template bank sufficiently covers the parameter space of the \ac{SSM} search.

\subsubsection{Simulation studies with \ac{BNS} and \ac{NSBH} signals}
\label{subsubsection:results_banksim_bnsnsbh}

\begin{figure*}[htbp]
	\centering
	\begin{subfigure}{0.24\textwidth}
		\includegraphics[height=3cm]{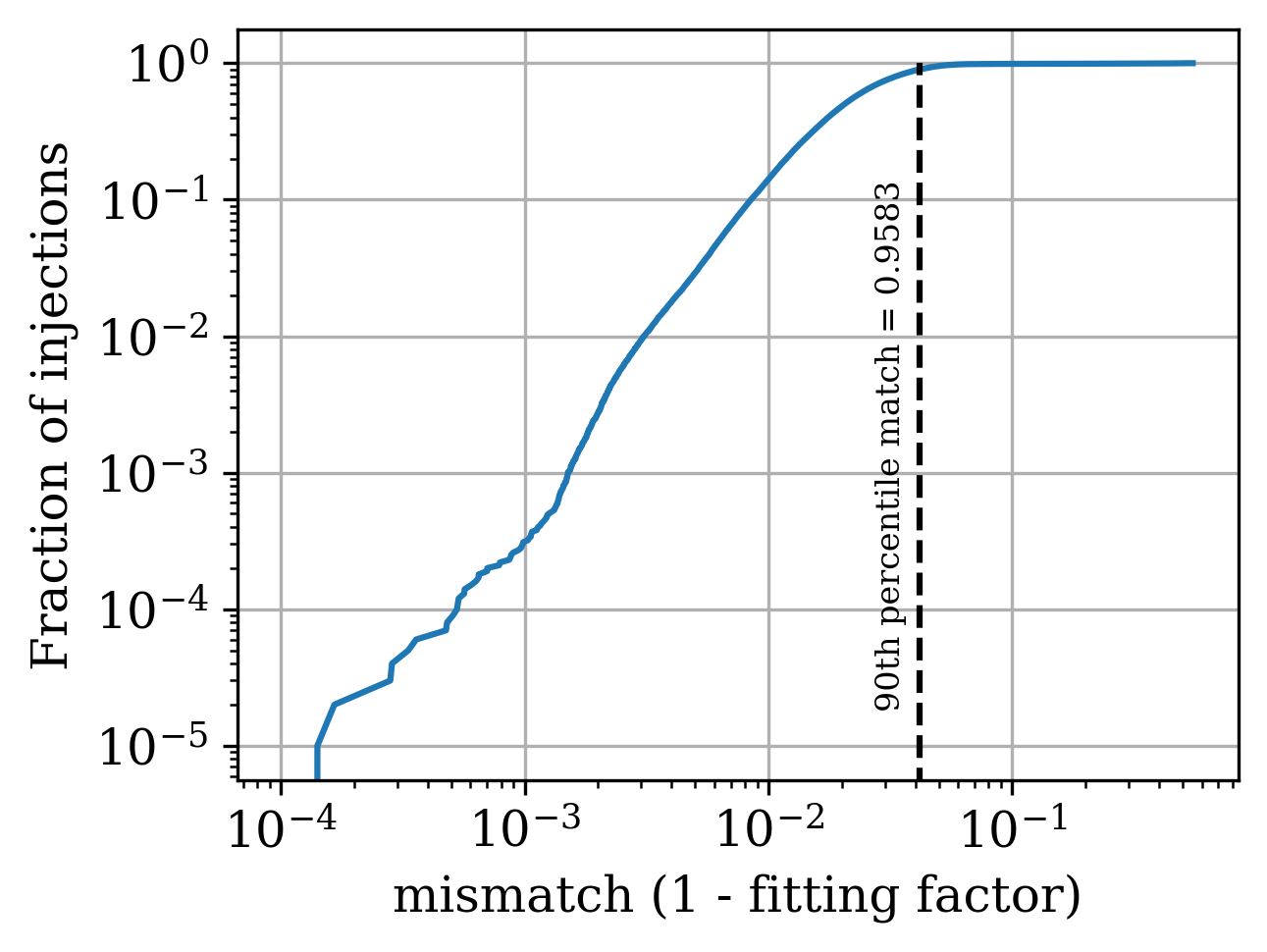}
		\caption{\protect Cumulative histogram of mismatches.}
		\label{fig:results_banksim_bnslow_cumulative}
	\end{subfigure}%
	\begin{subfigure}{0.24\textwidth}
		 \includegraphics[height=3cm]{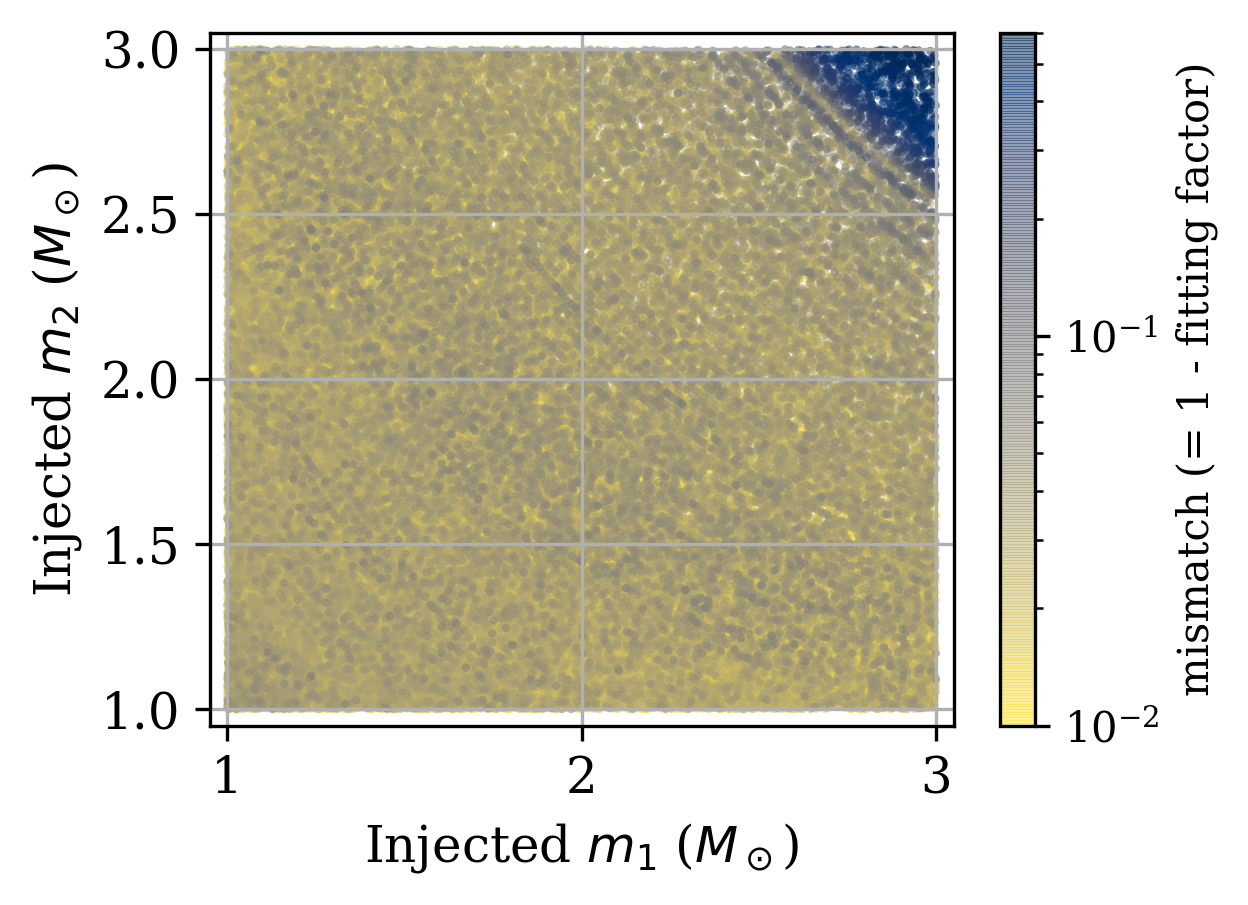}
		\caption{\protect Mismatches in the $m_1$-$m_2$ plane.}
		\label{fig:results_banksim_bnslow_m1m2}
	\end{subfigure}
	\begin{subfigure}{0.24\textwidth}
		\includegraphics[height=3cm]{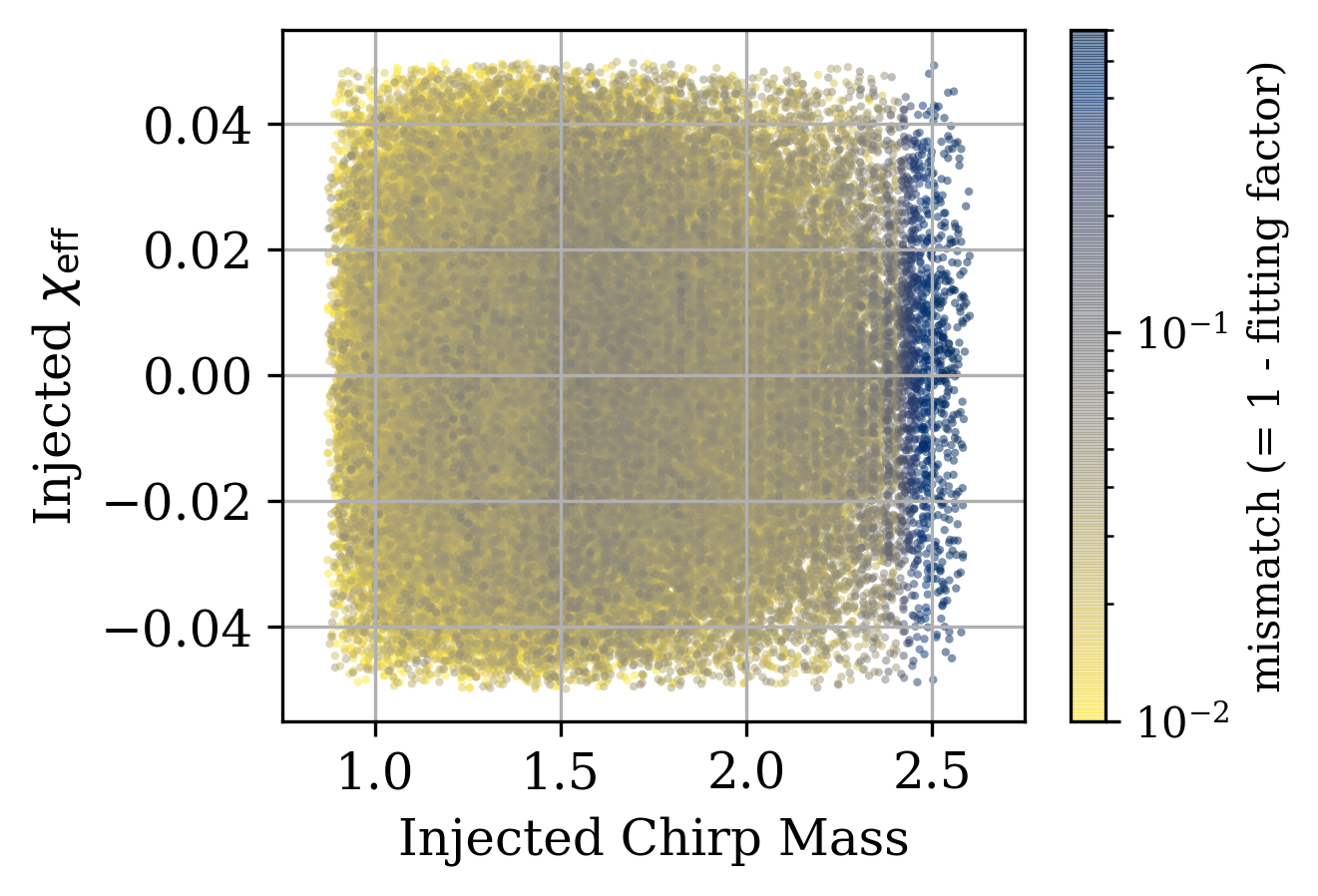}
		\caption{\protect Mismatches in the $\mathcal{M}$-$\chi_{\rm {eff}}$ plane.}
		\label{fig:results_banksim_bnslow_mcchi}
	\end{subfigure}%
	\begin{subfigure}{0.24\textwidth}
		\includegraphics[height=3cm]{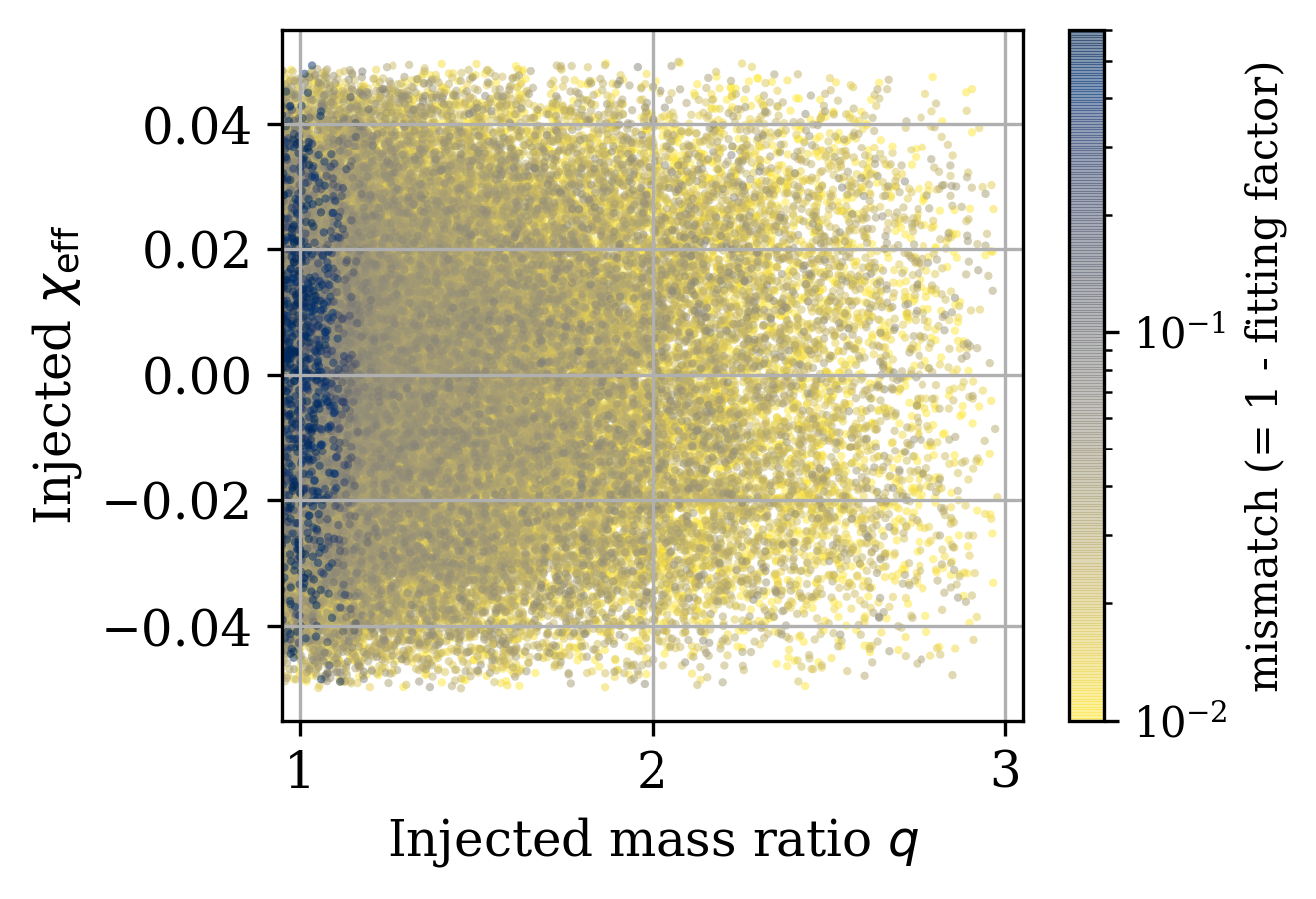}
		\caption{\protect Mismatches in the $q$-$\chi_{\rm {eff}}$ plane.}
		\label{fig:results_banksim_bnslow_qchi}
	\end{subfigure}
	\caption{\protect Plots for \ac{BNS} simulated signals with low-spins. For purposes of visually presenting the mismatches, mismatches smaller than $10^{-2}$ have been mapped to $10^{-2}$. (a) shows that $90 \%$ of the simulated signals are recovered with a match $\geqslant$ $95.83 \%$.}
	\label{fig:results_banksim_bnslow}
\end{figure*}

\begin{figure*}[htbp]
	\centering
	\begin{subfigure}{0.24\textwidth}
		\includegraphics[height=3cm]{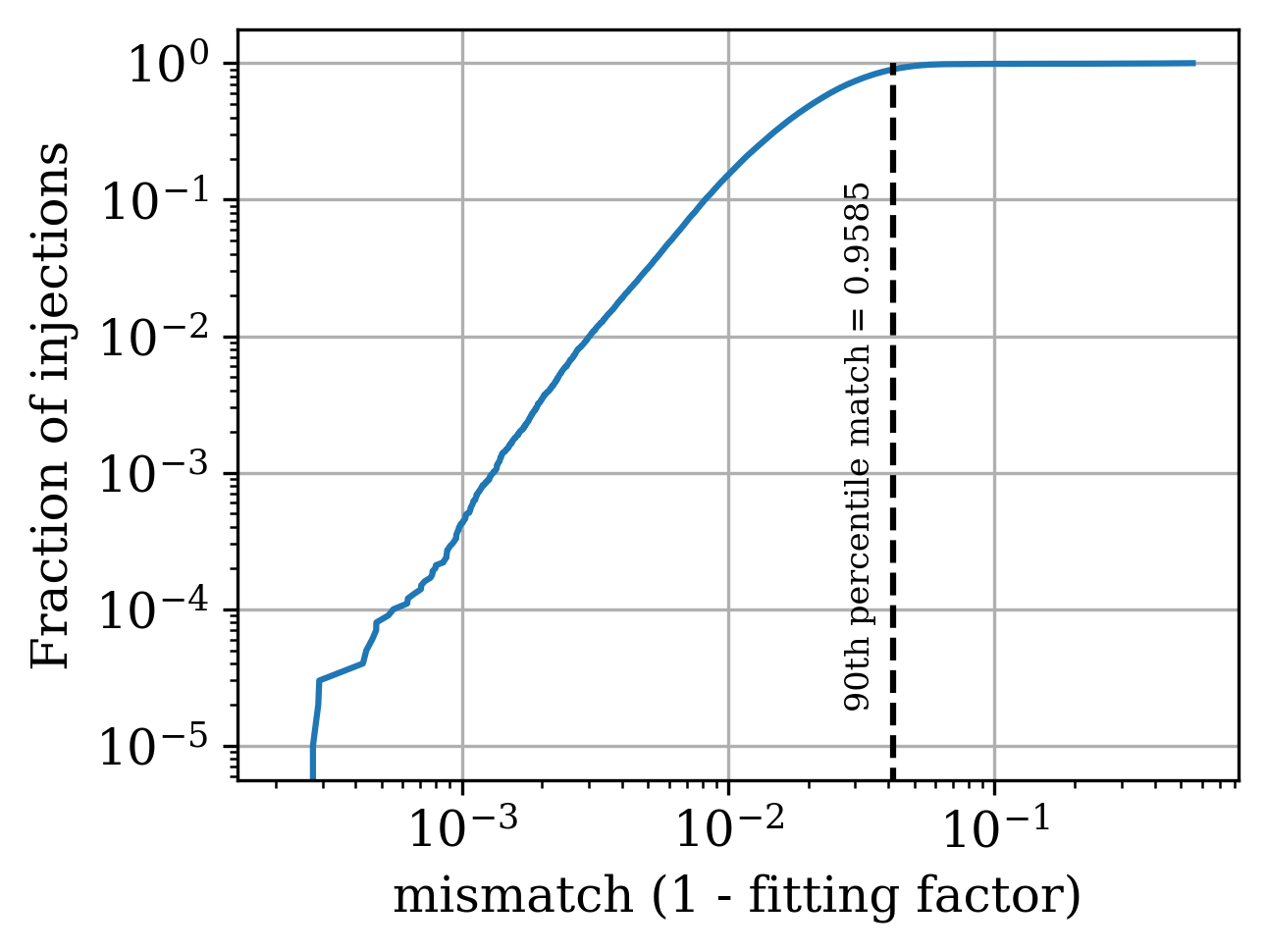}
		\caption{\protect Cumulative histogram of mismatches.}
		\label{fig:results_banksim_bnshigh_cumulative}
	\end{subfigure}%
	\begin{subfigure}{0.24\textwidth}
		 \includegraphics[height=3cm]{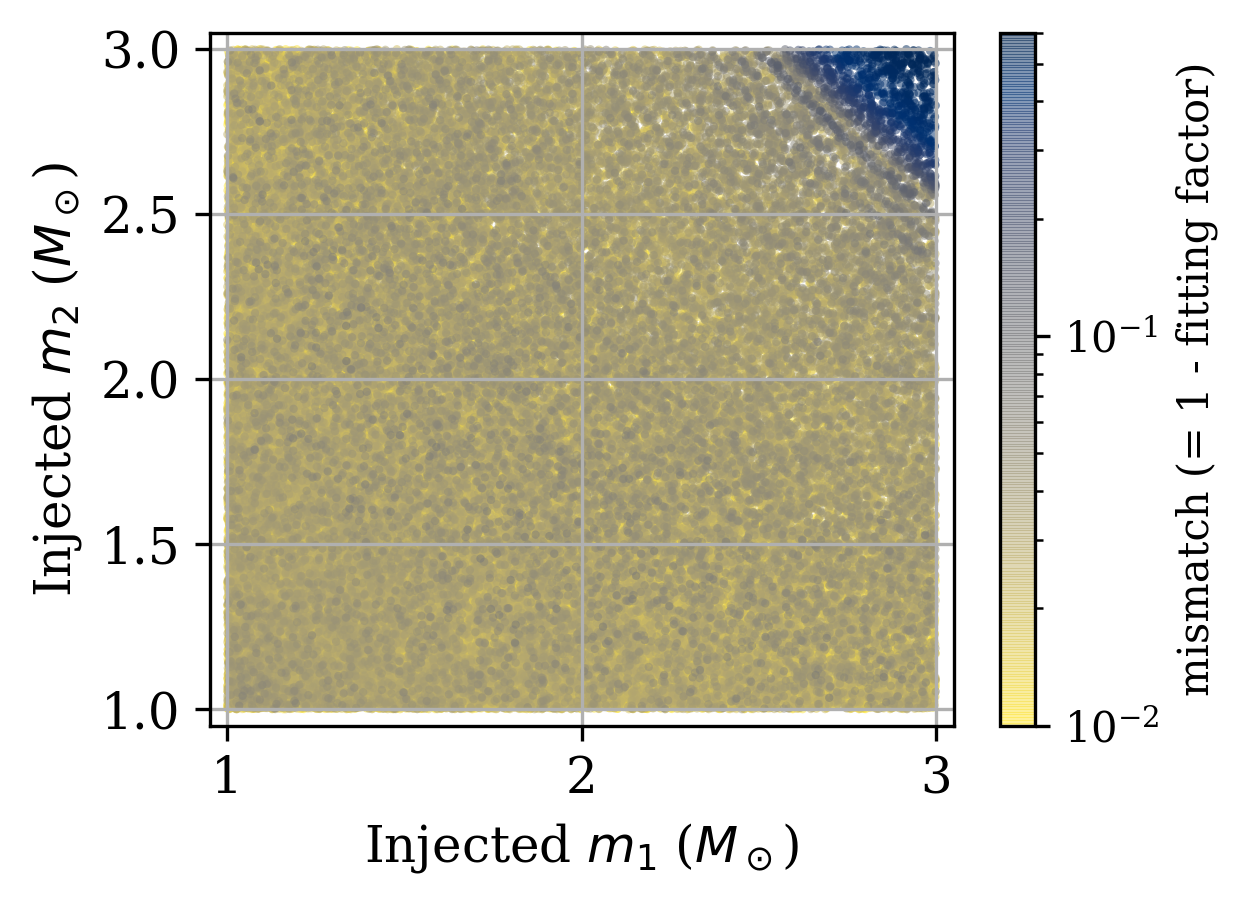}
		\caption{\protect Mismatches in the $m_1$-$m_2$ plane.}
		\label{fig:results_banksim_bnshigh_m1m2}
	\end{subfigure}
	\begin{subfigure}{0.24\textwidth}
		\includegraphics[height=3cm]{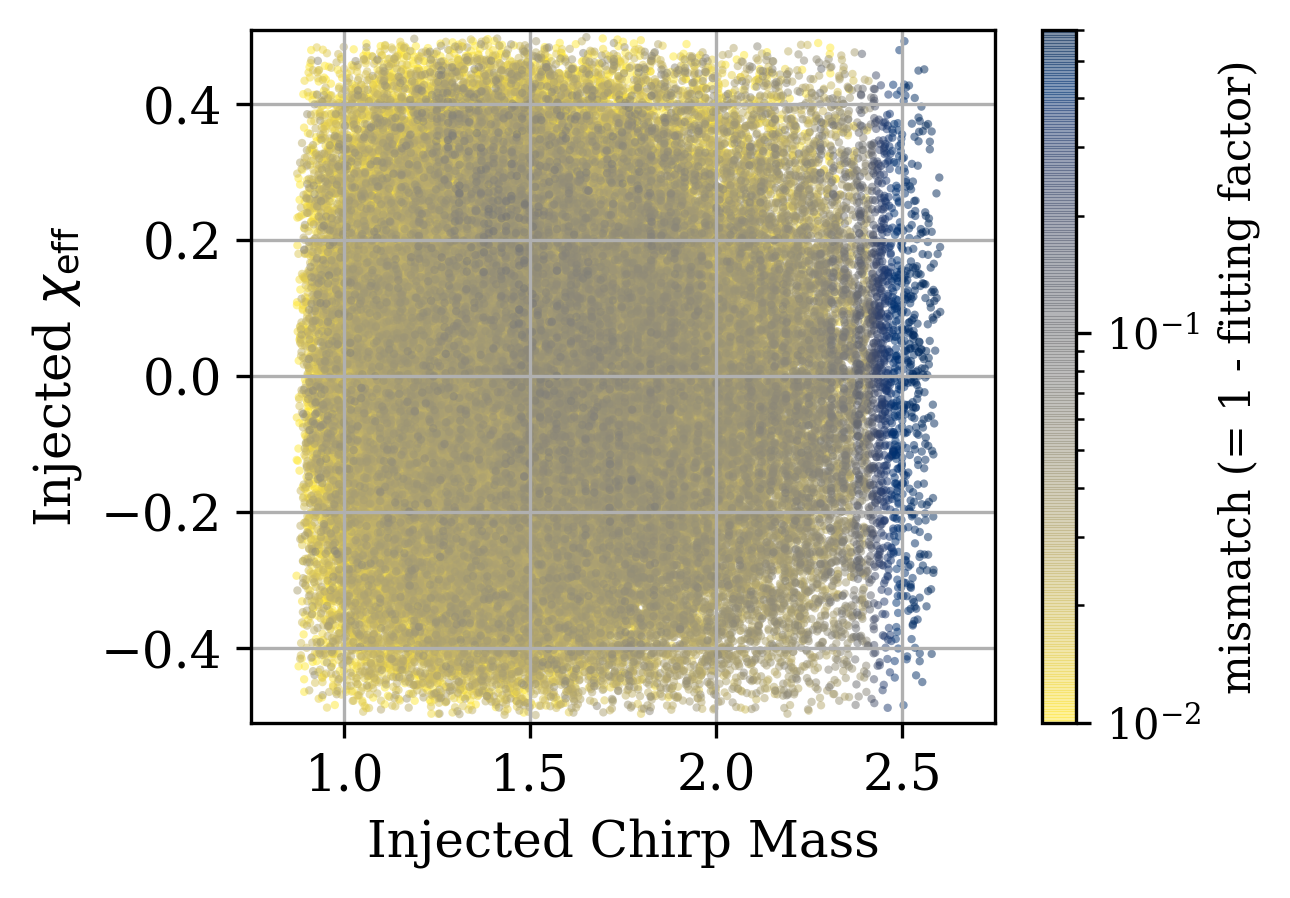}
		\caption{\protect Mismatches in the $\mathcal{M}$-$\chi_{\rm {eff}}$ plane.}
		\label{fig:results_banksim_bnshigh_mcchi}
	\end{subfigure}
	\begin{subfigure}{0.24\textwidth}
		\includegraphics[height=3cm]{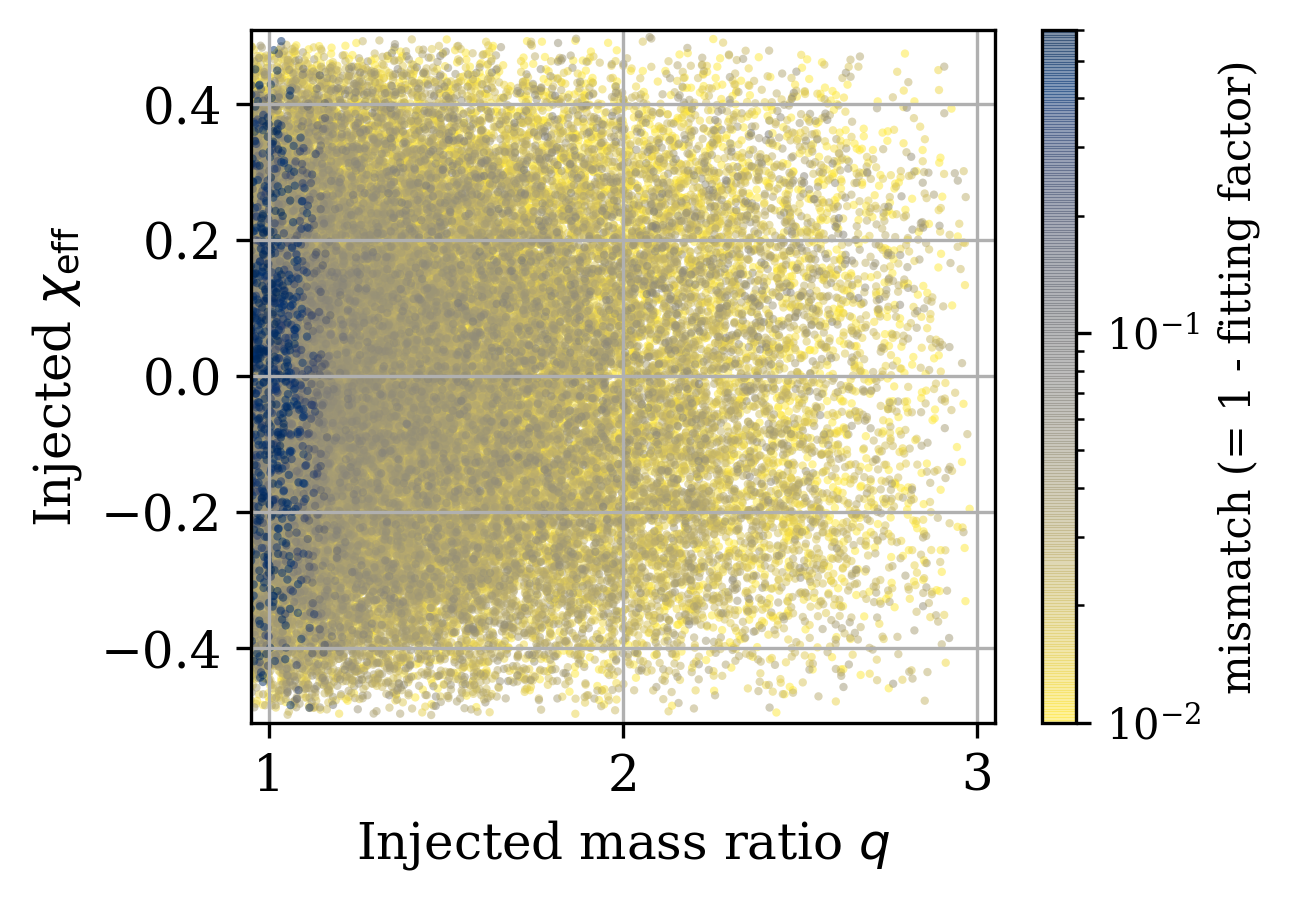}
		\caption{\protect Mismatches in the $q$-$\chi_{\rm {eff}}$ plane.}
		\label{fig:results_banksim_bnshigh_qchi}
	\end{subfigure}
	\caption{\protect Plots for \ac{BNS} simulated signals with high-spins. For purposes of visually presenting the mismatches, mismatches smaller than $10^{-2}$ have been mapped to $10^{-2}$. (a) shows that $90 \%$ of the simulated signals are recovered with a match $\geqslant$ $95.85 \%$.}
	\label{fig:results_banksim_bnshigh}
\end{figure*}

\begin{figure*}[htbp]
	\centering
	\begin{subfigure}{0.24\textwidth}
		\includegraphics[height=3cm]{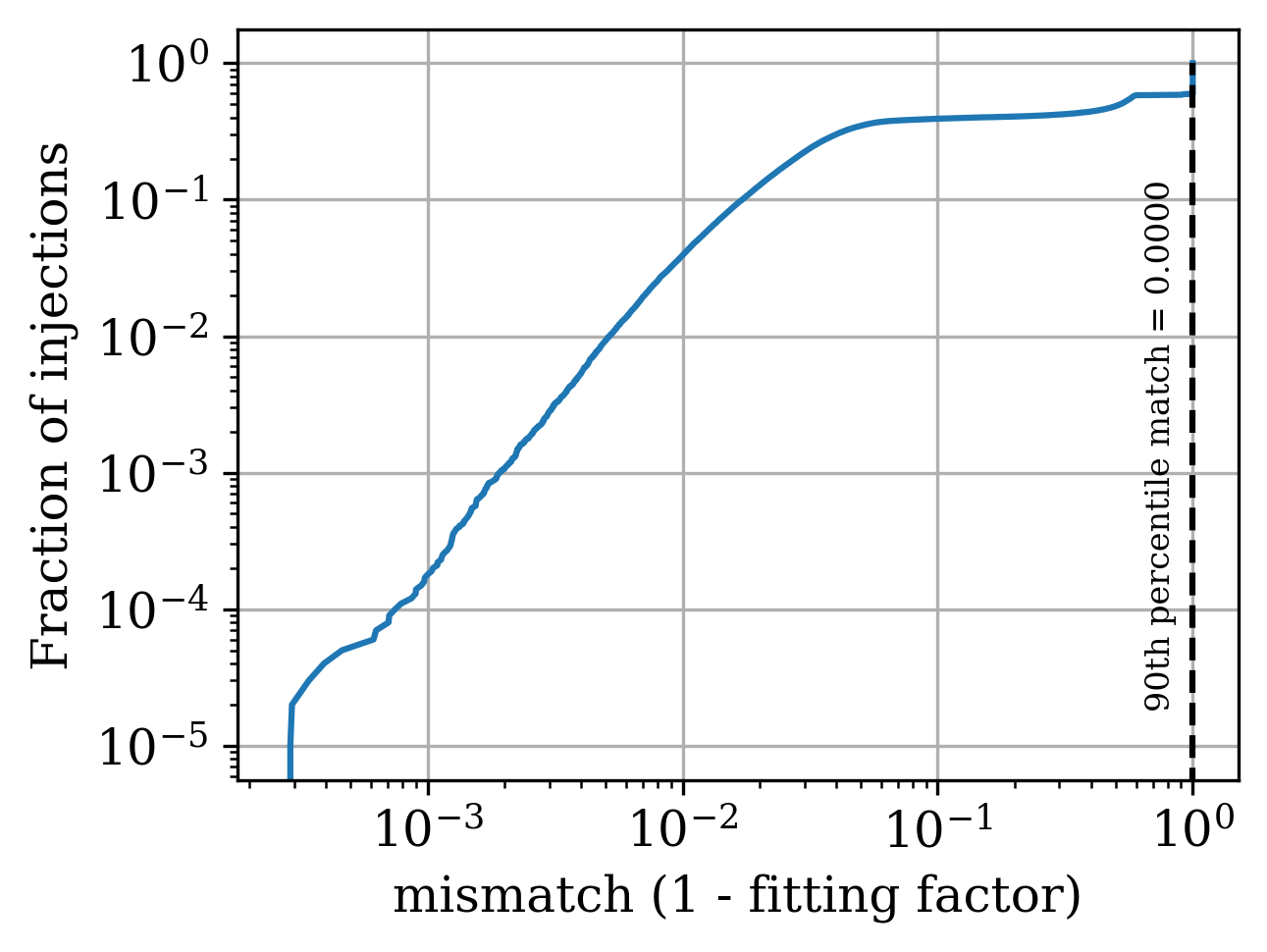}
		\caption{\protect Cumulative histogram of mismatches.}
		\label{fig:results_banksim_nsbhlow_cumulative}
	\end{subfigure}%
	\begin{subfigure}{0.24\textwidth}
		\includegraphics[height=3cm]{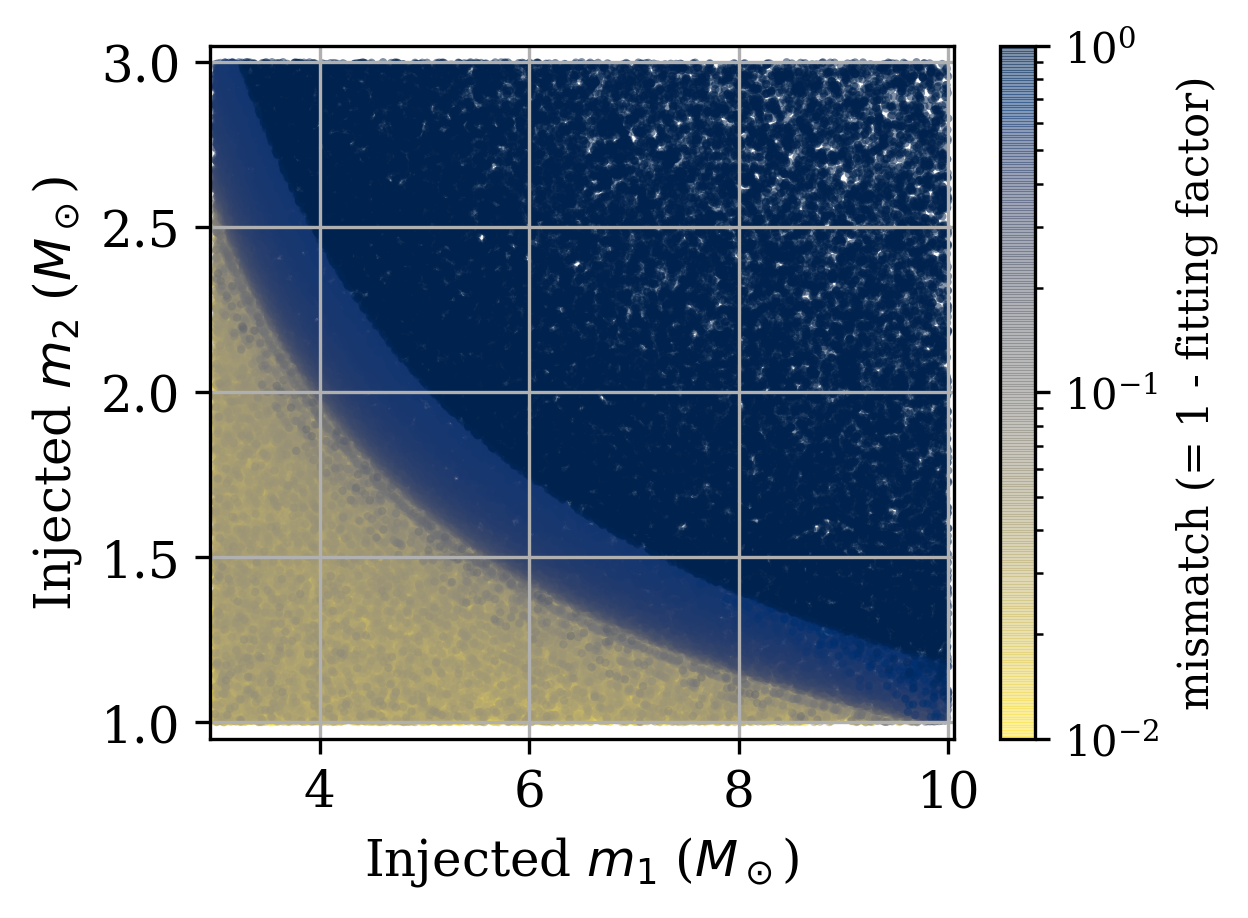}
		\caption{\protect Mismatches in the $m_1$-$m_2$ plane.}
		\label{fig:results_banksim_nsbhlow_m1m2}
	\end{subfigure}
	\begin{subfigure}{0.24\textwidth}
		\includegraphics[height=3cm]{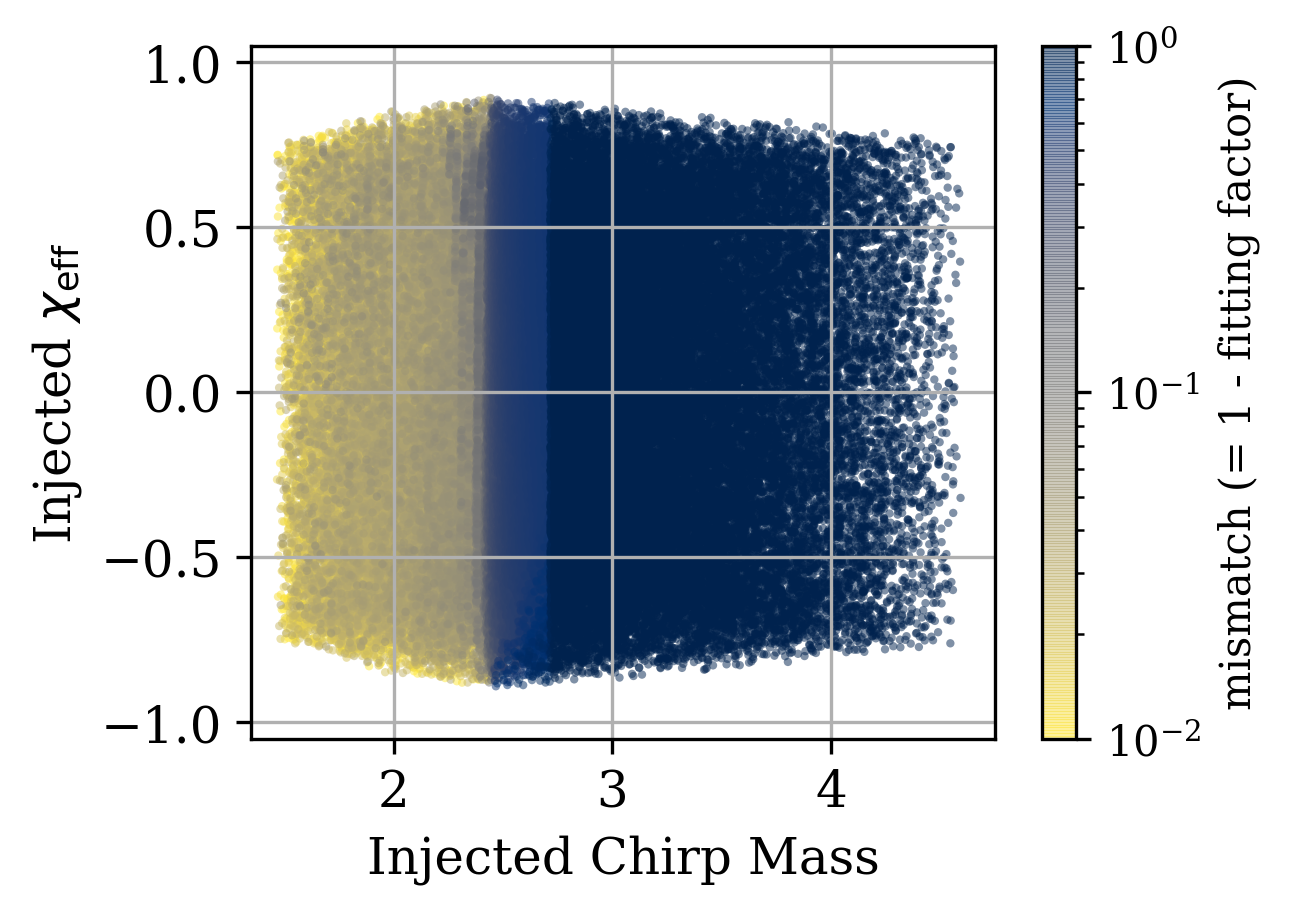}
		\caption{\protect Mismatches in the $\mathcal{M}$-$\chi_{\rm {eff}}$ plane.}
		\label{fig:results_banksim_nsbhlow_mcchi}
	\end{subfigure}%
	\begin{subfigure}{0.24\textwidth}
		\includegraphics[height=3cm]{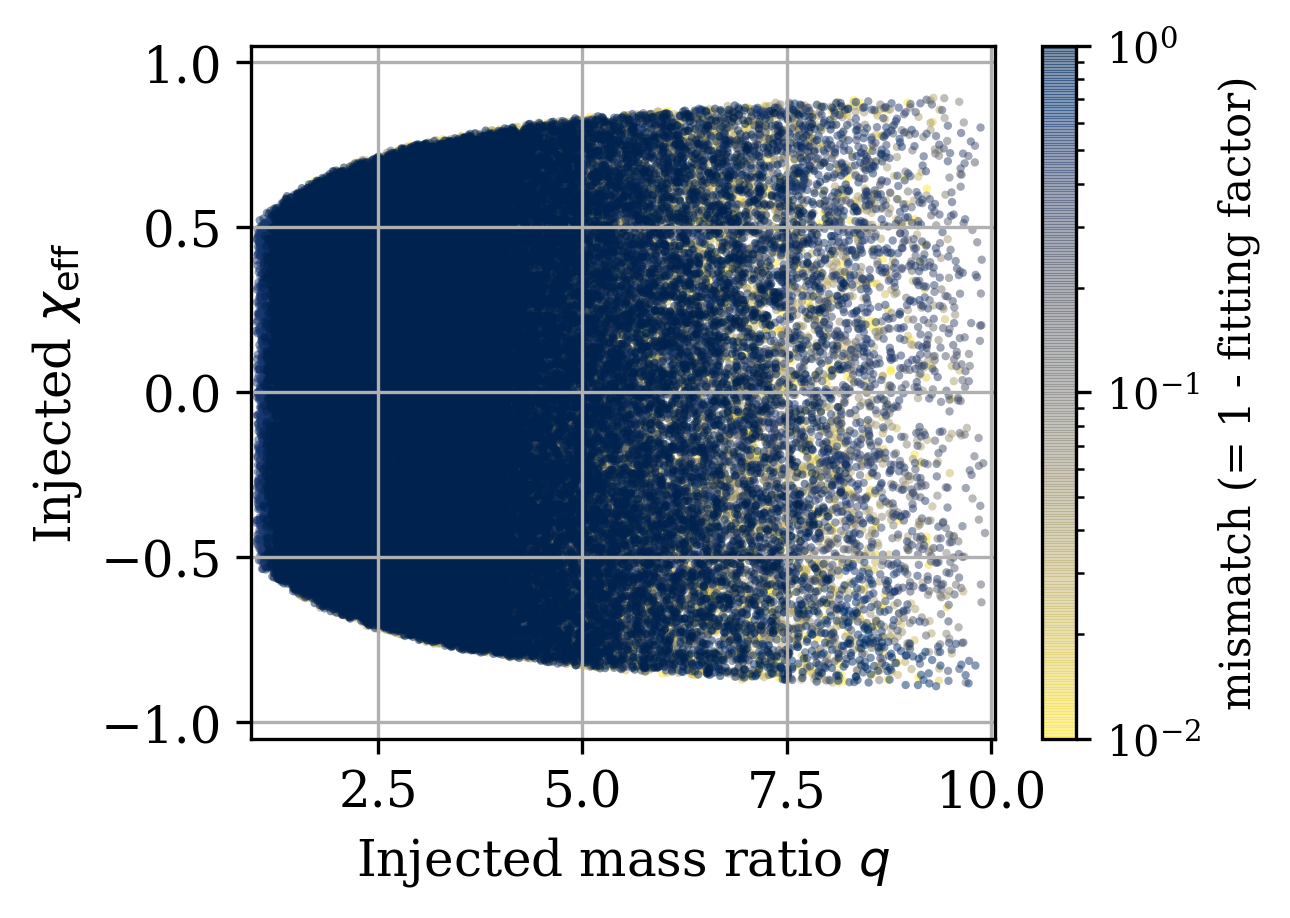}
		\caption{\protect Mismatches in the $q$-$\chi_{\rm {eff}}$ plane.}
		\label{fig:results_banksim_nsbhlow_qchi}
	\end{subfigure}
	\caption{\protect Plots for \ac{NSBH} simulated signals with low-spins. For purposes of visually presenting the mismatches, mismatches smaller than $10^{-2}$ have been mapped to $10^{-2}$. (a) shows that $90 \%$ of the simulated signals are recovered with a match $\geqslant$ $0.00 \%$.}
	\label{fig:results_banksim_nsbhlow}
\end{figure*}

\begin{figure*}[htbp]
	\centering
	\begin{subfigure}{0.24\textwidth}
		\includegraphics[height=3cm]{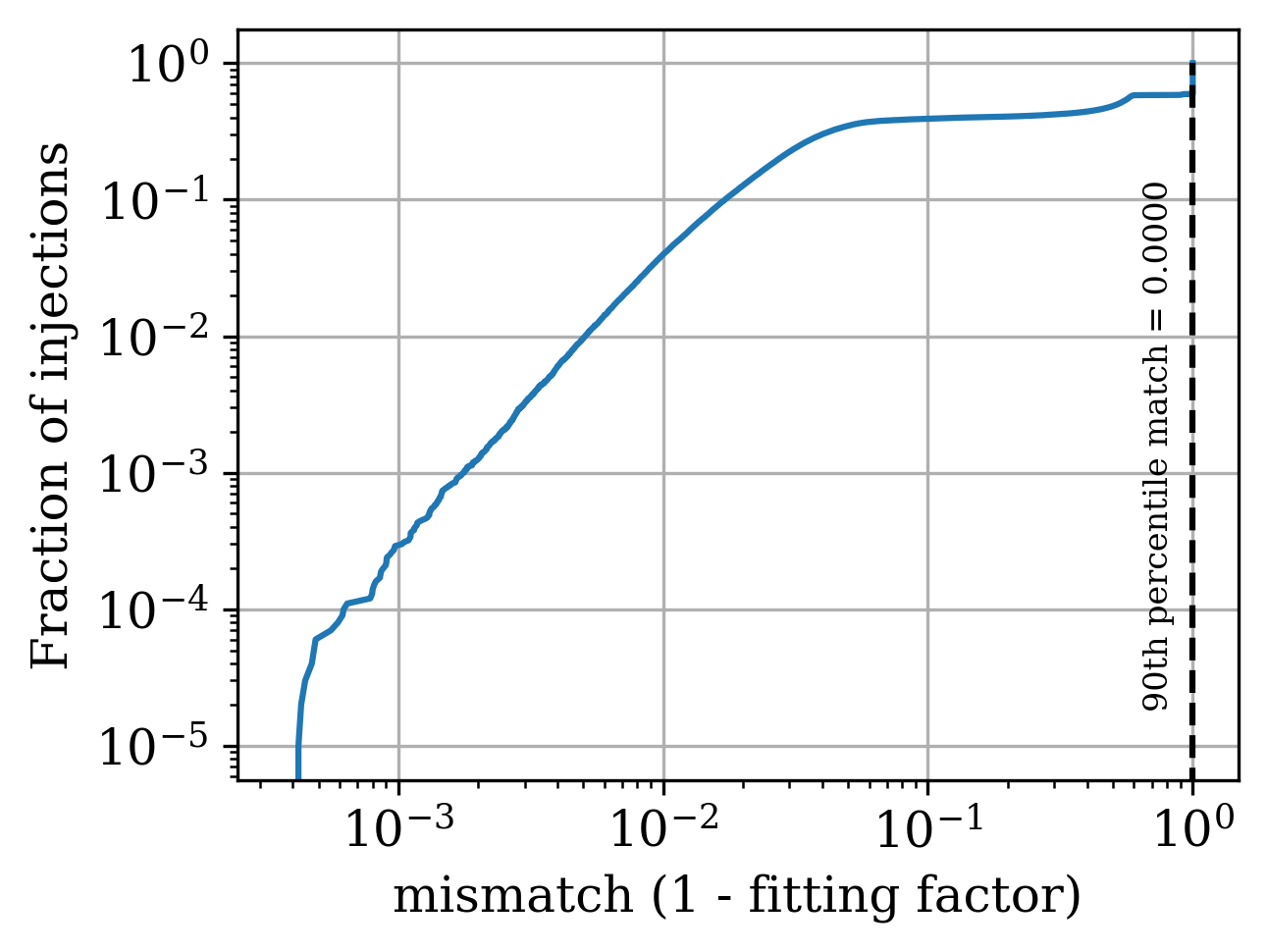}
		\caption{\protect Cumulative histogram of mismatches.}
		\label{fig:results_banksim_nsbhhigh_cumulative}
	\end{subfigure}%
	\begin{subfigure}{0.24\textwidth}
		 \includegraphics[height=3cm]{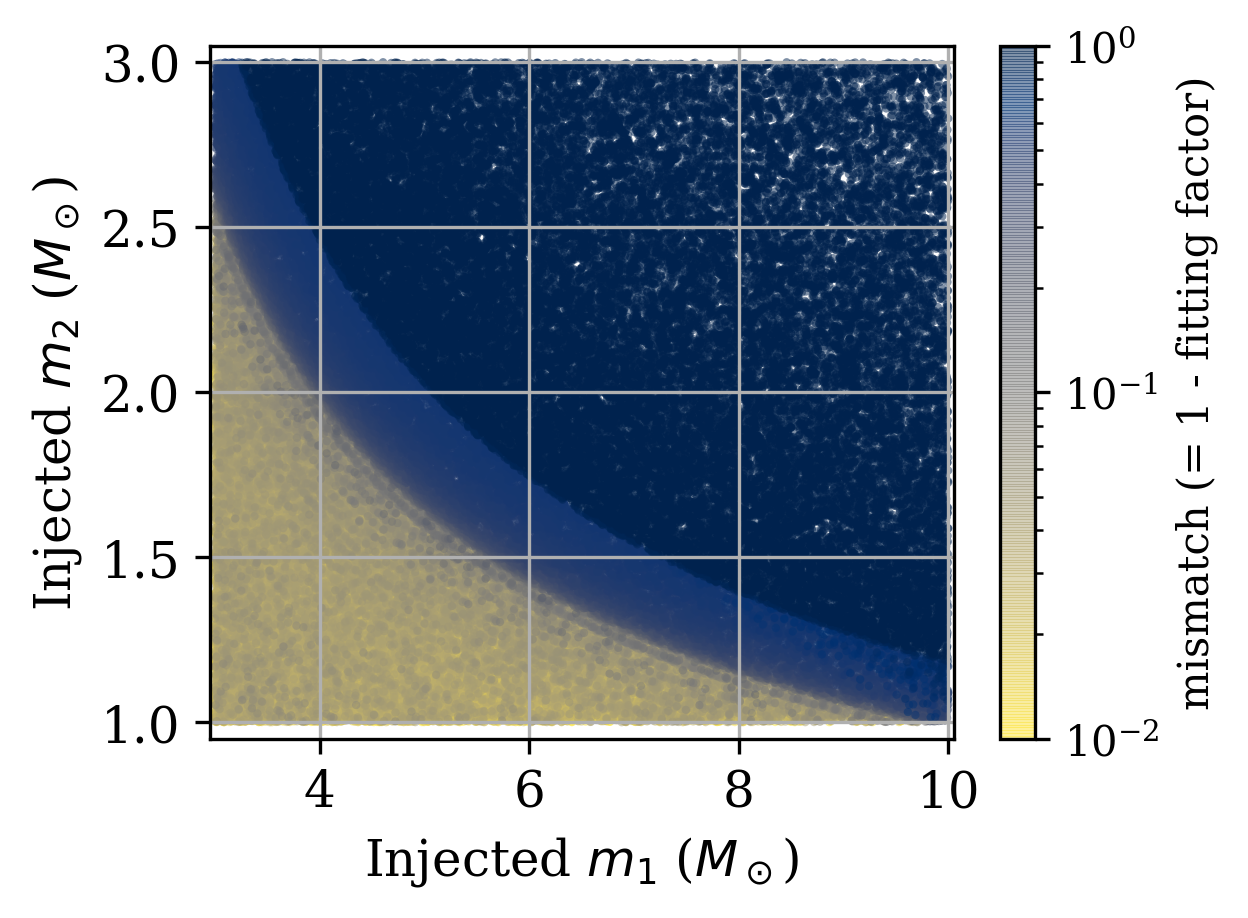}
		\caption{\protect Mismatches in the $m_1$-$m_2$ plane.}
		\label{fig:results_banksim_nsbhhigh_m1m2}
	\end{subfigure}
	\begin{subfigure}{0.24\textwidth}
		\includegraphics[height=3cm]{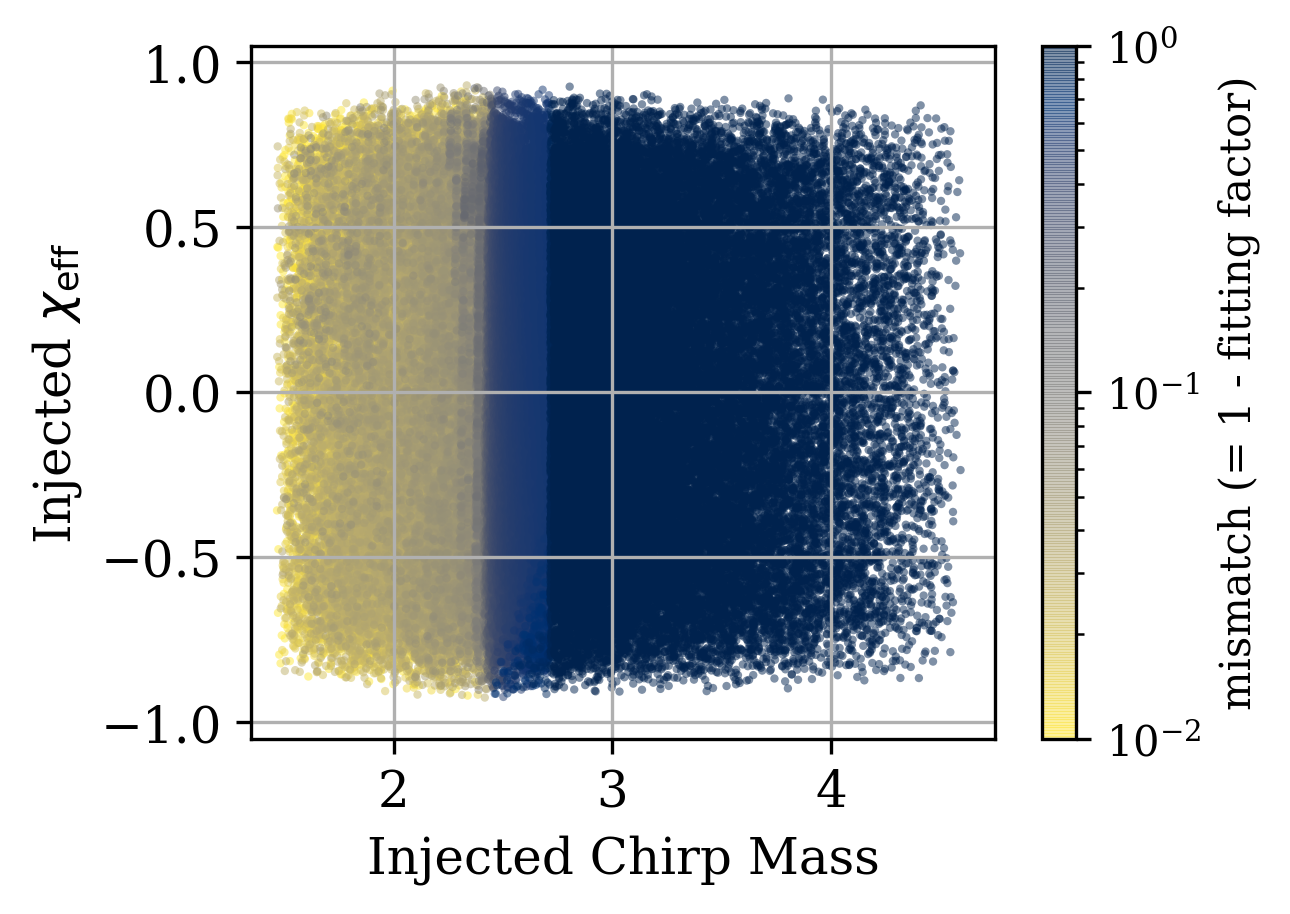}
		\caption{\protect Mismatches in the $\mathcal{M}$-$\chi_{\rm {eff}}$ plane.}
		\label{fig:results_banksim_nsbhhigh_mcchi}
	\end{subfigure}%
	\begin{subfigure}{0.24\textwidth}
		\includegraphics[height=3cm]{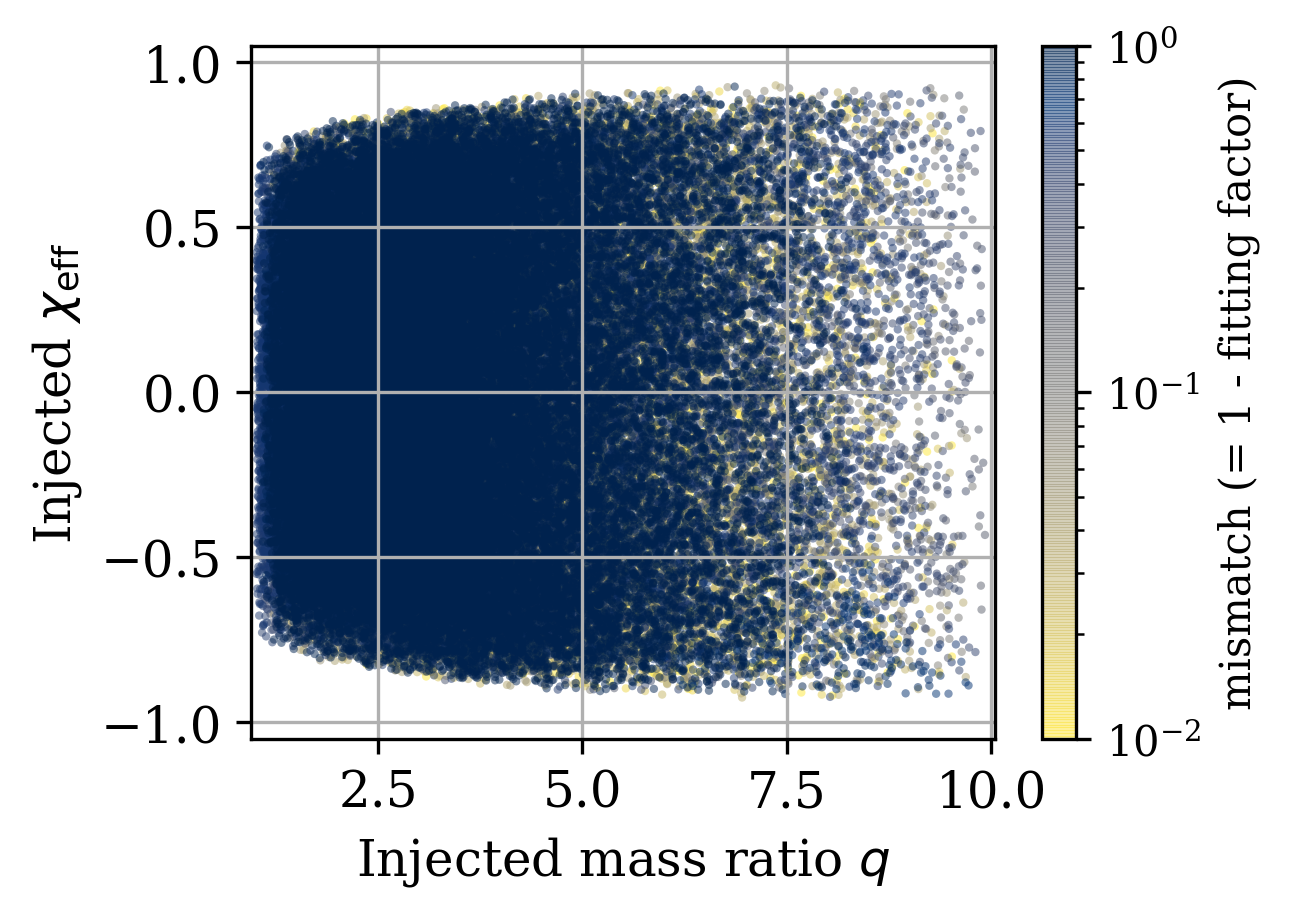}
		\caption{\protect Mismatches in the $q$-$\chi_{\rm {eff}}$ plane.}
		\label{fig:results_banksim_nsbhhigh_qchi}
	\end{subfigure}
	\caption{\protect Plots for \ac{NSBH} simulated signals with high-spins. For purposes of visually presenting the mismatches, mismatches smaller than $10^{-2}$ have been mapped to $10^{-2}$. (a) shows that $90 \%$ of the simulated signals are recovered with a match $\geqslant$ $0.00 \%$.}
	\label{fig:results_banksim_nsbhhigh}
\end{figure*}

In the $m_1$-$m_2$ space, chirp mass forms an ambiguity contour leading to a degeneracy where binaries with different component masses can have the same chirp mass. Due to this, the chirp mass $\mathcal{M}$ of templates in the \ac{SSM} bank coincides with typical chirp masses of \ac{BNS} and \ac{NSBH} signals. Therefore, we perform efficacy tests in the \ac{BNS} and \ac{NSBH} parameter space in addition to the studies discussed in section~\ref{subsubsection:results_banksim_tempbankregion} to assess the response of this bank towards electromagnetically-bright events, which we define to be those in which at least one component is \ac{NS}.
The parameters of the \ac{BNS} and \ac{NSBH} simulated signals are summarized in Table \ref{table:em_banksim_params}. The motivations for limiting \ac{NS} masses to $3.0 M_\odot$ can be found in ~\cite{Sakon:2024}. As done for the bank simulation tests presented in Section ~\ref{subsubsection:results_banksim_tempbankregion}, we sample component masses from a log distribution.

\begin{table}
\begin{center}
	\begin{tabular}{ l | c | c | c | c}
		\hline
		Parameter & \BNS Low & \BNS High & NSBH Low & NSBH High \\
		\hline
		\hline
		$m_1$ ($M_\odot$) & $\left[1.0, 3.0 \right]$ & $\left[1.0, 3.0 \right]$ & $\left[3.0, 10.0 \right]$ & $\left[3.0, 10.0 \right]$  \\
		$m_2$ ($M_\odot$) & $\left[1.0, 3.0 \right]$ & $\left[1.0, 3.0 \right]$ & $\left[1.0, 3.0 \right]$ & $\left[1.0, 3.0 \right]$  \\
		$s_{1,z}$  & $\left[-0.05, 0.05\right]$ & $\left[-0.5, 0.5\right]$ & $\left[-0.99, 0.99\right]$ & $\left[-0.99, 0.99\right]$  \\
		$s_{2,z}$  & $\left[-0.05, 0.05\right]$ & $\left[-0.5, 0.5\right]$ & $\left[-0.05, 0.05\right]$ & $\left[-0.5, 0.5\right]$  \\
		Count & 100,000 & 100,000 & 100,000 & 100,000  \\
		\hline
	\end{tabular}
	\caption{Parameter space constraints for electromagnetically-bright parameter regions.}
	\label{table:em_banksim_params} 
\end{center}
\end{table}

We check the effectualness of the bank towards low-spinning and high-spinning \ac{BNS} and \ac{NSBH} populations.
We use a lower frequency cutoff of $45$Hz for these bank simulation results.
Figs. \ref{fig:results_banksim_bnslow}, \ref{fig:results_banksim_bnshigh},  \ref{fig:results_banksim_nsbhlow}, and \ref{fig:results_banksim_nsbhhigh} show the mismatches between the templates in the template bank and the \ac{BNS} and \ac{NSBH} simulated signals.
With templates in the \ac{SSM} bank, $90$ $\%$ of the \ac{BNS} simulated signals with low-spin constraints have a match of $95.83$ $\%$ or higher, $95.85$ $\%$ 
or higher for \ac{BNS} simulated signals with high-spin constraints, and $0$ $\%$ for 
low-spinning and high-spinning \ac{NSBH} simulated signals.  
Figs. \ref{fig:results_banksim_bnslow}, \ref{fig:results_banksim_bnshigh},  \ref{fig:results_banksim_nsbhlow}, and \ref{fig:results_banksim_nsbhhigh} show that the \ac{BNS} and \ac{NSBH} simulated signals have low mismatches with the template bank in low $\mathcal{M}$ regions, but the mismatches spike above $\mathcal{M} \gtrapprox 2.4 M_\odot$. 
This is expected given that the highest $\mathcal{M}$ in the bank is $2.4 M_\odot$.
From these studies, we find that the \ac{O4} offline \ac{SSM} template bank may recover \ac{GW} signals 
from electromagnetically-bright events like \acp{BNS} and \acp{NSBH} that have $\mathcal{M} \lessapprox 2.4 M_\odot$.

\section{Conclusion}

We have presented template banks intended for use in an offline and low-latency \SSM searches as well as bank simulation results that show the banks to be sufficient for achieving their respective design objectives. These banks will be used by the \GSTLAL search pipeline in both low-latency (the bank in App. \ref{supp:ll_bank}) and archive (the ``offline" bank in \ref{section:design}) searches. Similar to the bank produced for the O4 low-latency \GSTLAL search \cite{Sakon:2024}, the banks presented in this paper make use of computationally efficient, geometric techniques from the \MANIFOLD software. The offline bank for \SSM searches presented in this paper spans the mass parameter space of $0.2$ to $10$ $M_\odot$ in the larger component and $0.2$ to $1.0$ $M_\odot$ in the smaller component, the spin parameter space of $-0.9$ to $0.9$ in the larger component and $-0.05$ to $0.05$ in the smaller component, and the mass ratio parameter space of $1$ to $10$. The \PSD used was from LIGO O4a and the low frequency cutoff was set to $45$ Hz with a maximum waveform duration of $128$ seconds. The bank simulations have shown that the banks presented in this paper have sufficient efficacy for use in their respective searches. These banks will be used on data from the fourth observing run of Advanced LIGO, Virgo, and KAGRA.

Further, we have also shown required methodological improvements to the treebank algorithm in \MANIFOLD necessary to produce a template bank for \SSM searches, including enhanced numerical stability of metric estimation at exceedingly low masses and boundary padding for improving the recovery of the lowest mass signals in the search. The neighborhood metric estimation was effective in reducing metric estimation failures, as measure by the presence of negative eigenvalues in $g$. The boundary padding was effective in improving the lower matches along the low-mass boundary, as measured by the fitting factor during low-low simulations. These techniques have been added to \MANIFOLD and will be available for future use should subsequent searches also want to cover lower-mass regions of the parameter space.

\begin{acknowledgments}
This research has made use of data, software and/or web tools obtained from the
Gravitational Wave Open Science Center (https://www.gw-openscience.org/ ), a
service of \ac{LIGO} Laboratory, the \ac{LSC} and the Virgo
Collaboration. 
This material is based upon work supported by \ac{NSF}'s \ac{LIGO} Laboratory which 
is a major facility fully funded by the \ac{NSF}. 
We especially made heavy use of the \ac{LVK} Algorithm
Library~\cite{LAL, lalsuite}. 
\ac{LIGO} was constructed by the California Institute of Technology and the 
Massachusetts Institute of Technology with funding from the United States 
\ac{NSF} and operates under cooperative agreements 
PHYS-$0757058$ and PHY-$0823459$.
In addition, the Science and Technology Facilities Council (STFC) of the United 
Kingdom, the Max-Planck-Society (MPS), and the State of Niedersachsen/Germany 
supported the construction of Advanced \ac{LIGO} and construction and operation of the 
GEO600 detector. 
Additional support for Advanced \ac{LIGO} was provided by the Australian Research Council.  
Virgo is funded, through the European Gravitational Observatory (EGO), by the 
French Centre National de Recherche Scientifique (CNRS), the Italian Istituto 
Nazionale di Fisica Nucleare (INFN) and the Dutch Nikhef, with contributions by 
institutions from Belgium, Germany, Greece, Hungary, Ireland, Japan, Monaco, 
Poland, Portugal, Spain.

The authors are grateful for computational resources provided by the 
the \ac{LIGO} Lab culster at the \ac{LIGO} Laboratory and supported by 
PHY-$0757058$ and PHY$-0823459$, the Pennsylvania State University's Institute 
for Computational and Data Sciences gravitational-wave cluster, 
and supported by 
OAC-$2103662$, PHY-$2308881$, PHY-$2011865$, OAC-$2201445$, OAC-$2018299$, 
and PHY-$2207728$.  
CH Acknowledges generous support from the Eberly College of Science, the 
Department of Physics, the Institute for Gravitation and the Cosmos, the 
Institute for Computational and Data Sciences, and the Freed Early Career Professorship.
We thank Ines Bentara for their helpful comments and suggestions.

\end{acknowledgments}

\appendix
\section{Metric Hyperellipses}
\label{supp:ellipse}

\subsubsection{Necessary Assumptions for Hyperellipsoids}
In general, a level surface of a Riemannian metric cannot be assumed to be an ellipsoid, as variations in curvature over the manifold could result in an arbitrarily warped, smooth, closed surface. However, within a local ball of radius $r$ such that $r^2 \ll \frac{1}{R}$ where $R$ is the Ricci curvature, we assume the metric to be locally constant. In this context, the metric is constant, symmetric, and positive definite, which is sufficient to describe a hyperellipsoid. The canonical form of the equation for a hyperellipsoid is given as
\begin{equation}
\label{ref:eq-ellipse-canonical}
    \frac{x_1^2}{a_1^2} + \frac{x_2^2}{a_2^2} + \cdots + \frac{x_n^2}{a_n^2} = 1
\end{equation}
Similarly, a hyperellipsoid $\mathcal{E}$ can be constructed by deforming an $n$-sphere using linear transformations along each principle axis. We generalize the above canonical description to more broadly define a hyperellipsoid as
\begin{equation}
\label{ref:eq-ellipse}
    \mathcal{E} \equiv \{p\ |\ A_{\mu\nu} x^\mu(p) x^\nu(p) = C\}
\end{equation}
 where $A$ is a constant matrix that is positive semi-definite and $C \in \mathbb{R}$ is a constant, often chosen as $C=1$ for convenience. In cases related to metric distance, this constant is chosen to be the defining distance squared of the surface. Note that for bank creation, the mismatch is already a squared-distance.

\subsubsection{Eigenvalue Decomposition \& Principal Axes}

The eigenvalue decomposition finds solutions to the eigenvalue equation, defined below, of the given matrix $A$. 
\begin{equation}
A_{\mu\nu}v^\nu = \lambda v^\mu   
\end{equation}
This is achieved by diagonalizing the matrix $A$, which results in a diagonal matrix of eigenvalues $\Lambda$. Note that $\Lambda_{ij} = \delta_{ij}\lambda_j$, where $\lambda_i$ is the eigenvalue corresponding to the $i$-th solution of the eigenvalue equation, $v_i^\mu$. In the basis of eigenvectors, the generalized ellipsoid definition in Eq. \ref{ref:eq-ellipse} reduces to the canonical form in Eq. \ref{ref:eq-ellipse-canonical}. 
\begin{equation}
\begin{split}
    1 &= \Lambda_{\mu\nu} x^\mu x^\nu = \delta_{\mu\nu} x^\mu x^\nu \\
    &= \lambda_1 (x^1)^2 + \cdots + \lambda_n (x^n)^2 \\
    &= \frac{(x^1)^2}{(a^1)^2} + \cdots + \frac{(x^n)^2}{(a^n)^2}
\end{split}
\end{equation}
where $a^i \equiv \sqrt{1 / \lambda^i}$ is a vector of scale factors with index $i$ referring to the $i$-th eigenvector $v_i^\mu$. Using these scale factors, it is possible to recover the principal axis vectors $p_i^\mu$ of the ellipsoid, where $p_i^\mu \equiv a_i v_i^\mu$. The principal axis vectors have the virtue of being orthogonal in the ambient space, which can be more natural or intuitive when dealing with ellipsoid geometry. They provide a natural basis for reconstructing ellipsoid points efficiently in code by requiring fewer matrix operations.

%
%
%
%
%

\section{Template Bank for low-latency \ac{SSM} search}\label{supp:ll_bank}

In previous observing runs ~\cite{Abbott:2018, Abbott:2019, Abbott:2022, Abbott:2023}, the \GSTLAL pipeline searched for binaries with at least one \ac{SSM} component after archival data was available.
In \ac{O4}, \GSTLAL performs \ac{SSM} searches in near real-time, i.e., low-latency, motivated by the discovery of a previously unobserved class of compact objects along with the possibility of detecting electromagnetically-bright mergers that lie outside the search parameter space of typical searches for compact binary mergers. This search is the first low-latency \ac{SSM} search performed by the \GSTLAL pipeline. In this section, the template bank used for this low-latency search is discussed. 

For the low-latency search, we continue to target systems with at least one \ac{SSM} compact object. 
Since points in the mass parameter space become more correlated at lower masses, a higher number of templates is needed to optimally cover the target region as the minimum component masses are lowered.
To mitigate this, the smallest component masses are restricted to $0.5~M_\odot$ for the low-latency search, unlike in the archival searches, i.e., offline searches, where component masses go down to $0.2~M_\odot$. 
Since the sensitivity of the search increases with the chirp mass of \acp{BBH} ~\cite{Abbott:2023}, the choice of a higher minimum component mass restricts the search to higher VT regions of the parameter space.

Like all archival \ac{SSM} searches, only spin-aligned templates are included.
Therefore, each template only has non-zero z-component spins for each object which contribute to the effective spin parameter, $\chi_{\rm {eff}}$.
Additionally, \MANIFOLD ~\cite{Hanna:2022} assigns equal z-component spins to both components such that $\chi_{\rm {eff}} = s_{\rm {i,z}}$.   
The effective spin parameter, $\chi_{\rm {eff}} (=s_{\rm {i,z}})$, is further restricted to $\pm 0.3$ based on ~\cite{Hanna:2022} where it is shown that short duration non-stationarities in strain data have $\chi_{\rm {eff}} < -0.3$.
 
Table~\ref{table:llssmbank} lists the parameter space and configuration that was used in the construction of this bank.

\begin{table}
\begin{center}
	\begin{tabular}{ l | l }
		\hline
		Parameter & \GSTLAL \\
		\hline
		\hline
		Primary mass, $m_1$ & $\in [0.5, 10 M_\odot]$  \\
		Secondary Mass, $m_2$ & $\in [0.5, 1 M_\odot]$  \\
		Mass ratio, $q=m_1/m_2$ & $\in [1, 10]$  \\
		Dimensionless spin, $s_{i, z}$ & $\lvert s_{i, z} \rvert <  0.3$ \\
		Lower frequency cut-off & 45 Hz \\
		Higher frequency cut-off & 1024 Hz \\
		Waveform approximant & IMRPhenomD \\
		Minimum match & 0.97 \\ 
		Maximum waveform duration & 128 seconds \\
		\hline
		Total number of templates & 1069651 \\
		\hline
	\end{tabular}
	\caption{Paramter space of the template bank used in \ac{O4a} low-latency \ac{SSM} search by \GSTLAL.}
	\label{table:llssmbank}
\end{center}
\end{table}

\begin{figure}
    \centering
    \begin{subfigure}[b]{9cm}
        \includegraphics[width=\linewidth]{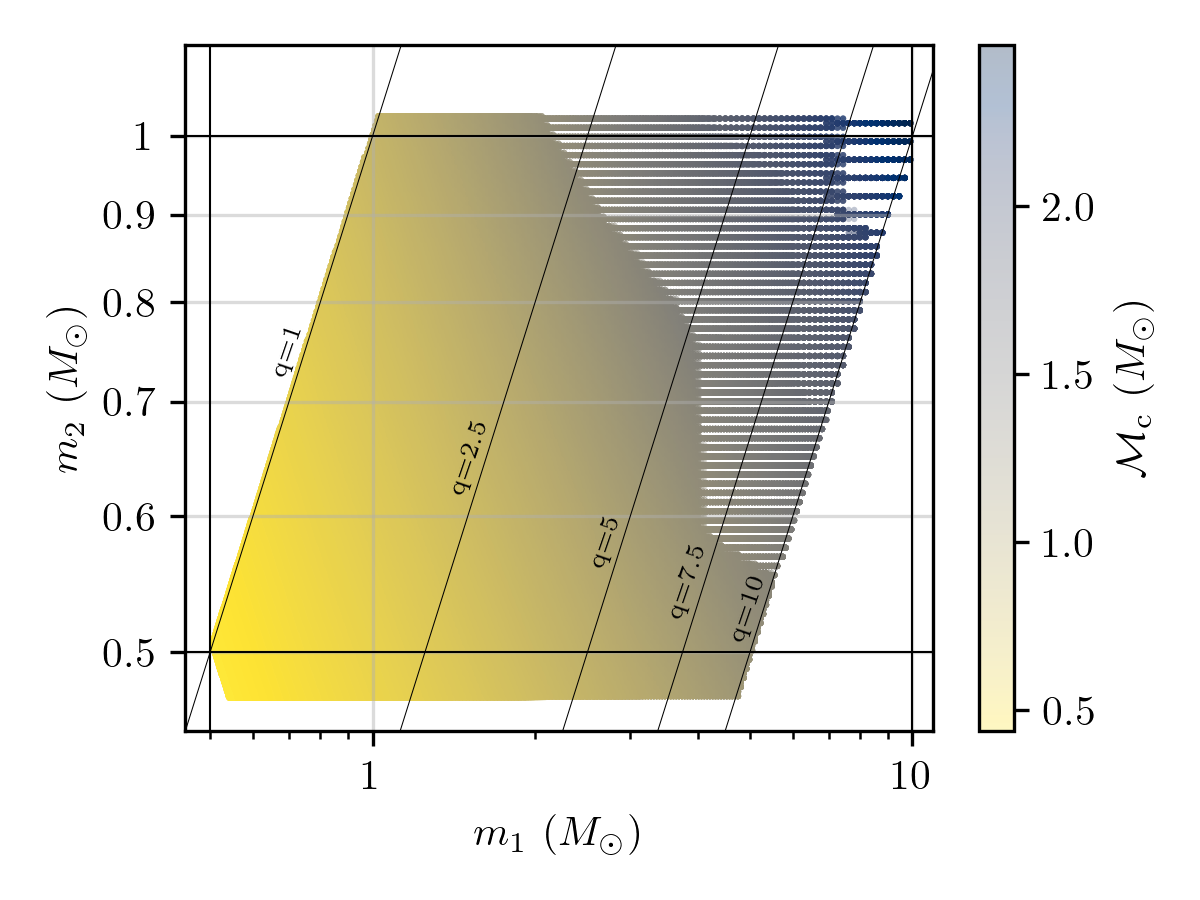}
        \caption{Template placement in the $m_1 - m_2$ space.}
    \end{subfigure}
    ~ 
    \begin{subfigure}[b]{9cm}
        \includegraphics[width=\linewidth]{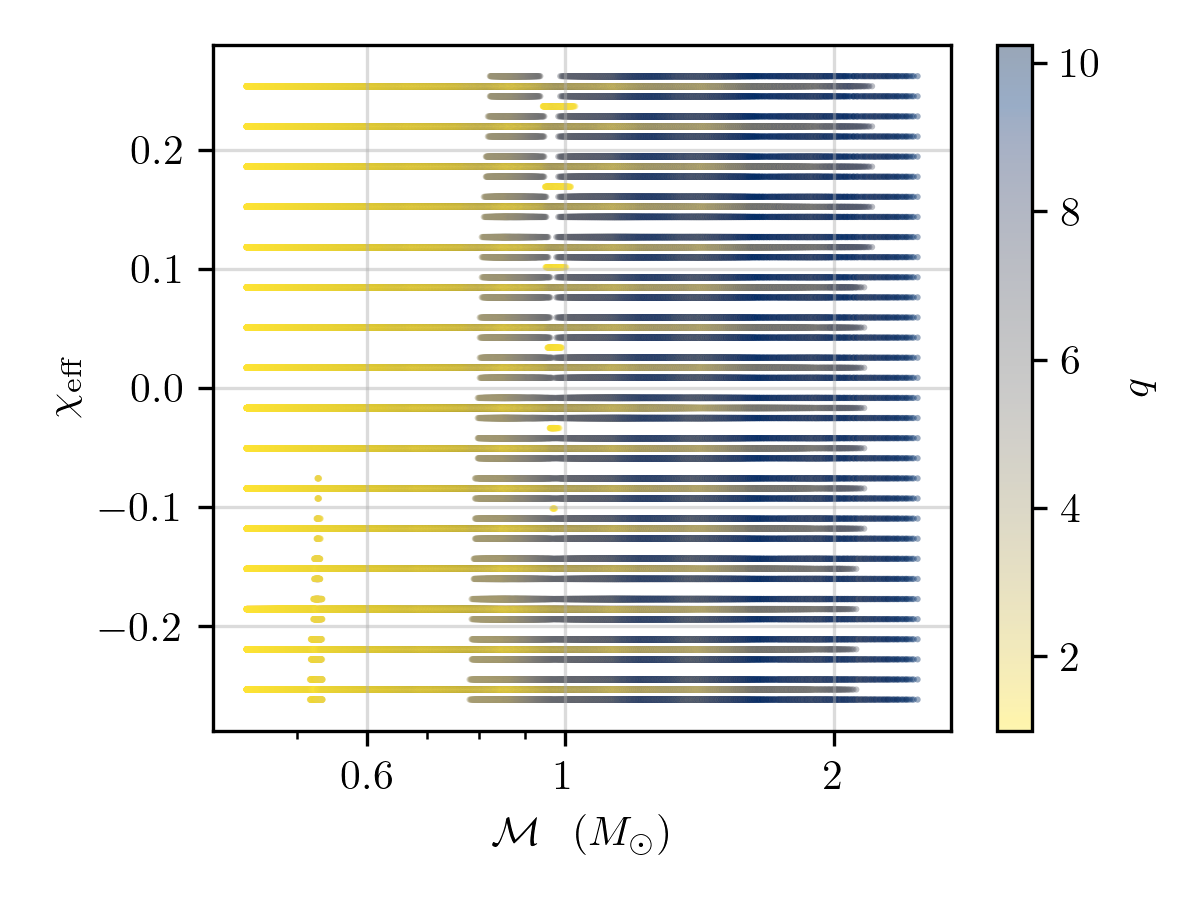}
        \caption{Template placement in the $\mathcal{M} - \chi_{\rm {eff}}$ space.}
    \end{subfigure}
\caption{Template placement in the \ac{O4} low-latency \ac{SSM} bank. The colormaps in (a) and (b) show trends in chirp mass, $\mathcal{M}$ and mass-ratio, $q$ respectively.}
\end{figure}

\subsection{Effectualness of template bank}\label{subsec:banksim}

Bank simulation studies were performed to determine if the constructed bank efficiently covers the target search space.
Since the mass boundaries extend above a solar mass for the primary, this search is sensitive to 
\ac{BNS} and \ac{NSBH} mergers in addition to mergers of \ac{SSM} black holes.
Therefore, we perform bank simulations for the following distinct populations
\begin{itemize}
	\item \textbf{SSM-BBH}: primary target of this search -- \ac{BBH} mergers with $m_1 \in [0.5,10 ~M_\odot]$, $m_2 \in [0.5,1 ~M_\odot]$ and $|\chi_{\rm eff}| < 0.3$,
	\item \textbf{BNS-LOW}: \ac{BNS} mergers with $m_{\rm NS} \in [1,3 ~M_\odot]$ and $|\chi_{\rm eff}| < 0.05$,
	\item \textbf{BNS-HIGH}: \ac{BNS} mergers with $m_{\rm NS} \in [1,3 ~M_\odot]$ and $|\chi_{\rm eff, NS}| < 0.5$,
	\item \textbf{NSBH-LOW}: \ac{NSBH} mergers with $m_{\rm NS} \in [1,3 ~M_\odot]$, $m_{\rm BH} \in [3,10~M_\odot]$, $|\chi_{\rm eff, NS}| < 0.05$ and $|\chi_{\rm eff, BH}| < 0.99$,
	\item \textbf{NSBH-HIGH}: \ac{NSBH} mergers with $m_{\rm NS} \in [1, 3 ~M_\odot]$, $m_{\rm BH} \in [3,10~M_\odot]$, $|\chi_{\rm eff, NS}| < 0.5$ and $|\chi_{\rm eff, BH}| < 0.99$.
\end{itemize}

\begin{figure*}[htbp]
    \centering
    \begin{subfigure}{0.24\textwidth}
        \includegraphics[height=3cm]{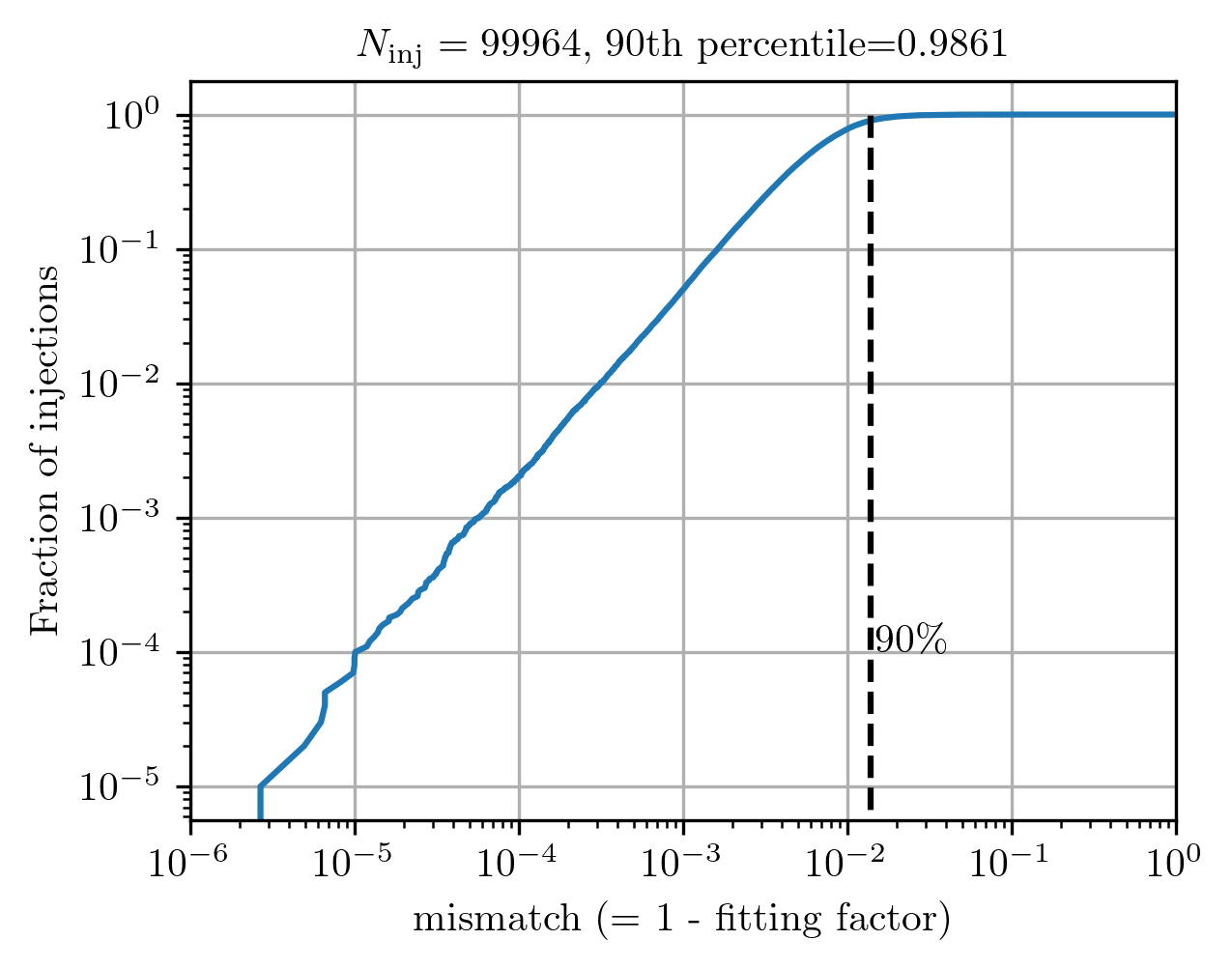}
        \caption{Cumulative histogram of mismatches.}\label{subfig:SSMBBHcumhist}
    \end{subfigure}%
    \begin{subfigure}{0.24\textwidth}
        \includegraphics[height=3cm]{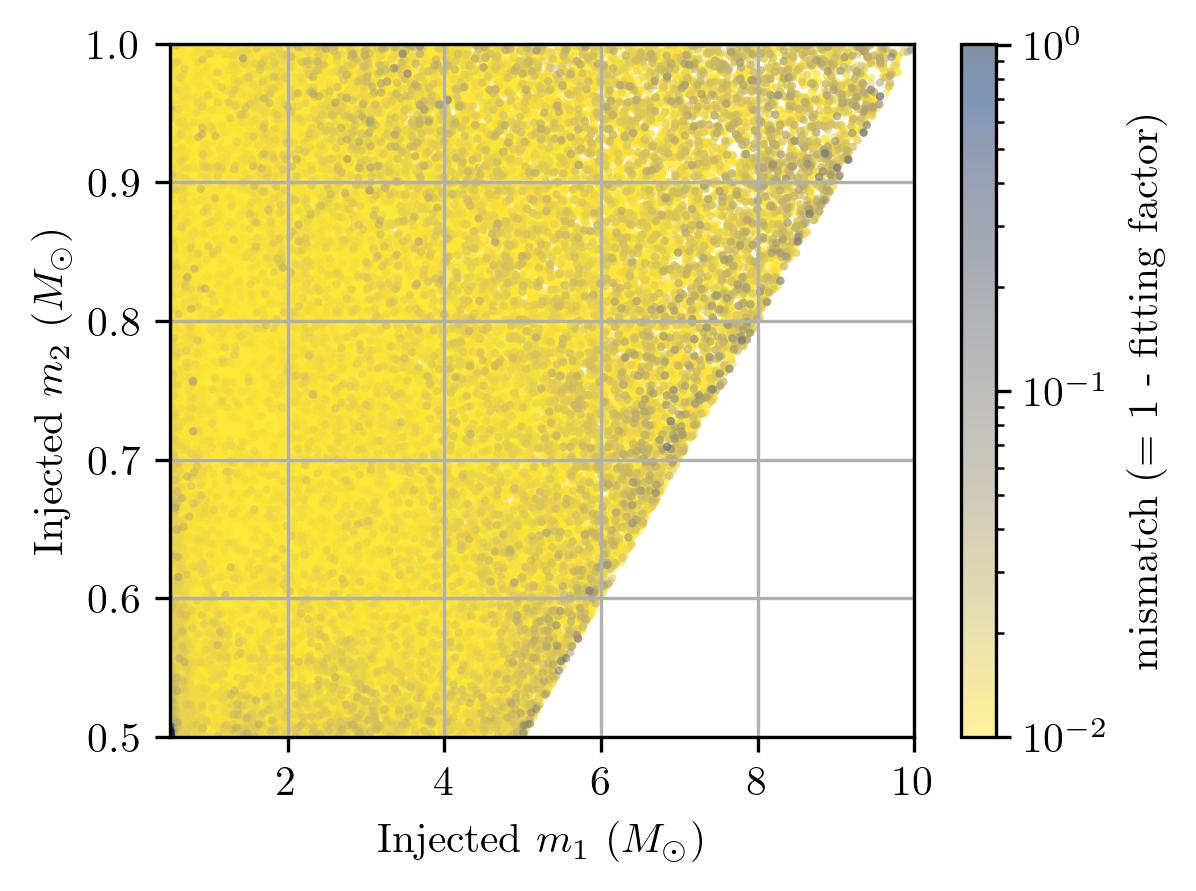}
        \caption{Mismatches in the $m_1 - m_2$ plane.}\label{subfig:SSMBBHm1m2}
    \end{subfigure}
	\begin{subfigure}{0.24\textwidth}
        \includegraphics[height=3cm]{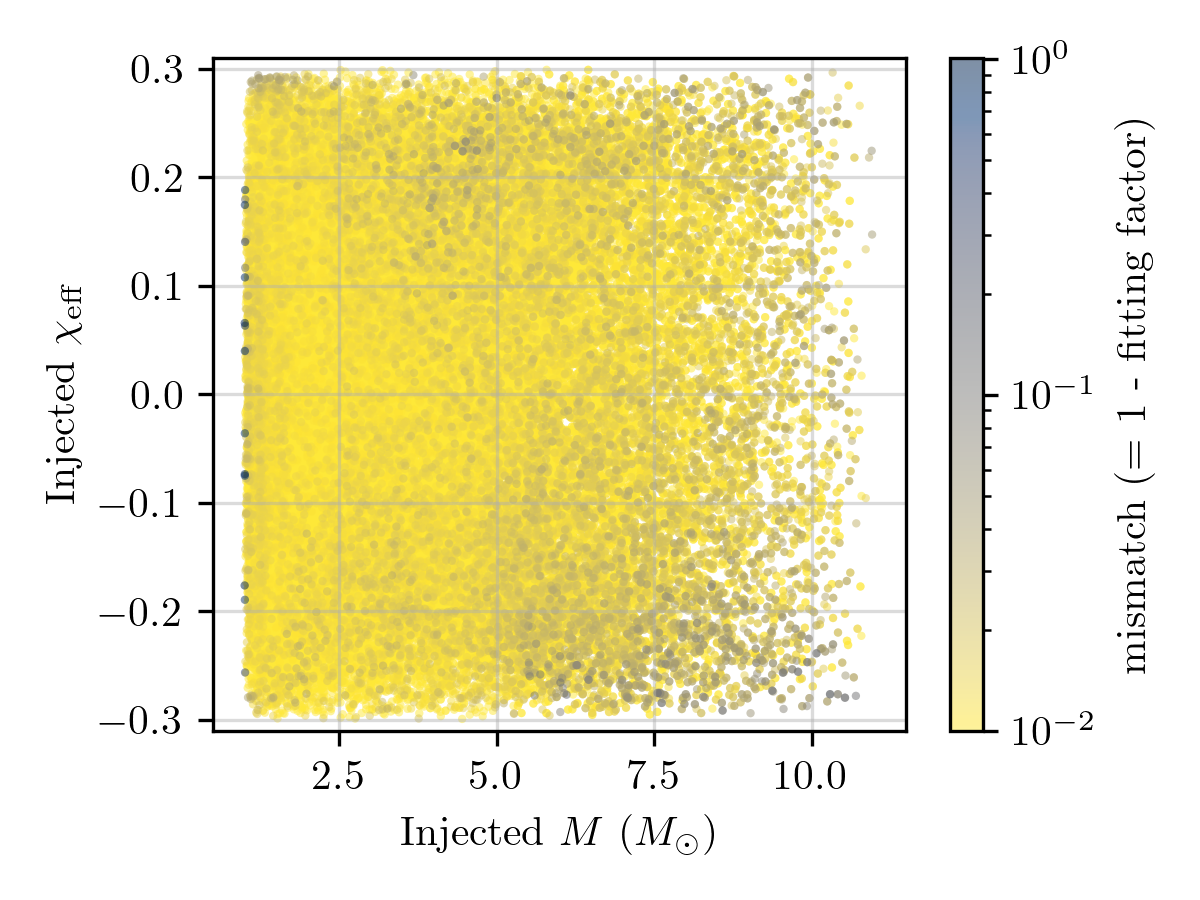}
        \caption{Mismatches in the $M - \chi_{\rm {eff}}$ plane.}\label{subfig:SSMBBHMchi}
    \end{subfigure}%
    \begin{subfigure}{0.24\textwidth}
        \includegraphics[height=3cm]{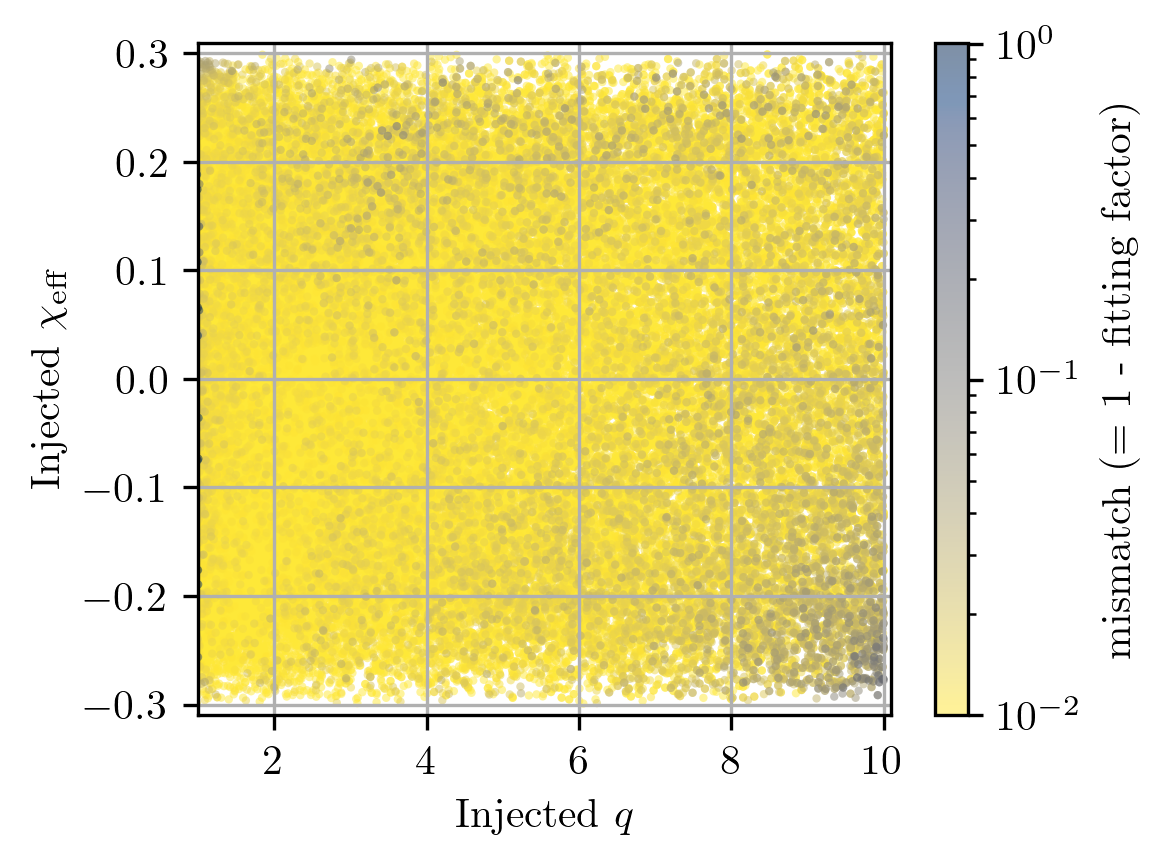}
        \caption{Mismatches in the $q - \chi_{\rm {eff}}$ plane.}\label{subfig:SSMBBHqchi}
    \end{subfigure}
    \caption{Plots for the \ac{SSM}-\ac{BBH} injections. (a) shows that $90 \%$ of the injections are recovered with a match $\geqslant 0.99$.}\label{fig:SSMBBHsims}
\end{figure*}

Since the bank is constructed with a required minimal match of $0.97$, the expected highest mismatch for most \ac{SSM}-\ac{BBH} signals is $\leq 0.03$. As shown in Fig. ~\ref{subfig:SSMBBHcumhist}, we find that $90 \%$ of the injections are recovered with a match of $0.99$. 
The bank is less effectual at the boundaries of the bank due to difficulties with template placement at the edges with \MANIFOLD. 
This is mitigated by extending the boundaries of the bank during construction beyond the desired target region.  

\begin{figure*}[htbp]
    \centering
    \begin{subfigure}{0.24\textwidth}
        \includegraphics[height=3cm]{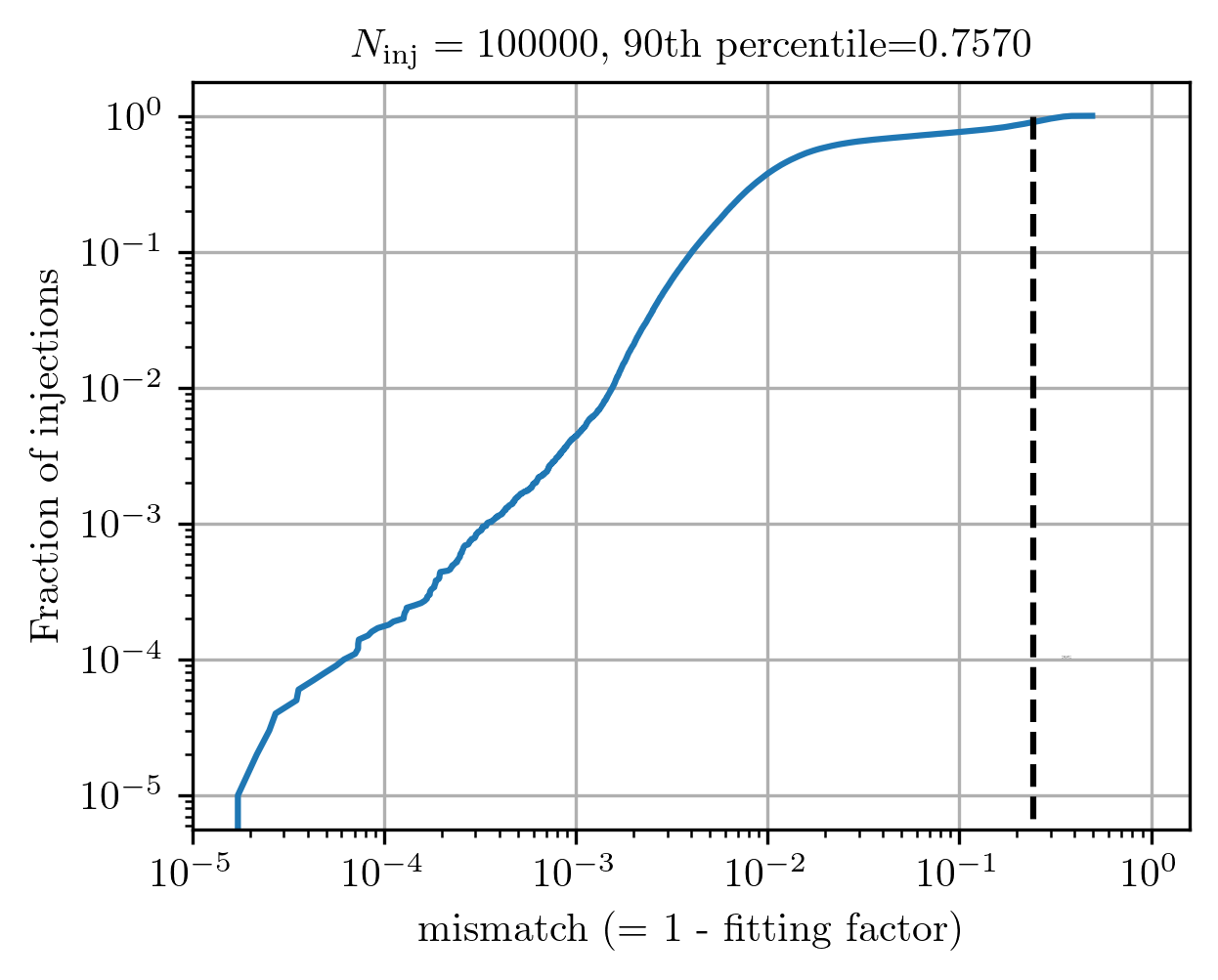}
        \caption{Cumulative histogram of mismatches.}\label{subfig:BNSLOWcumhist}
    \end{subfigure}%
    \begin{subfigure}{0.24\textwidth}
        \includegraphics[height=3cm]{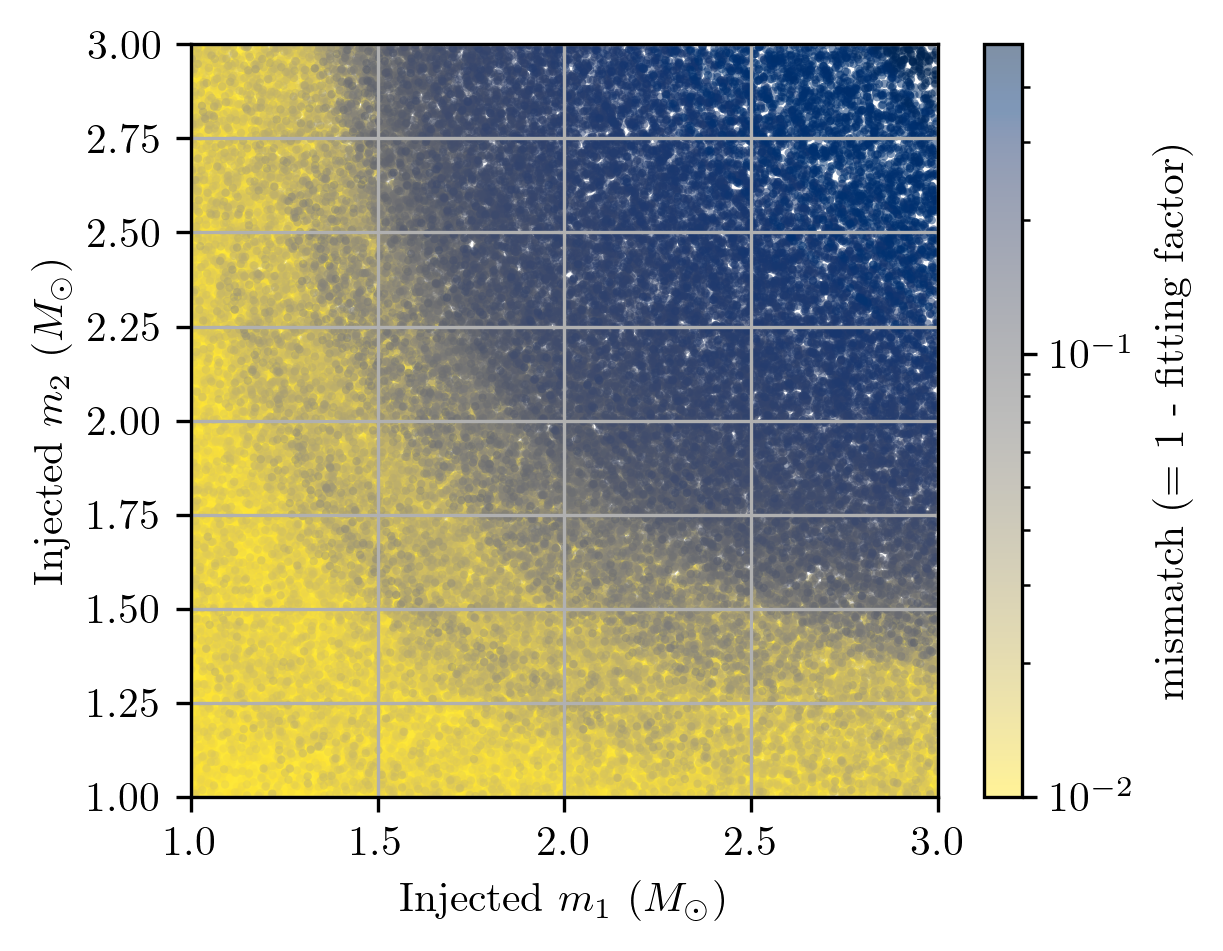}
        \caption{Mismatches in the $m_1 - m_2$ plane.}\label{subfig:BNSLOWm1m2}
    \end{subfigure}
	\begin{subfigure}{0.24\textwidth}
        \includegraphics[height=3cm]{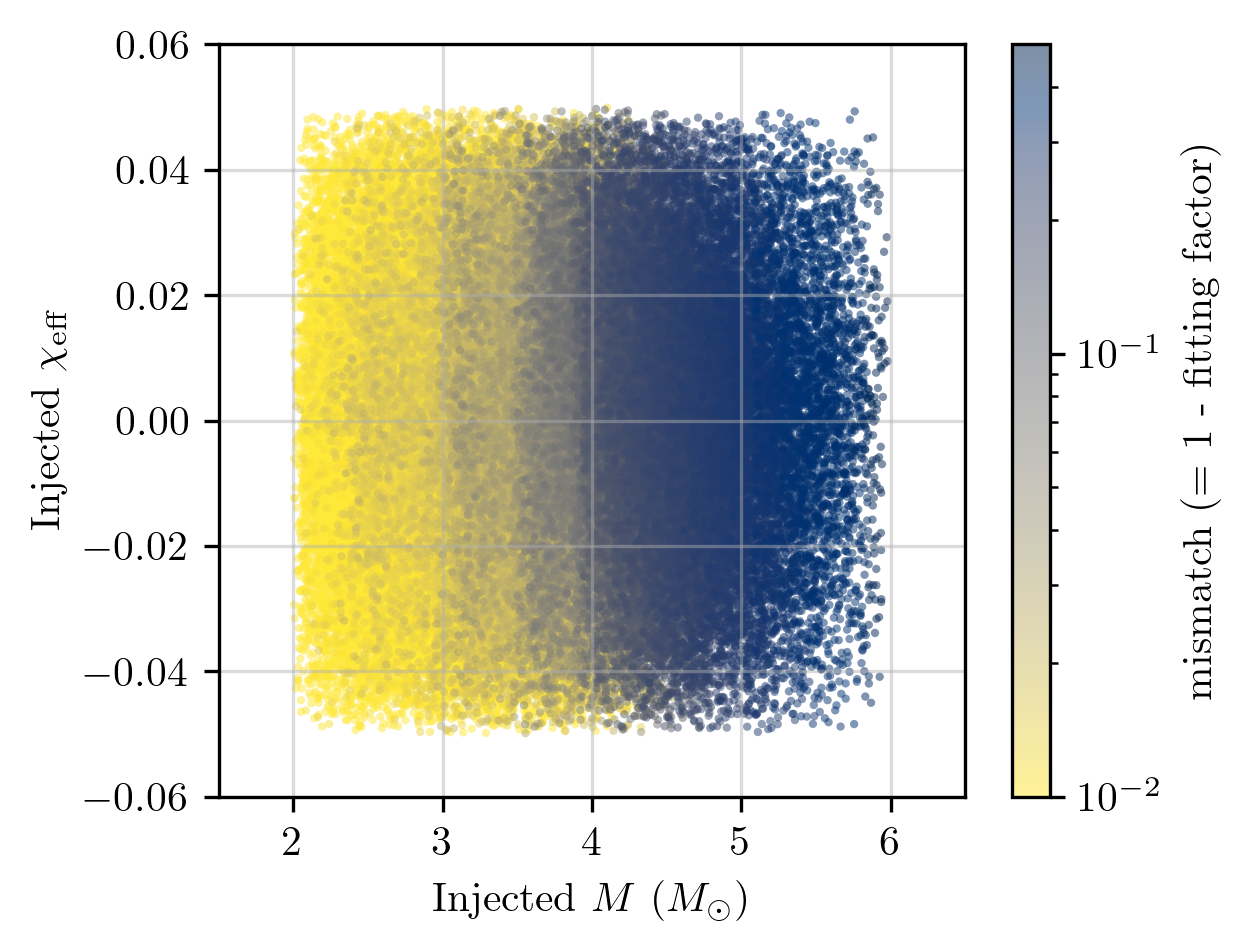}
        \caption{Mismatches in the $M - \chi_{\rm {eff}}$ plane.}\label{subfig:BNSLOWMchi}
    \end{subfigure}%
    \begin{subfigure}{0.24\textwidth}
        \includegraphics[height=3cm]{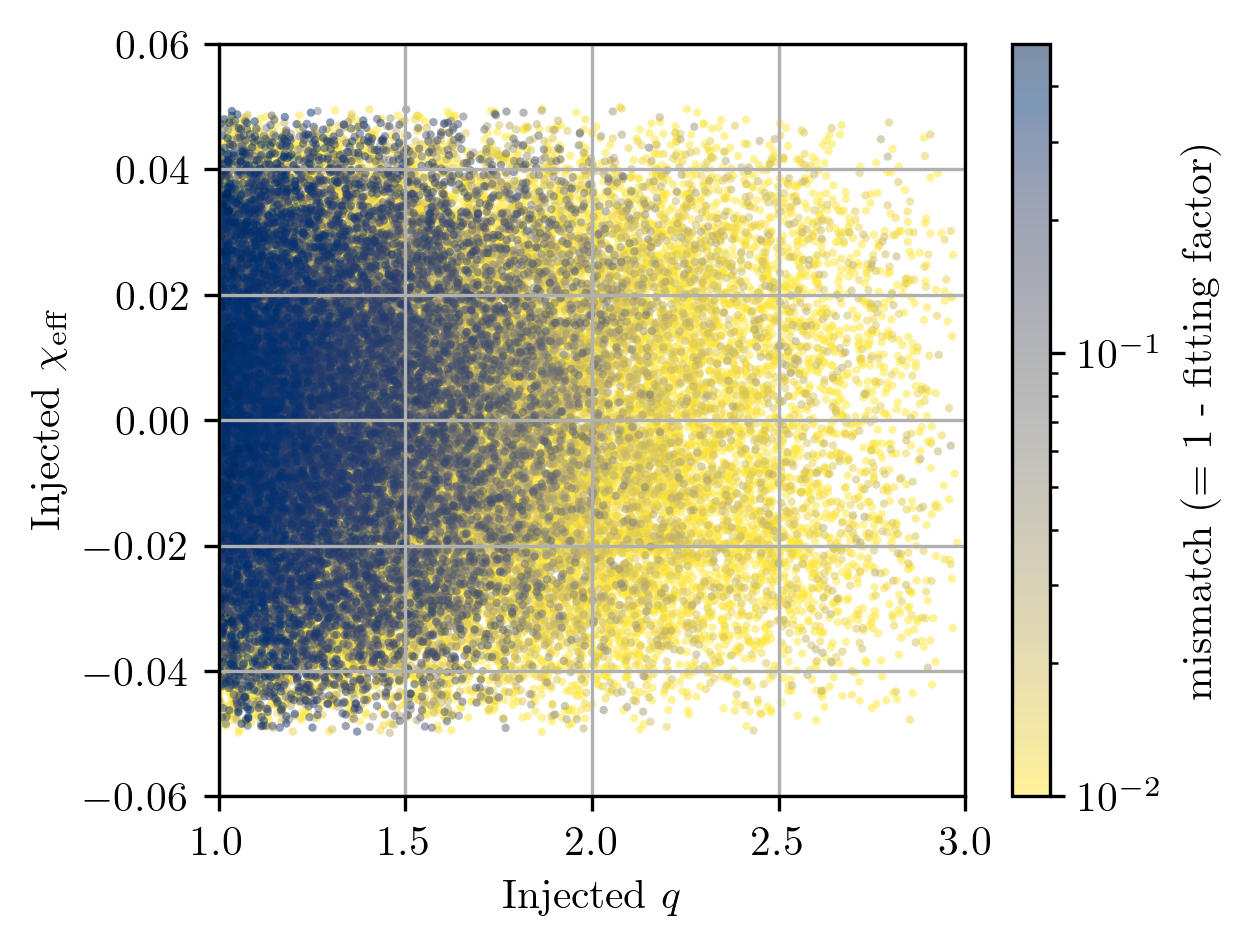}
        \caption{Mismatches in the $q - \chi_{\rm {eff}}$ plane.}\label{subfig:BNSLOWqchi}
    \end{subfigure}
    \caption{Plots for the \ac{BNS}-LOW injections. (a) shows that $90 \%$ of the injections are recovered with a match $\geqslant 0.76$.}\label{fig:BNSLOWsims}
\end{figure*}

\begin{figure*}[!htbp]
    \centering
    \begin{subfigure}{0.24\textwidth}
        \includegraphics[height=3cm]{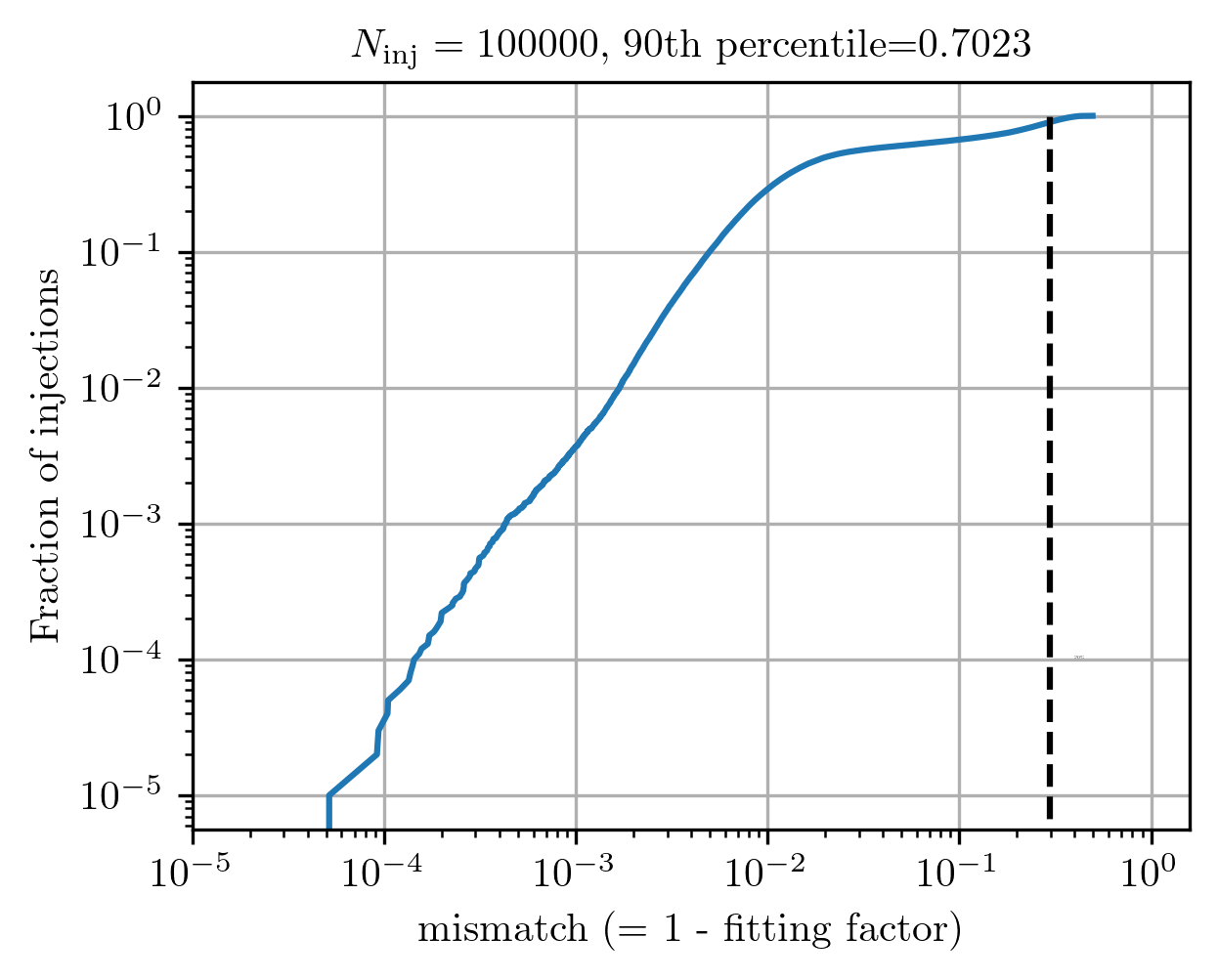}
        \caption{Cumulative histogram of mismatches.}\label{subfig:BNSHIGHcumhist}
    \end{subfigure}%
    \begin{subfigure}{0.24\textwidth}
        \includegraphics[height=3cm]{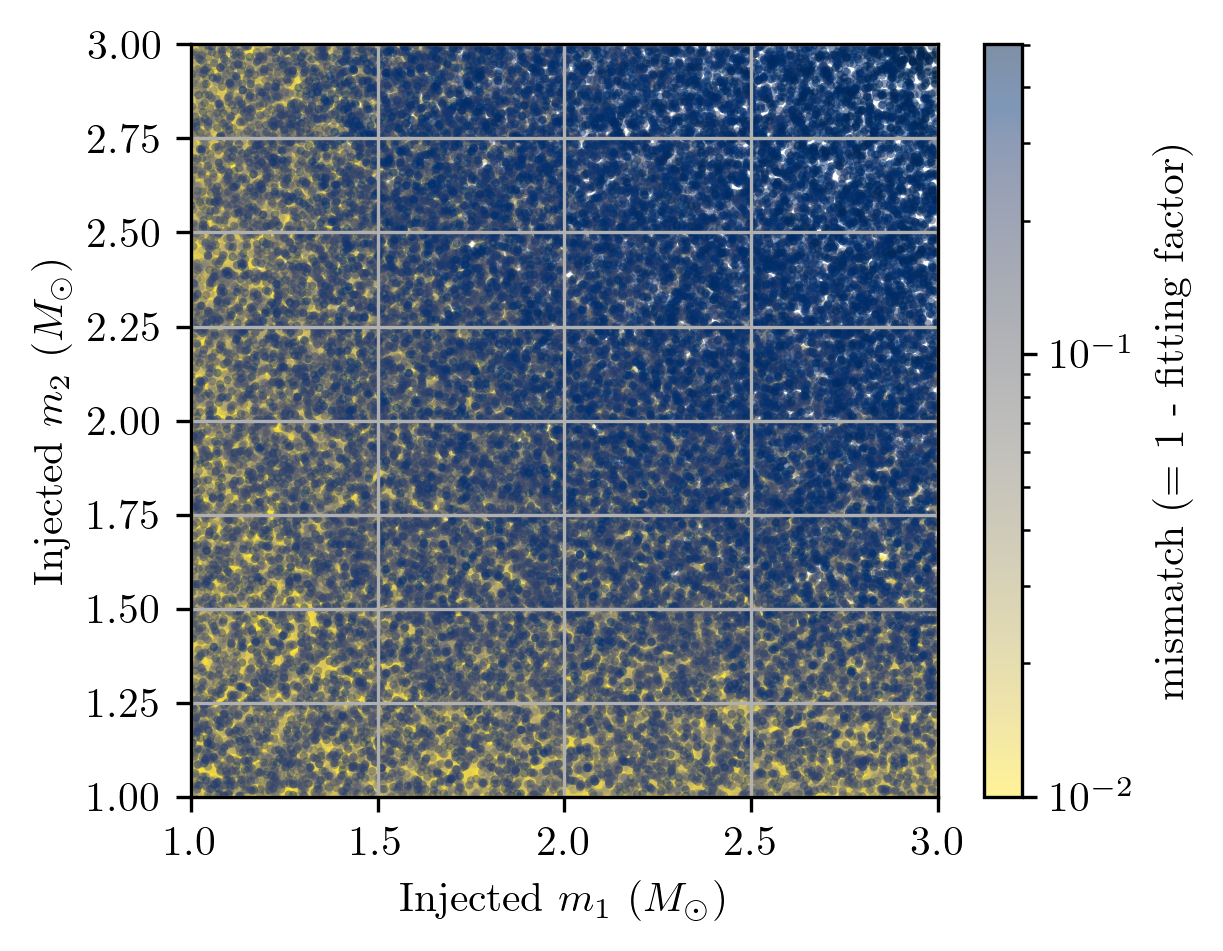}
        \caption{Mismatches in the $m_1 - m_2$ plane.}\label{subfig:BNSHIGHm1m2}
    \end{subfigure}
	\begin{subfigure}{0.24\textwidth}
        \includegraphics[height=3cm]{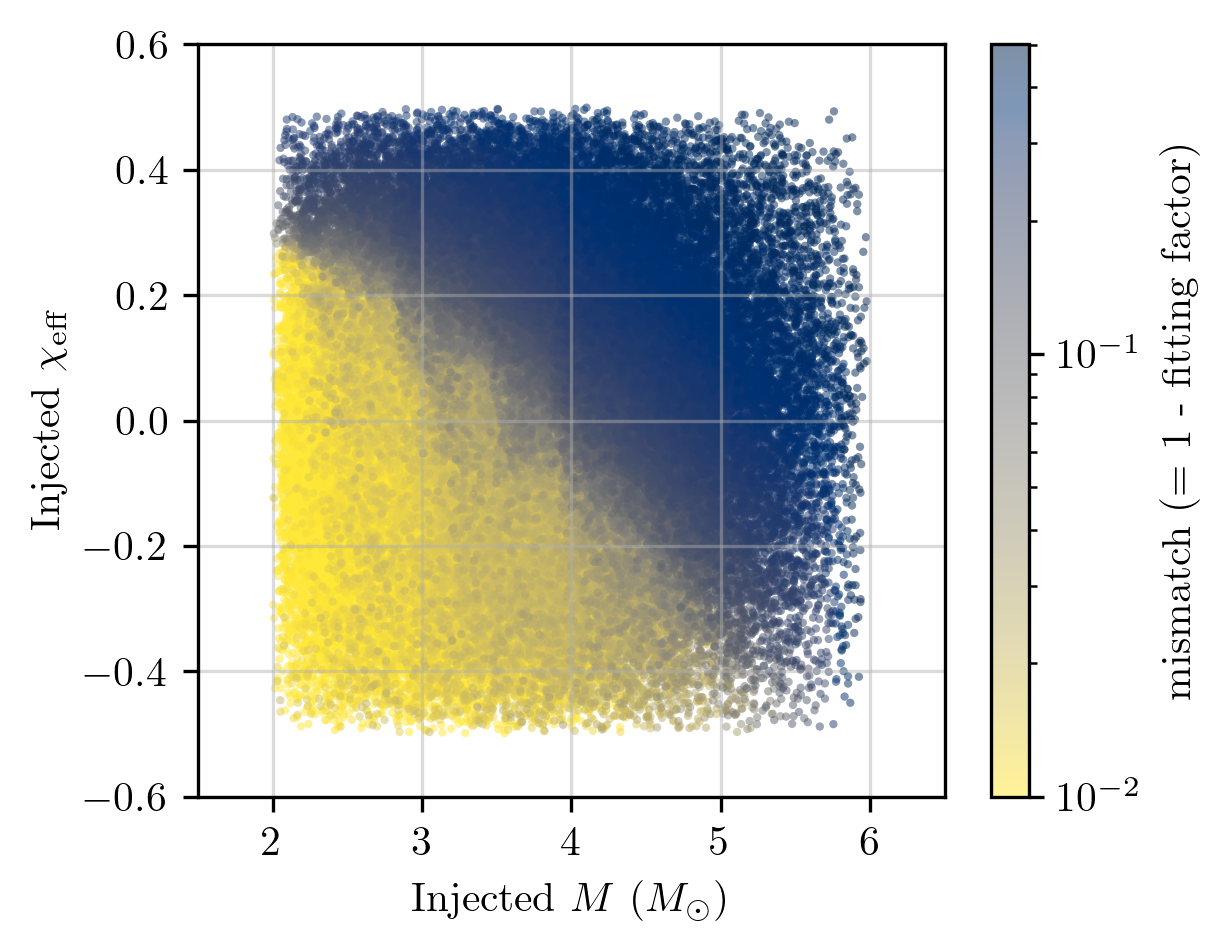}
        \caption{Mismatches in the $M - \chi_{\rm {eff}}$ plane.}\label{subfig:BNSHIGHMchi}
    \end{subfigure}%
    \begin{subfigure}{0.24\textwidth}
        \includegraphics[height=3cm]{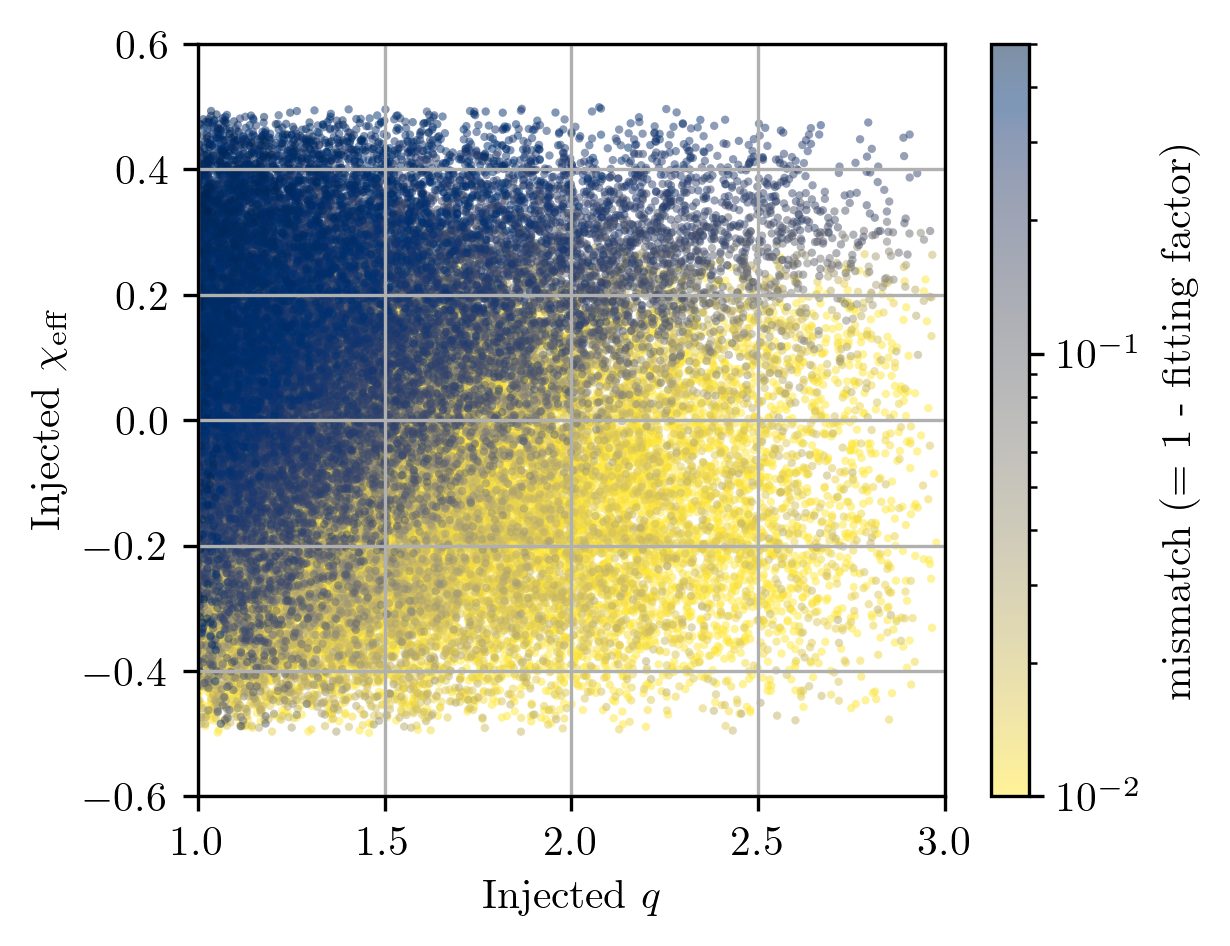}
        \caption{Mismatches in the $q - \chi_{\rm {eff}}$ plane.}\label{subfig:BNSHIGHqchi}
    \end{subfigure}
    \caption{Plots for the \ac{BNS}-HIGH injections. (a) shows that $90 \%$ of the injections are recovered with a match $\geqslant 0.70$.}\label{fig:BNSHIGHsims}
\end{figure*}
\begin{figure*}[!htbp]
    \centering
    \begin{subfigure}{0.24\textwidth}
        \includegraphics[height=3cm]{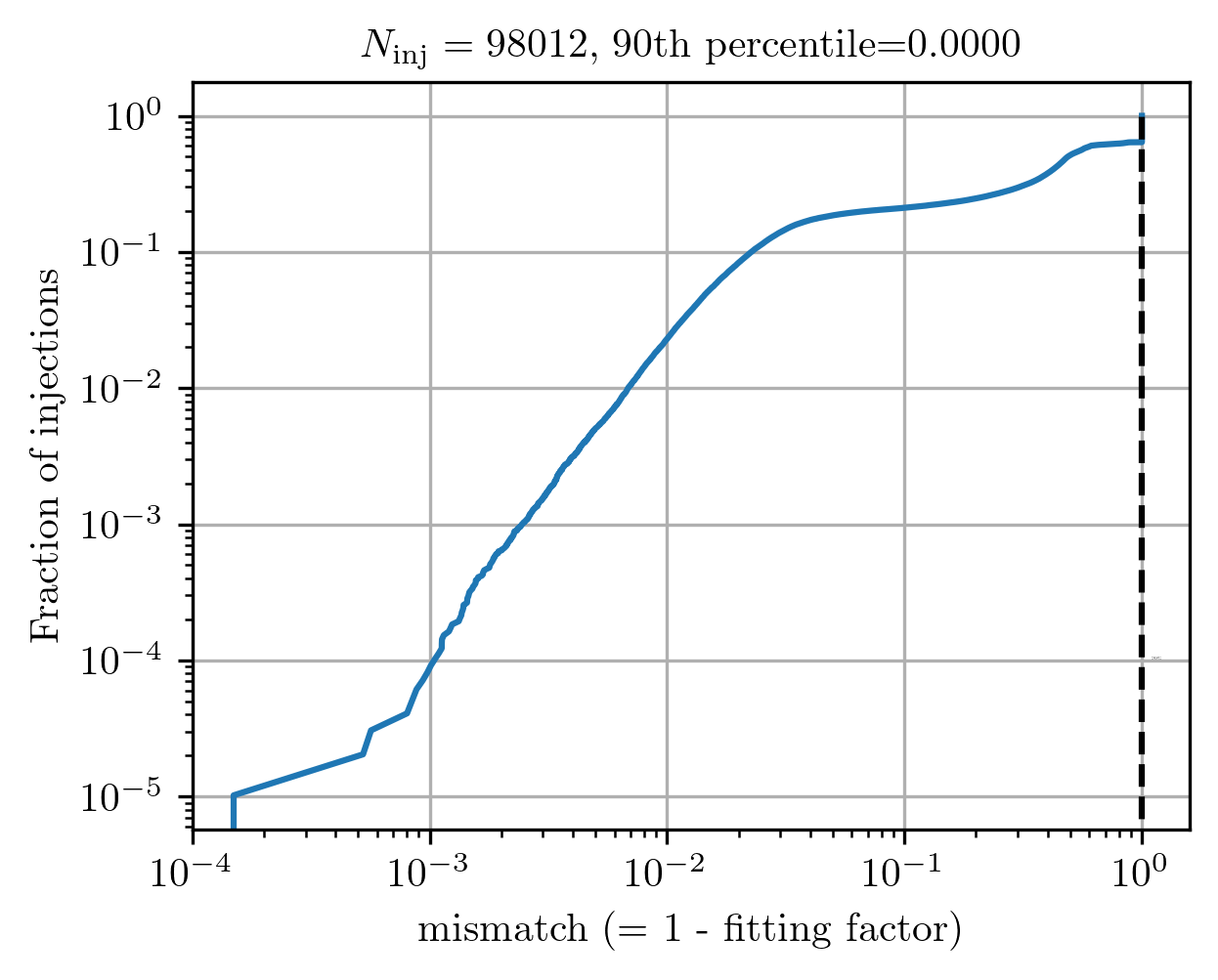}
        \caption{Cumulative histogram of mismatches.}\label{subfig:NSBHLOWcumhist}
    \end{subfigure}%
    \begin{subfigure}{0.24\textwidth}
        \includegraphics[height=3cm]{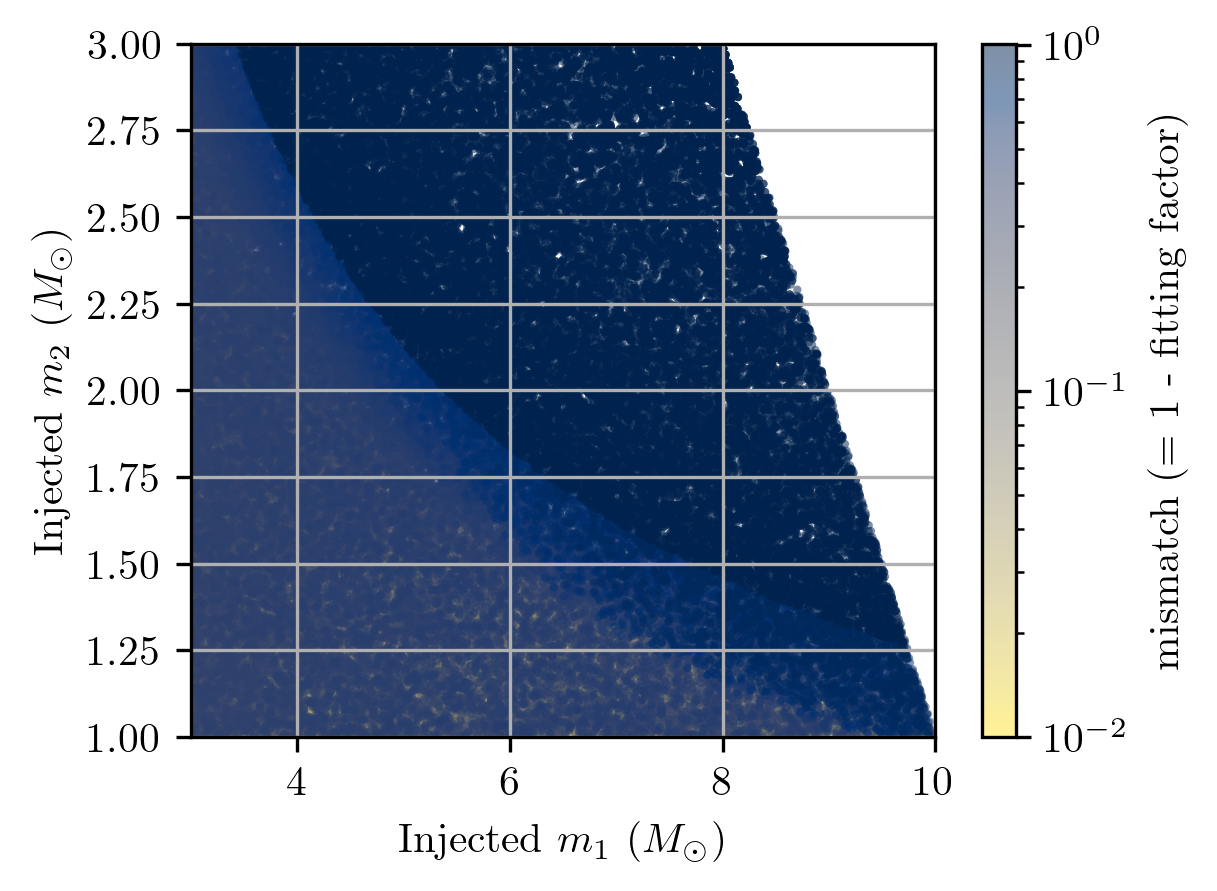}
        \caption{Mismatches in the $m_1 - m_2$ plane.}\label{subfig:NSBHLOWm1m2}
    \end{subfigure}
	\begin{subfigure}{0.24\textwidth}
        \includegraphics[height=3cm]{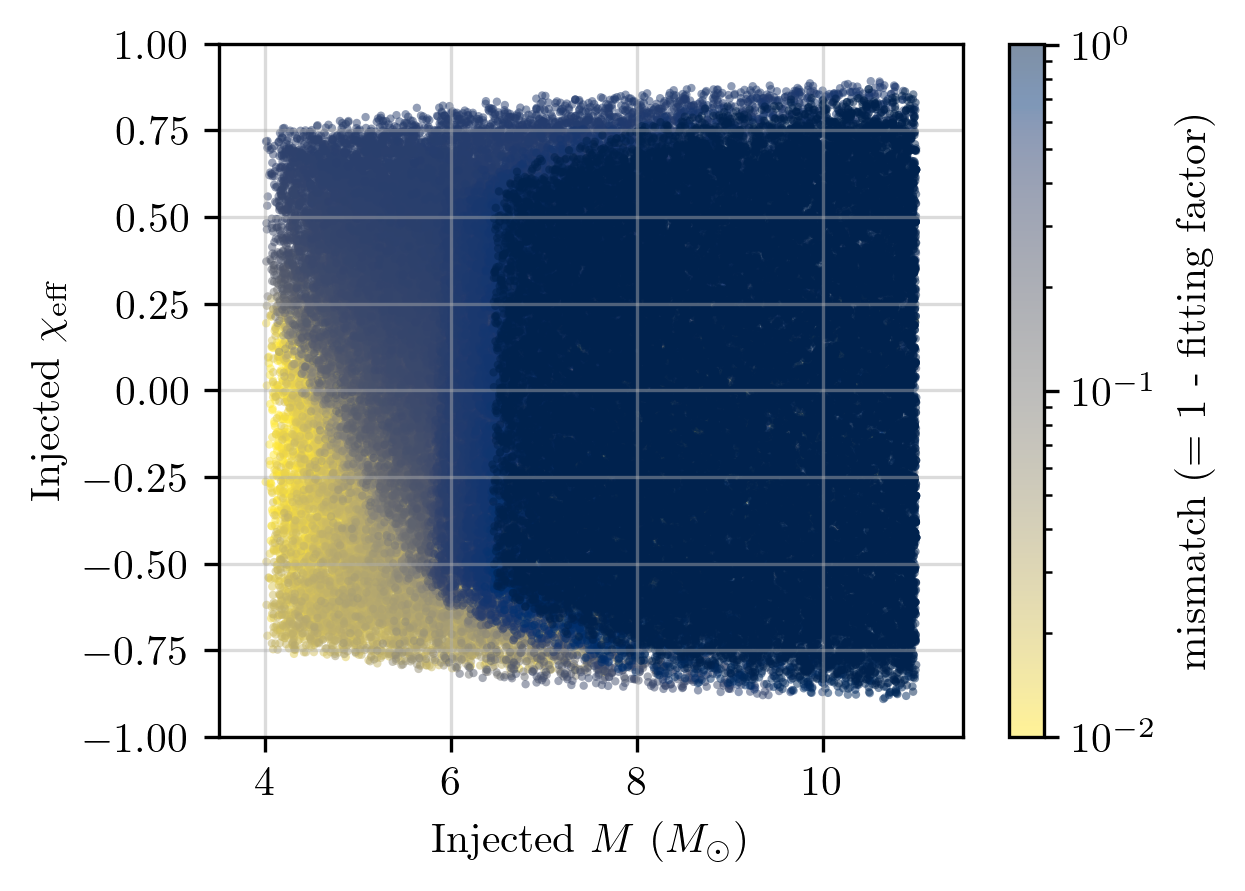}
        \caption{Mismatches in the $M - \chi_{\rm {eff}}$ plane.}\label{subfig:NSBHLOWMchi}
    \end{subfigure}%
    \begin{subfigure}{0.24\textwidth}
        \includegraphics[height=3cm]{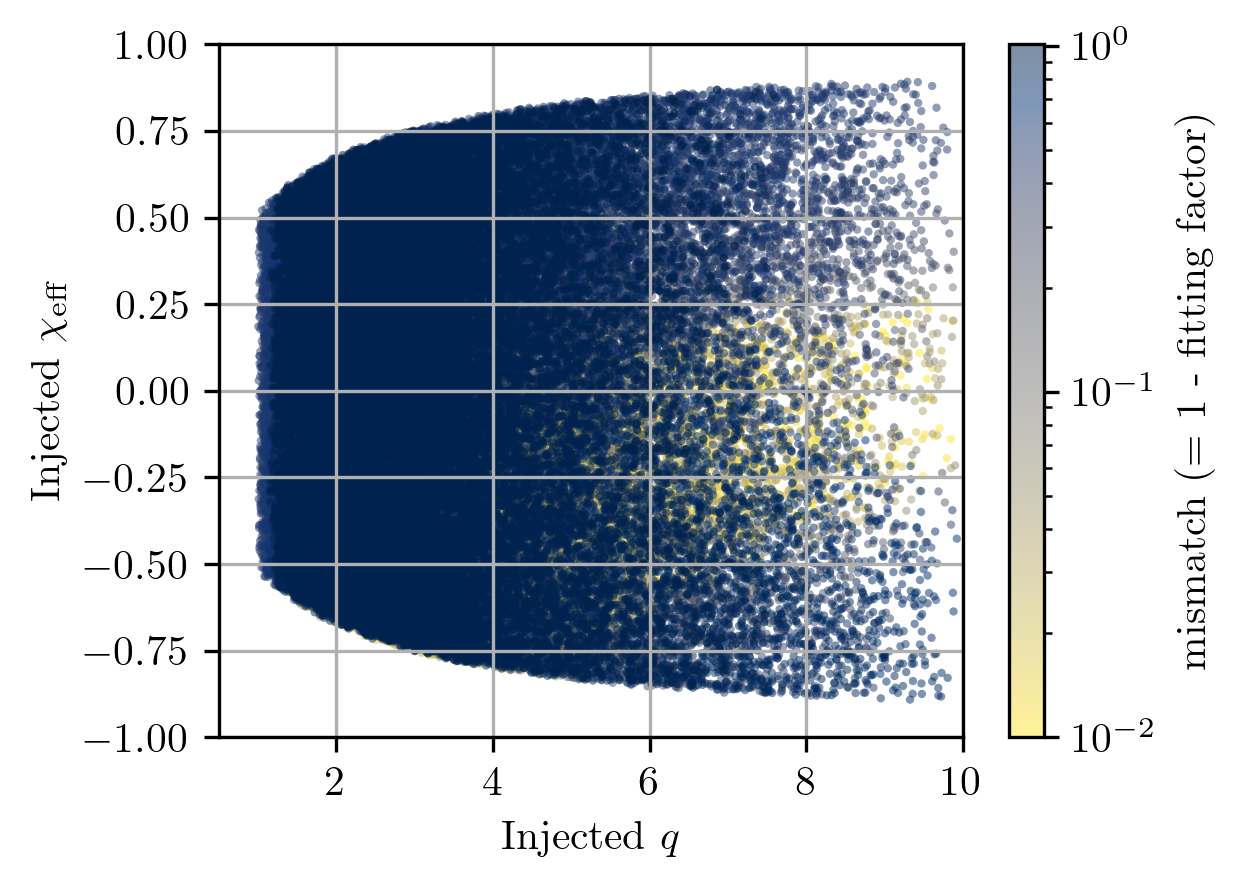}
        \caption{Mismatches in the $q - \chi_{\rm {eff}}$ plane.}\label{subfig:NSBHLOWqchi}
    \end{subfigure}
    \caption{Plots for the \ac{NSBH}-LOW injections. (a) shows that only $10\%$ injections are recovered with a mismatch $\leqslant 0.03$.}\label{fig:NSBHLOWsims}
\end{figure*}

\begin{figure*}[!htbp]
    \centering
    \begin{subfigure}{0.24\textwidth}
        \includegraphics[height=3cm]{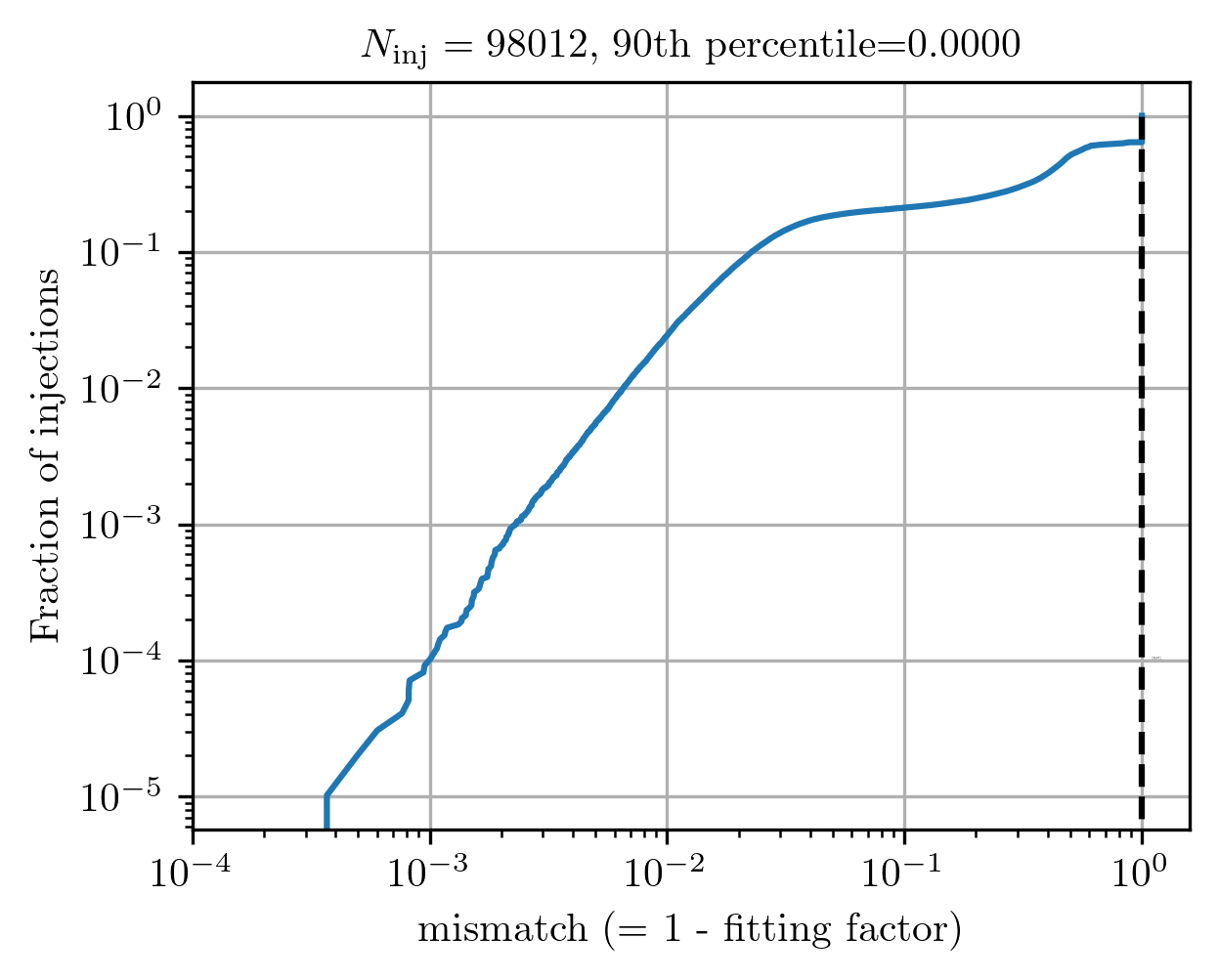}
        \caption{Cumulative histogram of mismatches.}\label{subfig:NSBHHIGHcumhist}
    \end{subfigure}%
    \begin{subfigure}{0.24\textwidth}
        \includegraphics[height=3cm]{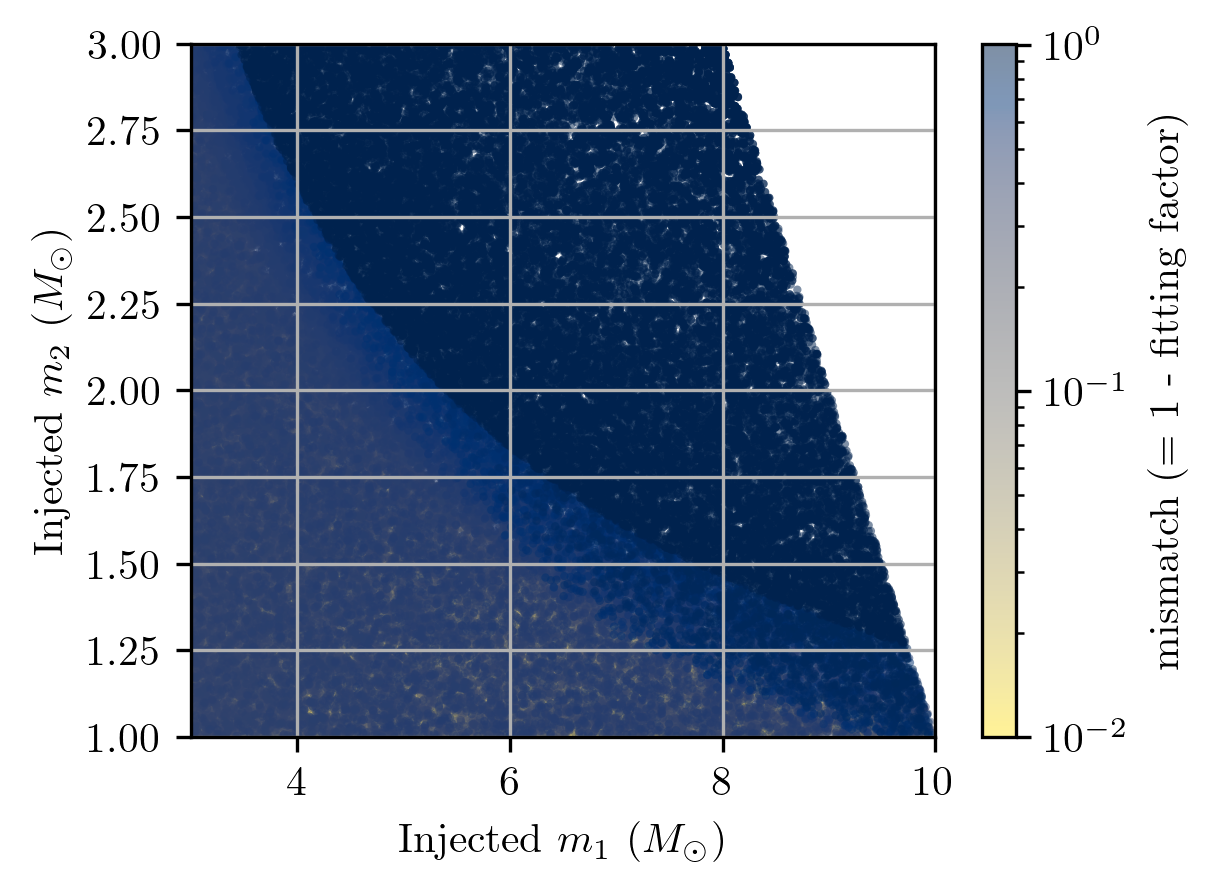}
        \caption{Mismatches in the $m_1 - m_2$ plane.}\label{subfig:NSBHHIGHm1m2}
    \end{subfigure}
	\begin{subfigure}{0.24\textwidth}
        \includegraphics[height=3cm]{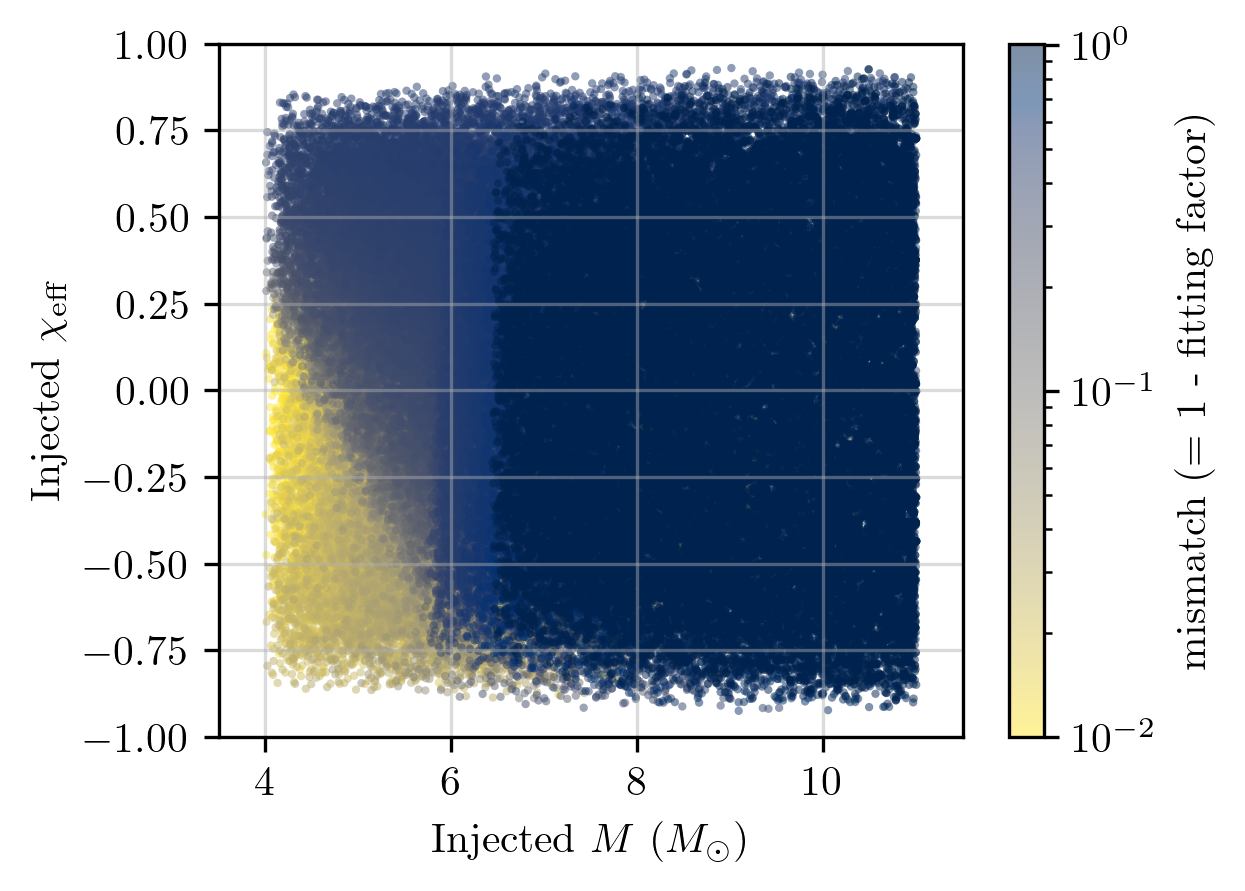}
        \caption{Mismatches in the $M - \chi_{\rm {eff}}$ plane.}\label{subfig:NSBHHIGHMchi}
    \end{subfigure}%
    \begin{subfigure}{0.24\textwidth}
        \includegraphics[height=3cm]{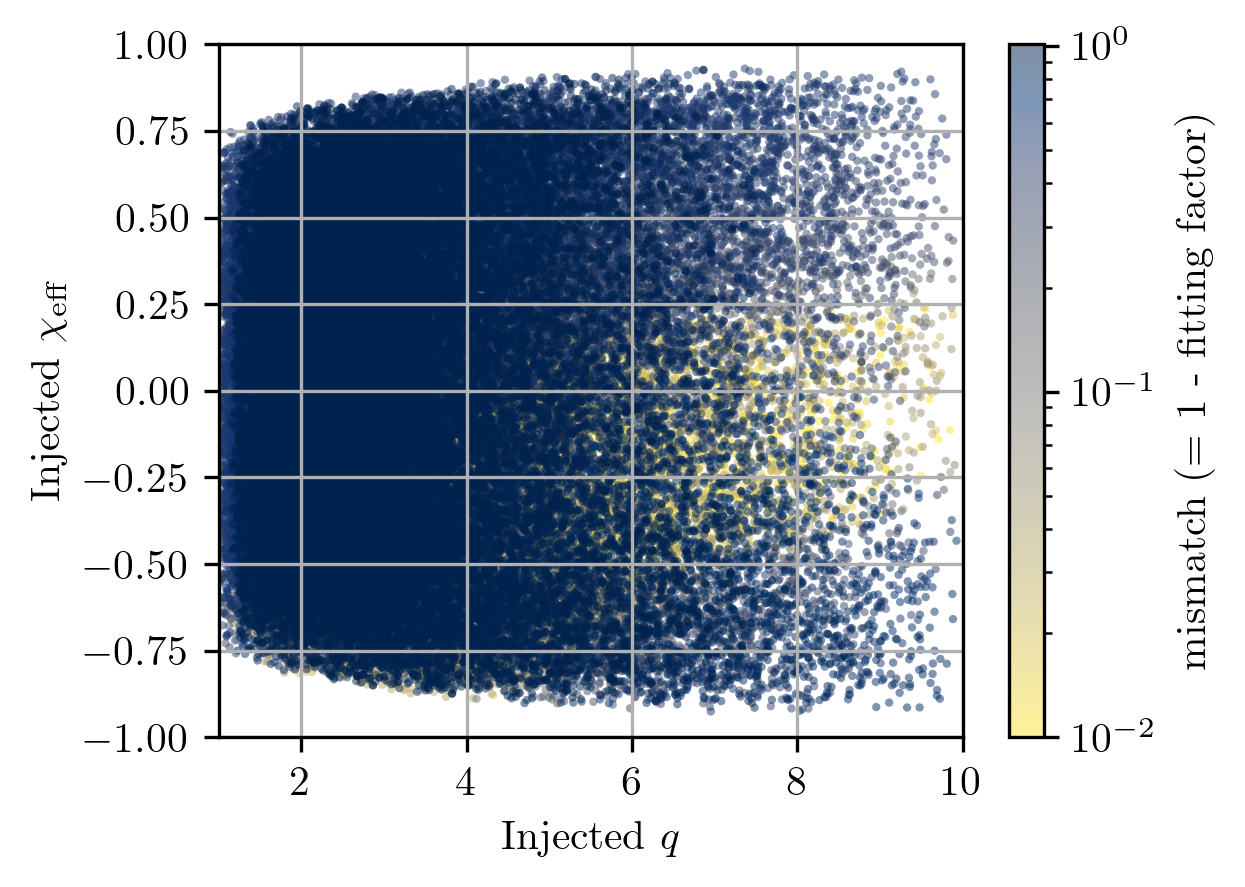}
        \caption{Mismatches in the $q - \chi_{\rm {eff}}$ plane.}\label{subfig:NSBHHIGHqchi}
    \end{subfigure}
    \caption{Plots for the \ac{NSBH}-HIGH injections. (a) shows that only $10\%$ injections are recovered with a mismatch $\leqslant 0.03$.}\label{fig:NSBHHIGHsims}
\end{figure*}

\ac{BNS}-LOW and \ac{NSBH}-LOW represent astrophysically expected binary systems where the \acp{NS} are extremely low-spinning when they occur in a binary. 
In the \ac{BNS}-LOW regime, this bank effectively recovers $90 \%$ of the injections with a match $\geq 0.76$. Fig. ~\ref{fig:BNSLOWsims} shows that more asymmetric \ac{BNS} with low total mass, $M$ are detectable with this template bank irrespective of $\chi_{\rm{eff}}$.
For typical \ac{NSBH} systems, \ac{NSBH}-LOW, the bank performs sub-optimally with most injections lying completely outside the target space of the search.
Only $10 \%$ of the injections have mismatches $\leqslant 0.03$, and lie mostly in $-0.75 \leq \chi_{\rm{eff}} \leq 0.25$ and $4~M_\odot \leq M \leq 7~M_\odot$ with a cut-off at $\mathcal{M} \approx 2.5~M_\odot$ which is the maximum chirp mass in the template bank.
Here, $M$ is the total mass, $M = m_1 + m_2$. 

The low spin limit for \acp{NS} in binaries is primarily informed by observations of the fastest-spinning pulsars in BNS systems which are expected to merge within Hubble time -- PSR J0737- 3039A ~\cite{Lyne:2006} and PSR J1946+2052 ~\cite{Stovall:2018}.
These systems are expected to evolve to $\chi < 0.04$ and $\chi < 0.05$ at the time of the merger.
However, it is within the realm of possibility that millisecond pulsars form binaries but these have not yet been observed. 
In ~\cite{Chattopadhyay:2021}, it is also shown that if the \ac{NS} forms before the \ac{BH} in an \ac{NSBH}, the resultant spins could be much higher than $0.05$.
Considering that the \ac{BNS}/\ac{NSBH}/\ac{BBH} searches do not target high-spinning \acp{NS} in binaries and the \ac{O4} template bank used by \GSTLAL is ineffectual in recovering such high spin \acp{BNS} ~\cite{Sakon:2024}, the \ac{SSM} search provides a unique opportunity to detect previously unexplored \acp{NS} with high spins. 
Therefore, we perform bank simulations with \ac{BNS}-HIGH and \ac{NSBH}-HIGH signals shown in Fig. ~\ref{fig:BNSHIGHsims} and Fig. \ref{fig:NSBHHIGHsims}.
We find that $90 \%$ of \ac{BNS}-HIGH injections are recovered with a match of $0.7$.
On the other hand, the bank is not effectual in the \ac{NSBH}-HIGH regime which is evident in Fig. ~\ref{subfig:NSBHHIGHcumhist}.
Similar to \ac{NSBH}-LOW, only $10 \%$ of the injections have mismatches $\leqslant 0.03$, which mostly have $-0.8 \leq \chi_{\rm{eff}} \leq 0.25$ and $4~M_\odot \leq M \leq 7~M_\odot$ with a cut-off at $\mathcal{M} \approx 2.5~M_\odot$ which is the maximum chirp mass in the template bank.

\subsection{Population model}\label{subsec:massmodel}

\begin{figure}[!h]
    \centering
    \includegraphics[width=\linewidth]{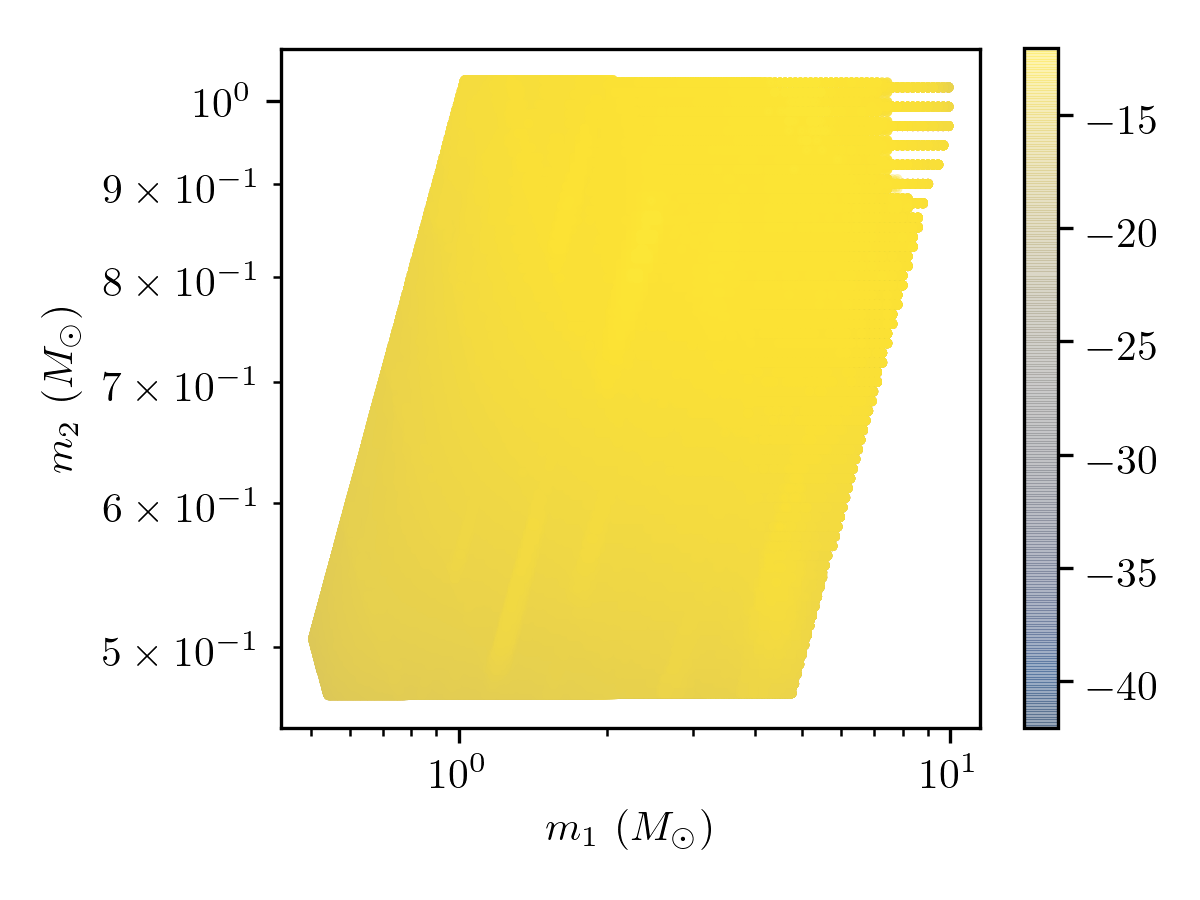}
	\caption{Template weights corresponding to the uniform population distribution, as a function of component masses. The color bars represent $\rm{log}~P(t_k,\rho_k | H_{\rm {ssm}})$ for $\rho_k = 10$.}
	\label{fig:massmodel}
\end{figure}

The population model was pregenerated using \MANIFOLD.
An uninformative prior is used to generate the population model ensuring that all templates in the bank carry uniform weight due to a lack of prior detections.
We disregard expected rates from models that predict \ac{SSM} compact objects to remove any bias from the search. 
The population model used in this search is shown in Fig. ~\ref{fig:massmodel}.
This model is only used in likelihood estimation here, since we do not yet assign source classification probabilities to events from this search.

\clearpage

\bibliography{references}

\end{document}